\documentclass[fleqn,usenatbib]{mnras}

\usepackage{newtxtext,newtxmath}

\usepackage[T1]{fontenc}

\DeclareRobustCommand{\VAN}[3]{#2}
\let\VANthebibliography\thebibliography
\def\thebibliography{\DeclareRobustCommand{\VAN}[3]{##3}\VANthebibliography}

\usepackage{graphicx}	
\usepackage{amsmath}	

\usepackage{multirow}
\usepackage{multicol}
\usepackage{supertabular}
\usepackage{longtable}

\def\msun{{\rm M}_{\sun}}

\title[MUSE strong-lensing models for 3 galaxy clusters]{Galaxy cluster cores as seen with VLT/MUSE: 
new strong-lensing analyses of RX\,J2129.4$+$0009, MS\,0451.6$-$0305 \& MACS\,J2129.4$-$0741}

\author[Jauzac et al.\ 2020]{
Mathilde Jauzac,$^{1,2,3,4,5}$\thanks{E-mail: mathilde.jauzac@durham.ac.uk}
Baptiste Klein,$^{5,6}$
Jean-Paul Kneib,$^{5,7}$
Johan Richard,$^{8}$
\newauthor
Markus Rexroth,$^{5}$
Christoph Schäfer,$^{5}$ and
Aurélien Verdier$^{5}$
\\
\\
$^{1}$Centre for Extragalactic Astronomy, Durham University, South Road, Durham DH1 3LE, UK\\
$^{2}$Institute for Computational Cosmology, Durham University, South Road, Durham DH1 3LE, UK\\
$^{3}$Astrophysics Research Centre, University of KwaZulu-Natal, Westville Campus, Durban 4041, South Africa\\
$^{4}$School of Mathematics, Statistics \& Computer Science, University of KwaZulu-Natal, Westville Campus, Durban 4041, South Africa\\
$^{5}$ Institute of Physics, Laboratory of Astrophysics, Ecole Polytechnique F\'ed\'erale de Lausanne (EPFL), Observatoire de Sauverny, 1290 Versoix, Switzerland\\
$^{6}$Sub-department of Astrophysics, Department of Physics, University of Oxford, Oxford OX1 3RH, UK\\
$^{7}$Aix Marseille Universit\'e, CNRS, LAM (Laboratoire d'Astrophysique de Marseille) UMR 7326, 13388, Marseille, France\\
$^{8}$Univ Lyon, Univ Lyon1, Ens de Lyon, CNRS, Centre de Recherche Astrophysique de Lyon UMR5574, F-69230, Saint-Genis-Laval, France
}

\date{Accepted XXX. Received YYY; in original form ZZZ}

\pubyear{2015}

\begin{document}
\label{firstpage}
\pagerange{\pageref{firstpage}--\pageref{lastpage}}
\maketitle

\begin{abstract}
We present strong-lensing analyses of three galaxy clusters, RX\,J2129.4$+$0009 (z$=$0.235), MS\,0451.6$-$0305 (z$=$0.55), and MACS\,J2129.4$-$0741 (z$=$0.589), using the powerful combination of \emph{Hubble Space Telescope} (\emph{HST}) multi-band observations, and Multi-Unit Spectroscopic Explorer (MUSE) spectroscopy.
In RX\,J2129, we newly spectroscopically confirm 15 cluster members. Our resulting mass model uses 8 multiple image systems as we include a galaxy-galaxy lensing system North-East of the cluster, and is composed of 71 halos including one dark matter cluster-scale halo and 2 galaxy-scale halos optimized individually.
For MS\,0451, we report the spectroscopic identification of 2 new systems of multiple images in the Northern region, and 112 cluster members. Our mass model uses 16 multiple image systems, and 146 halos, including 2 large-scale halos, and 7 galaxy-scale halos independently optimized.
For MACS\,J2129, we report the spectroscopic identification of one new multiple image system at $z=4.41$, and newly measure spectroscopic redshifts for 4 cluster members.
Our mass model uses 14 multiple image systems, and is composed of 151 halos, including 2 large-scale halos and 4 galaxy-scale halos independently optimized. 
Our best models have rms of 0.29\arcsec,0.6\arcsec, 0.74\arcsec\ in the image plane for RX\,J2129, MS\,0451, and MACS\,J2129 respectively. 
This analysis presents a detailed comparison with the existing literature showing excellent agreements, and discuss specific studies of lensed galaxies, e.g. a group of submilimeter galaxies at $z=2.9$ in MS\,0451, and a bright $z=2.1472$ red singly imaged galaxy in MACS\,J2129.
\end{abstract}



\section{Introduction}
Clusters of galaxies are the most spectacular strong lenses in the Universe. Due to the high mass density in their cores, it is not uncommon to observe giant arcs and multiple images of sources located behind them. 
This gravitational lensing effect distorts, and magnifies the light emitted by background galaxies, transforming these clusters into cosmic telescopes \citep[for a review see e.g.][]{massey2010,kneib2011,hoekstra2013,treu2015,kilbinger2015,bartelmann2017}.
Such strong-lensing features are extremely useful to map the total mass distribution within the central regions of clusters \citep[e.g.][]{richard2014,coe14,jauzac2014,johnson2014,caminha2017a,williams18,diego2018,mahler2018,lagattuta2019,sharon2020}. These mass models can then be used to constrain the physics of Dark Matter, such as its self-interaction cross section \citep[see][]{harvey2014,harvey2015,counterdavid}, or test the cosmological paradigm \citep[e.g.][]{jullo2010,acebron,natarajan2017,jauzac2018b}, but also to probe the early Universe and thus the epoch of reionization \citep[e.g.][]{atek2015,earlyuni2,livermore2017,atek2018,ishigaki2018}. Hence the need to have accurate and precise mass models for a large number of galaxy clusters.

Over the past two decades, the precision of strong-lensing mass modeling of galaxy clusters has dramatically increased. This is mainly due to the combination of powerful post-processing algorithms and high resolution imaging. On the one hand, the use of the Markov Chain Monte Carlo (MCMC) sampling of the parameter space in the Bayesian framework allowed for robust estimations of the most likely mass models for a given set of constraints \citep[e.g. our team uses \textsc{Lenstool} which is presented in][]{kneib1996,jullo2007,jullo2011,niemec2020}. On the other hand, the quality of observations with the \emph{Hubble Space Telescope} (\emph{HST}) provided astronomers with the deepest and highest resolution images of strong-lensing clusters \citep[e.g.][see also the webpages of \emph{Hubble} Frontier Fields\footnote{\url{https://frontierfields.org/}}, GLASS\footnote{\url{http://glass.astro.ucla.edu/}}, RELICS\footnote{\url{https://relics.stsci.edu/}} and BUFFALO\footnote{\url{https://buffalo.ipac.caltech.edu/}} surveys]{postmanclash,schmidt2014,treu_glass,hff,steinhardt2020}. This resulted in mass models with an unrivalled precision for numerous galaxy clusters \citep[e.g.][]{zitrin2011,richard2014,johnson2014,jauzacmacsj0416,diego2017}.
High resolution imaging allows precise measurements of the location of multiple images. However, if their exact distance, i.e. their redshift, is not measured, then mass models are highly degenerate, and the resulting mass distribution is biased, hence the strong need for measurements of spectroscopic redshifts \citep[$\Delta z < 0.01$, see ][]{lagatutta,richard2015,grillo2016,jauzac2016a,jauzac2019,mahler2018,mahler19,lagattuta2019,gonzalez2019}. As shown in \cite{johnson2016}, \cite{cerny2018}, and \cite{gonzalez2019}, the spectroscopic redshift information is mandatory in order to obtain precise strong-lensing mass models.

\begin{table*}
\centering
\caption{\label{cluster_infos} Summary of MUSE observations for all three clusters. Columns~1 to~3 indicate respectively the name of the cluster, its average redshift, and the ID of the ESO programme (PI: Kneib for all observations). For each pointing, we give the observation date in column~4, the total exposure time in column~5, the right ascension (R.\,A.) and declination (Decl.) of the center of the field of view in columns~6 and~7, and the FWHM of the seeing during the observations in column~8.}
\begin{tabular}{cccccccc}
\hline
\hline
Cluster & $z$ & ESO program & Observation date & Exposure time  &  R.\,A. & Decl. & Seeing \\
               & & & & [s] & [J2000] & [J2000] & [\arcsec] \\
\hline
\hline
\multirow{2}{*}{RX\,J2129} & \multirow{2}{*}{0.235} & \multirow{2}{*}{097.A-0909(A)} & 2016-08-05 & \multirow{2}{*}{8940} &  $322.42149$ & $0.09310$ & \multirow{2}{*}{0.5} \\ 
& & & 2016-09-04 &  & $322.41189$ & $0.08682$ &  \\
\hline
\multirow{2}{*}{MS\,0451} & \multirow{2}{*}{0.55} & \multirow{2}{*}{096.A-0105(A)} & 2016-01-10 & \multirow{2}{*}{8682} &  $73.55165$ & $-3.01837$ & \multirow{2}{*}{0.8} \\ 
& & & 2016-01-11 &  & $73.54106$ & $-3.00965$ & \\
\hline
\multirow{2}{*}{MACS\,J2129} & \multirow{2}{*}{0.589} & \multirow{2}{*}{095.A-0525(A)} & \multirow{2}{*}{2015-06-17} & \multirow{2}{*}{8772} &  $322.36602$ & $-7.69040$ & \multirow{2}{*}{0.9} \\ 
& & & & & $322.35192$ & $-7.69040$ &  \\
\hline
\hline
\end{tabular}
\end{table*}


\begin{table}
\caption{\label{hst_infos}Summary of the \emph{HST} observations used in this analysis to carry out the source identifications for RX\,J2129. In the context of the CLASH survey (Proposal ID\ 12457), observations in the UV were also carried but we do not list them as they are not used in this analysis.}
\begin{center}
\begin{tabular}{ccccc}
\hline
\hline
Band & PID & P.\,I. & Exp. time  & Obs. date \\
                & & & [s] & \\
\hline
\hline
ACS/F435W & 12457 & Postman & 1023 & 2012-05-31  \\
 & -- & -- & 932 & 2012-06-30 \\
\hline
 ACS/F475W & -- & -- & 932 & 2012-05-23 \\
 & -- & -- & 932 & 2012-07-09 \\
\hline
 ACS/F606W & -- & -- & 1003 & 2012-05-01 \\
 & -- & -- & 932 & 2012-06-12 \\
\hline
 ACS/F625W & -- & -- & 932 & 2012-04-03 \\
 & -- & -- & 932 & 2012-05-23 \\
\hline
 ACS/F775W & -- & -- & 932 & 2012-05-01\\
 & -- & -- & 1018 & 2012-05-23\\
\hline
 ACS/F814W & -- & -- & 932 & 2012-05-31 \\
  & -- & -- & 989 & 2012-06-12  \\
  & -- & -- & 1022 & 2012-06-30\\
  & -- & -- & 990 & 2012-07-20\\
\hline
 ACS/F850LP & -- & -- & 1022 & 2012-04-03  \\
 & -- & -- & 1022 & 2012-05-23 \\
 & -- & -- & 1006 & 2012-07-09 \\
 & -- & -- & 932 & 2012-07-20 \\
 & 12461 & Reiss & 1780 & 2012-07-23  \\
 & -- & -- & 1780 & 2012-07-30 \\
\hline
 WFC3/F105W & 12457 & Postman & 1206 & 2012-05-31 \\
  & -- & -- & 1006 & 2012-06-13 \\
\hline
 WFC3/F110W & -- & -- & 1409 & 2012-05-23 \\
 & -- & -- & 1006 & 2012-07-20 \\
\hline
 WFC3/F125W & -- & -- & 1409 & 2012-04-03 \\
  & -- & -- & 1006 & 2012-06-27  \\
  & -- & -- & 1006 & 2012-07-09  \\
\hline
 WFC3/F140W & -- & -- & 1306 & 2012-05-31  \\
  & -- & -- & 1006 & 2012-06-13  \\
\hline
 WFC3/F160W & -- & -- & 1006 & 2012-04-03\\ 
  & -- & -- & 1006 & 2012-05-23  \\ 
  & -- & -- & 1409 & 2012-06-27  \\ 
  & -- & -- & 1409 & 2012-07-09  \\ 
\hline
\hline
\end{tabular}
\end{center}
\end{table}

\begin{table}
\caption{\label{hst_infos_ms0451}Summary of the \emph{HST} observations used in this analysis to carry out the source identifications for MS\,0451.}
\begin{center}
\begin{tabular}{ccccc}
\hline
\hline
Band & PID & P.\,I. & Exp. time  & Obs. date \\
                & & & [s] & \\
\hline
\hline
 ACS/F555W & 9722 & Ebeling & 4410 & 2002-01-15 \\
 \hline
 ACS/F775W & 9292 & Ford & 2440 & 2002-04-09 \\
 \hline
 ACS/F814W & 9836 & Ellis & 2036 & 2004-01-27 \\
 & 10493 & Gal-Yam & 2162 & 2005-07-31  \\
 & 11591 & Kneib & 7240 & 2011-02-07  \\
 \hline
 ACS/F850LP & 9292 & Ford & 2560 & 2002-04-10  \\
 \hline
 WFC3/F110W & 11591 & Kneib & 2612 & 2010-01-13 \\
 \hline
 WFC3/F160W & -- & -- & 2412 & 2010-01-13 \\
\hline
\hline
\end{tabular}
\end{center}
\end{table}

\begin{table}
\caption{\label{hst_infos_m2129}Summary of the \emph{HST} observations used in this analysis to carry out the source identifications for MACS\,J2129.
In the context of the CLASH survey (PID 12100), observations in the UV were also carried out but we do not list them as they are not used for this analysis.}
\begin{center}
\begin{tabular}{ccccc}
\hline
\hline
Band & PID & P.\,I. & Exp. time  & Obs. date \\
                & & & [s] & \\ 
\hline
\hline
 ACS/F435W & 12100 & Postman & 932 & 2011-07-14 \\ 
 \hline
 ACS/F475W & -- & -- & 1110 & 2011-06-03\\ 
 & -- & -- & 1110 & 2011-07-14 \\ 
 \hline
 ACS/F555W & 9722 & Ebeling & 4440 & 2003-09-11 \\ 
 \hline
 ACS/F606W & 12100 & Postman & 932 & 2011-05-15\\ 
 & -- & -- & 932 & 2011-06-24\\ 
 \hline
 ACS/F625W & -- & -- & 932 & 2011-05-16 \\
 & -- & -- & 991 & 2011-06-24 \\
 \hline
 ACS/F775W & -- & -- & 1029 & 2011-05-16 \\ 
 & -- & -- & 995 & 2011-07-14 \\ 
 \hline
 ACS/F814W & 9722 & Ebeling & 4530 & 2003-09-11 \\
 & 10493 & Gal-Yam & 2168 & 2005-06-18 \\
 & 12100 & Postman & 932 & 2011-06-24 \\
 \hline
 ACS/F850LP & -- & -- & 1020 & 2011-05-15 \\ 
  & -- & -- & 932 & 2011-06-03 \\
  & -- & -- & 1020 & 2011-06-24 \\
  & -- & -- & 932 & 2011-07-14 \\
  \hline
 WFC3/F105W & -- & -- & 1006 & 2011-05-16 \\
 & -- & -- & 1409 & 2011-08-03 \\
 & 13459 & Treu & 812 & 2014-05-28 \\
 & -- & -- & 356 & 2014-05-29 \\
 & -- & -- & 356 & 2014-08-14\\
 & -- & -- & 812 & 2014-08-15\\
 \hline
 WFC3/F110W & 12100 & Postman & 1409 & 2011-05-15  \\
 & -- & -- & 1006 & 2011-07-20 \\
 \hline
 WFC3/F125W & -- & -- & 1409 & 2011-05-16  \\
 & -- & -- & 1006 & 2011-08-03 \\
 \hline
 WFC3/F140W & -- & -- & 1006 & 2011-06-03 \\
 & -- & -- & 1306 & 2011-06-24 \\
 & 13459 & Treu & 812 & 2014-05-29 \\
 & -- & -- & 1574 & 2014-08-14  \\
 \hline
 WFC3/F160W & 12100 & Postman & 1006 & 2011-05-15 \\
 & -- & -- & 1409 & 2011-06-03 \\
 & -- & -- & 1206 & 2011-06-24 \\
 & -- & -- & 1409 & 2011-07-20 \\
\hline
\hline
\end{tabular}
\end{center}
\end{table}

The Multi-Unit Spectroscopic Explorer \citep[MUSE;][]{bacon2010} is a second generation integral field spectrograph at the Very Large Telescope (VLT). MUSE large field of view of 1\,arcmin$^{2}$ is perfectly adapted to the observation of the core of galaxy clusters \citep{richard2015,richard2020,grillo2016,caminha2017a,caminha2017b,chirivi2018,rescigno2020}, where most multiple images are likely to be observed \citep[e.g.][]{kneib2011}. Its high sensitivity between 4750\,{\AA} and 9350\,{\AA} enables the detection of sources with redshifts up to $z=6$ \citep{bacon2015}. Over the past 6 years, strong cluster lenses have been commonly observed with MUSE, leading to the measurement of spectroscopic redshifts for multiple images, their multiplicity confirmation, as well as the identification of new multiple image systems which are not even detected in \emph{HST} observations \citep[e.g.][]{richard2015,jauzac2016a,caminha2017b,caminha2019,lagattuta2019}. 

In this paper, we present MUSE observations, and their subsequent strong-lensing analyses, for three well-known galaxy clusters: RX\,J2129.4$+$0009, MS\,0451.6$-$0305, and MACS\,J2129.4$-$0741. These clusters have been observed with \emph{HST}, and already have strong-lensing mass models published in the literature \citep[more details are given bellow, but e.g.][]{rxj2129_model,zitrin2011,zitrin2015,mackenzie,monna,caminha2019} which are used as references in this analysis, and referred to as \emph{fiducial models} in the rest of the paper).

\textbullet \ RX\,J2129.4+0009 ($z=0.235$, RX\,J2129 hereafter) was observed as part of the CLASH survey, and was first modeled with \textsc{Lenstool} for the Local Cluster Substructure Survey \citep[LoCuSS, PI: G. P. Smith, see][]{rxj2129_model}. This model relied on a single system of multiple images which redshift was updated by \citet{belli2013}. Then, \citet{zitrin2015} published a model which uses 4 multiple image systems, two being spectroscopically confirmed. \cite{desprez2018} presented a revised model, including one galaxy-galaxy lensing system located $\sim$80\arcsec\ from the cluster center, in the vicinity of an isolated cluster galaxy. Finally, \cite{caminha2019} presented a strong-lensing mass model which takes advantage of the MUSE observations presented in this paper.

\textbullet \ MS\,0451.6$-$0305 (z$=$0.55 - MS\,0451 hereafter) is originally known for its large brightness in the X-rays \citep[e.g.][]{ellingson1998,molnar2002,laroque2003,gioia,donahue2003,geach2006}, and hosts several strongly lensed submillimetric sources with radio counterparts \citep[e.g.][]{takata2003,borys2004,berciano_alba2007,berciano_alba2010,mackenzie}. Increasingly precise strong-lensing mass models were obtained by \citet{borys2004}, \citet{berciano_alba2007}, \citet{zitrin2011}, and most recently by \citet{mackenzie}, where they included sub-millimeter detections. The latter relies on 8 multiple image systems located in the South of the cluster, leaving the North poorly-constrained. However, more recent imaging with \emph{HST}, and spectroscopy with VLT/X-Shooter and Keck/LRIS, allowed the identification of 8 new multiple image systems, including a quintuple image at redshift $z=6.7$ in the North \citep[][Richard et al.\ \emph{in prep.}]{knudsen2016}.
The giant arc identified in \citet{borys2004}, and the system at redshift $z=6.7$, are the only multiple image systems with confirmed spectroscopic redshifts. 

\textbullet \ MACS\,J2129.4$-$0741 \citep[$z=0.589$ - MACS\,J2129 hereafter,][]{ebeling2007} is part of the Cluster Lensing And Supernova survey with \emph{Hubble} \citep[CLASH][]{postmanclash}. It was modeled by \citet{zitrin2011_b}, and more recently by \citet{monna} using CLASH photometry \citep{zitrin2015}, and VLT-VIMOS spectroscopic data \citep{rosati2014}. Among the 9 multiple image systems used in the mass model presented by \citet{monna}, two systems are not spectroscopically confirmed. 
Then, \cite{caminha2019} presented a strong-lensing model using \textsc{Lenstool}, which takes advantage of the MUSE observations presented in this work. 

This paper is organized as follows. Section\,\ref{section2} presents the details of the pipeline used to extract the spectra from the MUSE datacubes. Section\,\ref{section3} describes redshift measurements, and presents our results for the three galaxy clusters. Section\,\ref{section4} details the strong-lensing mass models of RX\,J2129, MS\,0451, and MACS\,J2129. Section\,\ref{section5} presents our results and discuss them with regards to previous strong-lensing analyses of these clusters. We finally conclude in Section\,\ref{section6}. 
Throughout the paper, we assume a standard cosmological model with $\Omega_{M}=0.3$, $\Omega_{\Lambda}=0.7$, and $H_{0}=70$\,km\,s$^{-1}$\,Mpc$^{-1}$. 
At the redshift of RX\,J2129 (z$=$0.235), 1\arcsec\ covers a physical distance of 3.3734\,kpc.
At the redshift of MS\,0451 (z$=$0.55), 1\arcsec\ covers a physical distance of 6.412\,kpc. Finally, for MACS\,J2129 (z$=$0.589), 1\arcsec\ corresponds to 6.63\,kpc. All magnitudes are measured using AB system. 

\section{MUSE Observations \& Analysis}
\label{section2}

\subsection{Observations and data reduction}
RX\,J2129, MS\,0451, and MACS\,J2129 were observed with MUSE on the VLT. Table\,\ref{cluster_infos} gives the details of the observations, including dates, pointing positions, ID, seeing conditions, and total exposure time for each dataset. 
Observations were taken using MUSE WFM-NOAO-N mode, in good seeing conditions with full width at half maximum (FWHM) of $\approx$\,0.5\arcsec, 0.8\arcsec, and 0.9\arcsec\ for RX\,J2129, MS\,0451, and MACS\,J2129 respectively. At each pointing, three exposures were taken, slightly shifted (by $\sim 0.5$\arcsec) in order to mitigate systematics from the image slicer and detectors. 

The data were reduced with version 1.6.4 of the standard MUSE pipeline \citep[][]{pipeline2,pipeline}. We use a set of standard calibration exposures taken daily to produce bias, arcs and flat field master calibration files. Dark current is neglected due to its very low value with MUSE \citep[$\approx$ 1\,electron/h][]{bacon2015}. We first subtract the master bias exposures from each dataset, and perform an illumination correction using in combination the master flat field, the twilight sky exposures taken at the beginning of the night, and the illumination calibration taken soon before/after the science observations. We carry out wavelength, geometrical and astrometric calibrations in order to assign the World Coordinate System (WCS) right ascension and declination, and the wavelength to each pixel of the datacube. The flux calibration is carried out using standard star observations taken at the beginning of the observing night. For each pointing, the three individual exposures are then combined into a full datacube using a single interpolation step. 

We apply the \textsc{Zürich Atmosphere Purge} \citep[\textsc{zap};][]{zap} software version 1.0, which uses a principal components analysis to analyse objects-free regions in the datacube and subtract systematics due to sky subtraction residuals. To create the \textsc{zap} objects mask, we use the segmentation map obtained by applying the \textsc{Sextractor} software \citep{sextractor} on a white-light image collapsing the datacube along its wavelength axis.

The wavelength range of the final datacube stretches from 4750\,{\AA} to 9350\,{\AA} in steps of 1.25\,{\AA}, and the spaxel size is 0.2\arcsec.

\subsection{Spectrum extraction}
We combine MUSE observations with high resolution images from \emph{HST} to detect small and faint sources which remain invisible in the image obtained when the MUSE datacube is collapsed along the wavelength axis. This combination was notably used by \cite{bacon2015,bacon2017} for the analysis of MUSE observations of the \emph{Hubble} Deep Field South. 

For MACS\,J2129 and RX\,J2129, we use \emph{HST} data obtained with the \emph{Advanced Camera for Surveys} \citep[ACS;][]{acs} as part of the CLASH survey in the F475W, F625W, and F814W pass-bands. We also use imaging by the \emph{Wide Field Camera 3} (WFC3) in the F110W and F160W pass-bands to cover a larger wavelength range for the source identification. For MS\,0451, we use the \emph{HST} data available in the MAST website\footnote{\url{http://archive.stsci.edu/}}. We present in Table~\ref{hst_infos}, Table~\ref{hst_infos_ms0451}, and Table~\ref{hst_infos_m2129}, a summary of the \emph{HST} observations used for this work for RX\,J2129, MS\,0451, and MACS\,J2129 respectively, including the observation ID, the PI, the exposure time, and the observational date. 
For all three clusters, we applied standard data-reduction procedures. We used \textsc{HSTCAL} and the most recent calibration files. The co-addition of individual frames was done using \textsc{Astrodrizzle} after registration to a common reference image using \textsc{Tweakreg}. After an iterative process, we achieve an alignment accuracy of 0.1\,pixel. Our final stacked images have a pixel size of 0.03\arcsec.

We use the \textsc{ifs-redex} software to align the datacubes with the corresponding \emph{HST} high resolution images \citep{ifs}. We then run \textsc{sextractor} on the \emph{HST/ACS} F814W pass-band image for each cluster to automatically measure the position and FWHM of the sources in the MUSE field of view. \textsc{ifs-redex} uses the catalogue of detected sources to extract the signal in the datacube within a circle of radius 3 to 5\,pixels according to the FWHM measurement. Sources with ${\rm FWHM} < 2$\,pixels are considered spurious detections, and are rejected. 

In order to maximize the number of extracted spectra, we carry out a blind search in the datacube using \textsc{muselet}\footnote{\url{http://mpdaf.readthedocs.io/en/latest/muselet.html}}. This software is part of the MUSE \textsc{Python Data Analysis Framework} \citep[\textsc{mpdaf};][]{mpdaf}. It builds a new datacube, the narrow-band datacube, within which each wavelength plane is the mean of the 5 closest wavelength planes in the science datacube. \textsc{muselet} then uses \textsc{Sextractor} to extract a catalogue of sources at each wavelength in the narrow-band datacube. The latter are finally merged and sorted, providing a continuum and a single line emission catalogues. 

Finally, all catalogues are merged to provide a master catalogue, which is then displayed on the high resolution image so that the user can determine whether \textsc{muselet} and \textsc{Sextractor} detections are matching the same source. This results in a set of spectra that we then analyze to measure the associated redshifts.

\begin{figure*}
\begin{center}
\includegraphics[width=0.47\textwidth]{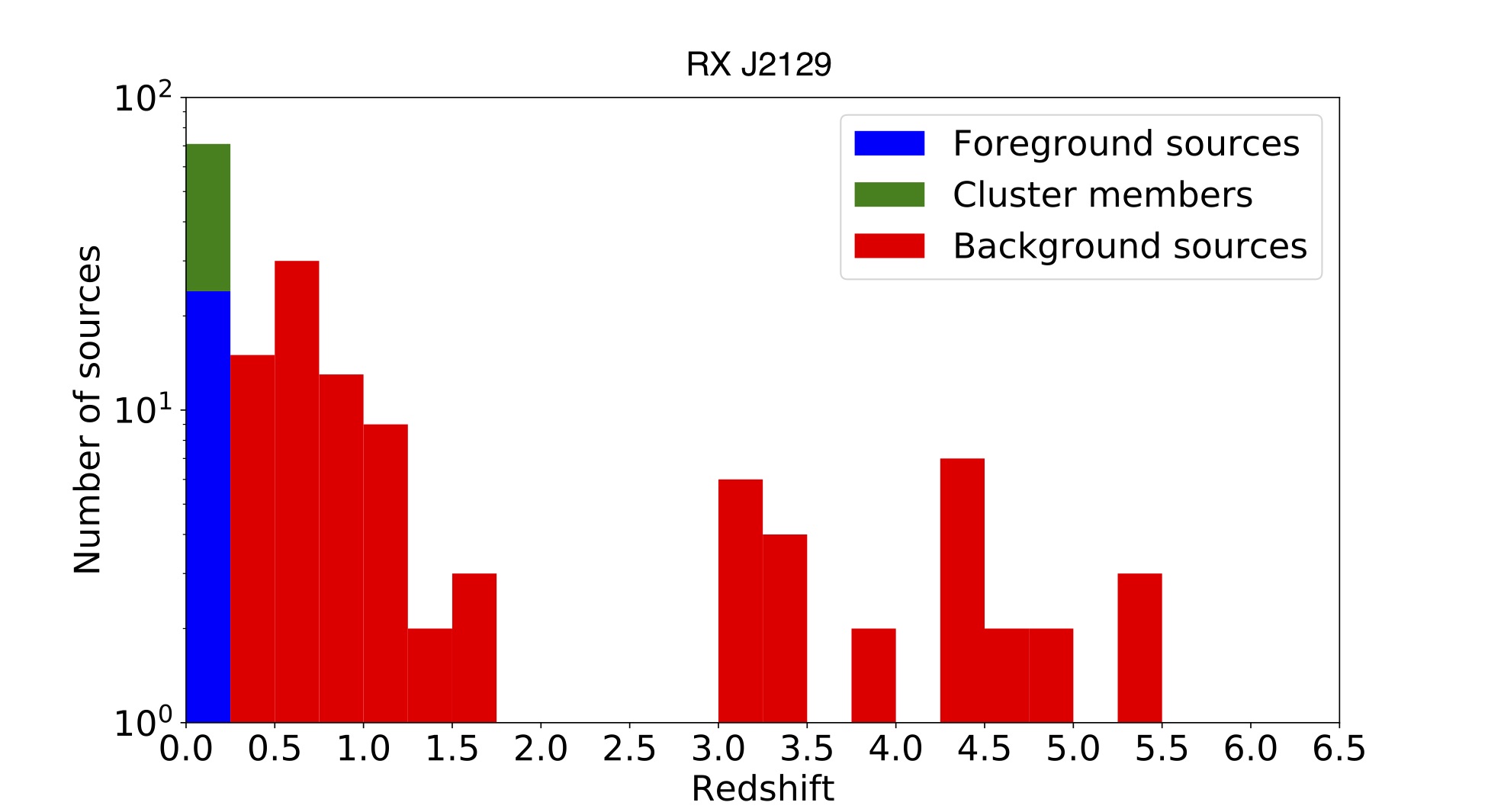}
\includegraphics[width=0.47\textwidth]{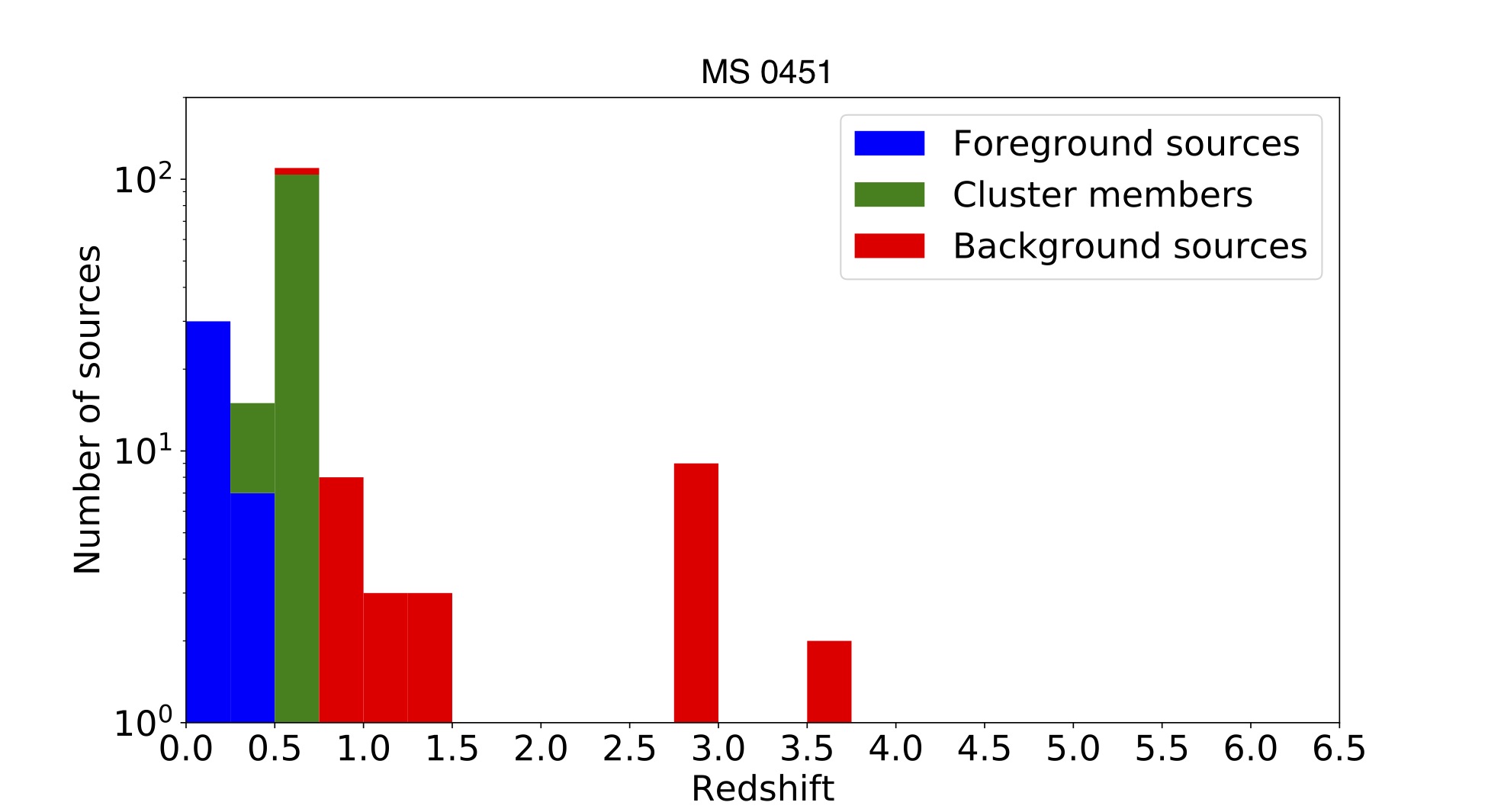}
\includegraphics[width=0.47\textwidth]{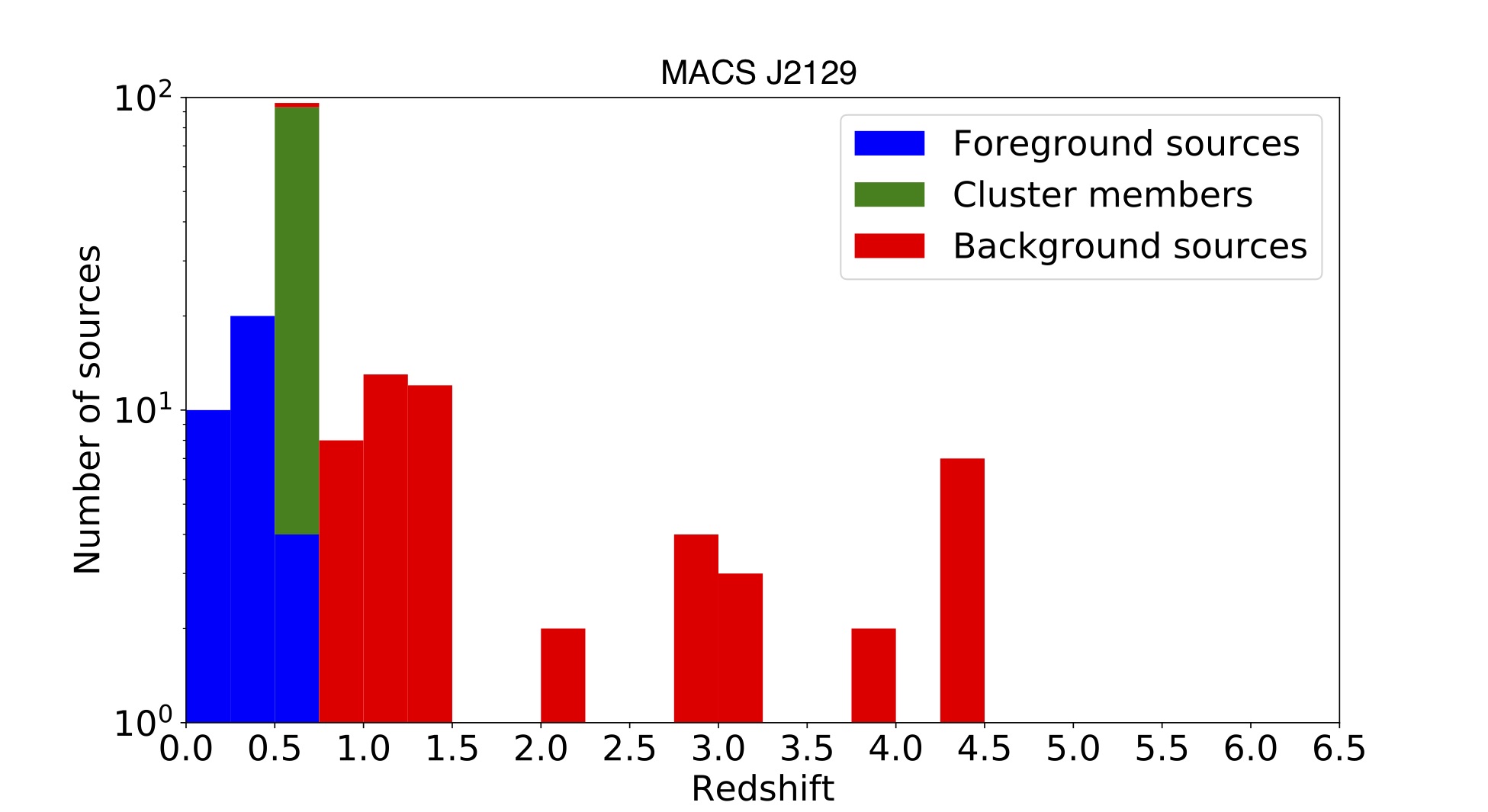}
\caption{Redshift distribution of the sources with QF $\geq$ 2 extracted within the MUSE datacubes for RX J2129, MS 0451, and MACS J2129.}
\label{histos}
\end{center}
\end{figure*}

\subsection{Redshift measurements}
\textsc{ifs-redex} has an interactive interface which displays each extracted spectrum, and its corresponding source in \textsc{saods9} \citep{ds9}. It allows the user to modify the source redshift to match the position of an emission/absorption line template to its most likely position in the spectrum. (The template contains $\sim$60 lines including notably Ly$\alpha$, [OII], [OIII], and H$\beta$ emission lines, and Ca\, H\&K, Mg, Fe, and Na absorption lines). To simplify the redshift identification, it is possible to smooth the signal with a Gaussian filter, and then perform a wavelet-based spectrum cleaning \citep{ifs}. The systematic error is calculated by quadratically adding the wavelength calibration error provided by the MUSE data reduction pipeline, and the error given by fitting a Gaussian to the most prominent line in the spectrum. For each redshift, we assign a quality flag (QF) of 3 if the redshift is secure, 2 if likely (e.g., only one characteristic line - for example the [OII] doublet or Ly$\alpha$ line with consistent photometric redshift), 1 if insecure, and 0 otherwise, i.e., in case of visually flat continuum or highly polluted spectrum.

We sequentially analyze all spectra extracted from the MUSE datacubes of the three clusters using the aforementioned method. The measured redshifts are sorted depending on whether they belong to a source located in the foreground of the cluster, in the cluster, or in the background.

\subsection{Results of the redshift extraction}
\label{section3}
Figure\,\ref{histos} shows the distribution of redshifts for the extracted sources for RX\,J2129, MS\,0451, and MACS\,J2129. Lists of the extracted redshifts with QF larger than 2 are given in the Appendix\,\ref{tab}, where Table\,\ref{rxj2129_z}, Table\,\ref{ms0451_z}, and Table\,\ref{macs2129_z} give the results for RX\,J2129, MS\,0451, and MACS\,J2129 respectively. The redshift intervals to consider galaxies as cluster members are established using the definition given in \cite{ma08}, combined with the criteria defined in \cite{ifs}. For each cluster, we include all galaxies with a redshift flagged as secure or likely (QF$=2$,$3$), and which present a continuum plus characteristic absorption lines. We here only summarize the redshift extraction.

\paragraph*{RX\,J2129 --}
We extracted 158 sources with redshifts ranging from 0.0 to 5.53 in RX\,J2129. Among them, 43 are identified as cluster members with $0.2145<z<0.2410$, 24 as foreground objects, and 91 as background sources (i.e., total number of sources without accounting for image multiplicity).When comparing our results to those reported in \citet{caminha2019}, we measure  22 new redshifts not reported in their analysis, and miss 34 of their identifications. Most of the sources we disagree on are faint cluster members. We attribute these differences to the different methods used for the extraction of spectra.

\paragraph*{MS\,0451 --}
We extracted 171 sources with redshifts ranging from 0.0 to 4.85 from the MUSE datacube. Among them, 112 are identified as cluster members, with $0.5307<z<0.5652$, 24 sources are identified as foreground objects, 35 are identified as background sources.

\paragraph*{MACS\,J2129 --}
We extracted 189 sources with redshifts ranging from 0.0 to 4.92. Among them, 89 are identified as cluster members with $0.5737<z<0.6032$, 39 as foreground objects, and 61 as background sources. The comparison of our measurements with those reported in \citet{caminha2019} yields similar results to RX\,J2129, with 16 new redshifts not reported in their analysis, and 25 of their measurements that we miss.

\section{Strong-lensing analyses}\label{section4}

We use the \textsc{Lenstool} software \citep{kneib1996,jullo2007} to perform the strong-lensing analysis of each cluster. We started from existing strong-lensing models, referred to as \emph{fiducial models} in the following, which were either already published, or shared privately with our team. Starting from these \emph{fiducial models}, we use the newly measured redshifts to carry out the identification of new cluster members, and multiple image systems. When possible, we also add the spectroscopic redshift information to already identified multiple image systems, and/or confirm counter-images of the same system.

\begin{table}
\caption{List of the multiple images used as constraints in our new \textsc{Lenstool} strong-lensing mass model of RX\,J2129. System\ 1 was identified by \citet{rxj2129_model}, and its spectroscopic redshift first measured by \citet{belli2013}. Systems\ 2 was studied in detail in \citet{desprez2018}. Systems\ 3 and 5 were reported in \citet{zitrin2015}. Systems\ 6, 7 and 8 are new identification from the MUSE observations. 
Column (1) is the ID of the image, columns (2) and (3) give the R.\,A. and Decl. in degrees (J2000) of each image, column (4) the spectroscopic redshift measured if available, and column (5) the redshift predicted by the best model when no spectroscopic information is available. Spectroscopic redshifts are highlighted in bold when confirmed/measured with MUSE. The rms for each image is given in column (6) in arcseconds.
$^\ast$We here fix the redshift of System\ 2 to the photometric redshift measured by \citet{desprez2018}. 
\label{mul_rxj2129}}
\begin{center}
\begin{tabular}{cccccc}
\hline
\hline
ID & R.\,A. & Decl. & $z_{spec}$ & $z_{m}$ & $rms$  \\
                & [J2000] & [J2000] & & & [\arcsec]\\
\hline
\hline 
1.1 & 322.42038 & 0.08832 & 1.522 & -- & 0.32\\
1.2 & 322.42017 & 0.08976 & 1.52 & -- & 0.19\\
1.3 & 322.41796 & 0.09327 & 1.52 & -- & 0.12\\
2.1 & 322.42900 & 0.10833 & $^\ast$1.61 & -- &  0.07\\
2.2 & 322.42856 & 0.10841 & $^\ast$1.61 & -- & 0.11\\
2.3 & 322.42912 & 0.10807 & $^\ast$1.61 & -- & 0.09\\
2.4 & 322.42867 & 0.10790 & $^\ast$1.61 & -- & 0.09\\
3.1 & 322.41843 & 0.08537 & 1.52 & -- & 0.35\\
3.2 & 322.41767 & 0.09027 & 1.52 & -- & 0.34\\
3.3 & 322.41572 & 0.09222 & 1.52 & -- & 0.10\\
4.1 & 322.41373 & 0.09208 & 3.427 & -- & 0.11 \\
4.2 & 322.41443 & 0.08863 & 3.427 & -- & 0.38 \\
4.3 & 322.41754 & 0.08386 & 3.427 & -- & 0.42 \\
5.1 & 322.41659 & 0.08774 & 0.916 & -- & 0.27\\
5.2 & 322.41627 & 0.08810 & 0.916 & -- & 0.17\\
5.3 & 322.41463 & 0.09236 & 0.916 & -- & 0.12 \\
6.1 & 322.41492 & 0.09038 & 0.679 & -- & 0.26\\
6.2 & 322.41663 & 0.08674 & 0.679 & -- & 0.27\\
6.3 & 322.41516 & 0.08898 & 0.679 & -- & 0.69\\
7.1 & 322.41675 & 0.08779 & 3.08 & -- & 0.02\\
7.2 & 322.41700 & 0.08739 & 3.08 & -- & 0.27\\
7.3 & 322.41376 & 0.09420 & 3.08 & -- & 0.69 \\
8.1 & 322.41592 & 0.09150 & 1.52 & -- & 0.31\\
8.2 & 322.41694 & 0.09031 & 1.52 & -- & 0.44\\
8.3 & 322.41854 & 0.08492 & 1.52 & -- & 0.25\\
\hline
\hline
\end{tabular}
\end{center}
\end{table}

\begin{table}
\caption{\label{mul_ms0451} List of the multiple images used as constraints in our new \textsc{Lenstool} strong-lensing mass model of MS\,0451. Systems\ A, B and C were reported by \citet{borys2004}. Systems\ E, F, G and I were identified by \citet{mackenzie}. Systems\ D, H, and J to P were detected by our team, and will be presented in an upcoming analysis (Richard et al.\ \emph{in prep.}). Systems\ R and S are new detections from this analysis. The table elements are the same as Table\,\ref{mul_rxj2129}.
$^{\ast}$System P was flagged as insecured, and thus not used in the model.}
\begin{center}
\begin{tabular}{cccccc}
\hline
\hline
ID & R.\,A. & Decl. & $z_{spec}$ & $z_{model}$ &  $rms$  \\
                & [J2000] & [J2000] & & & [\arcsec]\\
\hline 
\hline
A.1 & 73.55396 & -3.01482 & 2.91 & -- & 0.23\\
A.2 & 73.55389 & -3.01595 & 2.91 & -- &  0.18\\
A.3 & 73.54630 & -3.02404 & 2.91 & -- &  0.20\\
B.1 & 73.55335 & -3.01232 & -- & 2.9 $\pm$ 0.3 & 0.13\\
B.2 & 73.55285 & -3.01707 & -- & 2.9 $\pm$ 0.3  & 0.41\\
B.3 & 73.54553 & -3.02348 & -- & 2.9 $\pm$ 0.3  & 0.31\\
C.1 & 73.55339 & -3.01325 & -- & 2.8 $\pm$ 0.2  & 0.14\\
C.2 & 73.55304 & -3.01656 & -- & 2.8 $\pm$ 0.2 & 0.17\\
C.3 & 73.54545 & -3.02380 & -- & 2.8 $\pm$ 0.2 & 0.26\\
D.1 & 73.55409 & -3.01469 & -- & 2.9 $\pm$ 0.1 & 0.37\\
D.2 & 73.55399 & -3.01640 & -- & 2.9 $\pm$ 0.1 & 0.06\\
D.3 & 73.54658 & -3.02401 & -- & 2.9 $\pm$ 0.1 & 0.17\\
E.1 & 73.55481 & -3.01065 & -- & 2.8 $\pm$ 0.2 & 0.24\\
E.2 & 73.55241 & -3.01996 & -- & 2.8 $\pm$ 0.2 & 0.23\\
E.3 & 73.54911 & -3.02226 & -- & 2.8 $\pm$ 0.2 & 0.20\\
F.1 & 73.55435 & -3.01088 & -- & 2.9 $\pm$ 0.3 & 0.35\\
F.2 & 73.55282 & -3.01918 & -- & 2.9 $\pm$ 0.3 & 0.59\\
F.3 & 73.54775 & -3.02268 & -- & 2.9 $\pm$ 0.3 & 0.39\\
G.1 & 73.55593 & -3.01193 & 2.93 & -- & 0.41\\
G.2 & 73.55271 & -3.02124 & 2.93 & -- & 0.85\\
G.3 & 73.55071 & -3.02261 & 2.93 & -- & 0.02\\
H.1 & 73.53855 & -3.00589 & 6.7 & -- & 0.10\\
H.2 & 73.53687 & -3.00773 & 6.7 & -- & 0.10\\
H.3 & 73.53662 & -3.00807 & 6.7 & -- & 0.18\\
H.4 & 73.53647 & -3.00830 & 6.7 & -- & 0.22\\
I.1 & 73.55342 & -3.01089 & -- & 3.1 $\pm$ 0.3 & 1.17\\
I.2 & 73.55233 & -3.01807 & -- & 3.1 $\pm$ 0.3 & 1.16\\
I.3 & 73.54597 & -3.02285 & -- & 3.1 $\pm$ 0.3 & 0.55\\
J.1 & 73.54901 & -3.01848 & -- & 1.7 $\pm$ 0.2 & 0.03\\
J.2 & 73.54830 & -3.01930 & -- & 1.7 $\pm$ 0.2 & 0.04\\
K.1 & 73.55685 & -3.01410 & -- & 3.1 $\pm$ 0.2  & 0.76\\
K.2 & 73.55352 & -3.02183 & -- & 3.1 $\pm$ 0.2  & 0.77\\
K.3 & 73.55250 & -3.02276 & -- & 3.1 $\pm$ 0.2  & 0.24\\
L.1 & 73.54119 & -3.01469 & -- & 7.3 $\pm$ 0.8 & 0.53\\
L.2 & 73.54191 & -3.02009 & -- & 7.3 $\pm$ 0.8 & 0.29\\
L.3 & 73.55136 & -3.00437 & -- & 7.3 $\pm$ 0.8 & 0.44\\
M.1 & 73.54787 & -3.01719 & -- & 1.02 $\pm$ 0.07 & 0.38\\
M.2 & 73.54822 & -3.01656 & -- & 1.02 $\pm$ 0.07 & 0.21\\
M.3 & 73.54936 & -3.01404 & -- & 1.02 $\pm$ 0.07 & 0.39\\
O.1 & 73.54268 & -3.01943 & -- & 1.8 $\pm$ 0.1 & 0.54\\
O.2 & 73.54370 & -3.01400 & -- & 1.8 $\pm$ 0.1 & 1.12\\
O.3 & 73.54807 & -3.00859 & -- & 1.8 $\pm$ 0.1 & 1.63\\
$^{\ast}$P.1 & 73.54574 & -3.01966 & -- & -- & --\\
$^{\ast}$P.2 & 73.54872 & -3.01730 & -- & -- & --\\
R.1 & 73.53630 & -3.01234 & 3.765 & -- & 0.59\\
R.2 & 73.53618 & -3.01331 & 3.765 & -- & 0.23\\
R.3 & 73.54229 & -3.00485 & 3.765 & -- & 0.61\\
S.1 & 73.54723 & -3.01284 & 4.451 & -- & 1.39\\
S.2 & 73.54627 & -3.01263 & 4.451 & -- & 1.22\\
\hline
\hline
\end{tabular}
\end{center}
\end{table}

\begin{table}
\begin{center}
\caption{\label{mul_macs2129} List of the multiple images used as constraints in our new \textsc{Lenstool} strong-lensing mass model of MACS\,J2129. Systems\ 1 to 9 have been identified by \citet{monna}. Systems\ 10 is newly identified in this work. Systems\ 11 to 14 were initially identified by \citet{caminha2019}. Table elements are the same as in Table\,\ref{mul_rxj2129}.
$^\ast$ These images are not included in our model as their spectroscopic redshift is not considered secure. System\ 14 corresponds to System\ 5 in \citet{caminha2019}.
}
\begin{tabular}{cccccc}
\hline
\hline
ID & R.\,A. & Decl. & $z_{spec}$ & $z_{model}$ & $rms$  \\
                & [J2000] & [J2000] & & & [\arcsec]\\
\hline 
\hline
1.1 & 322.35797 & -7.68588 & 1.36 & -- & 0.35\\
1.2 & 322.35965 & -7.69082 & 1.36 & -- & 0.10\\
1.3 & 322.35925 & -7.69095 & 1.36 & -- & 0.24\\
1.4 & 322.35712 & -7.69109 & 1.36 & -- & 0.31\\
1.5 & 322.35764 & -7.69115 & 1.36 & -- & 0.11\\
1.6 & 322.35861 & -7.69489 & 1.36 & -- & 0.60\\
2.1 & 322.35483 & -7.6907 & 1.048 & -- & 0.736\\
2.2 & 322.35477 & -7.6916 & 1.048 & -- & 0.08\\
2.3 & 322.35538 & -7.69332 & 1.048 & -- & 0.53\\
3.1 & 322.35022 & -7.68886 & 2.24 & -- & 0.18\\
3.2 & 322.35011 & -7.68950 & 2.24 & -- & 0.76\\
3.3 & 322.35095 & -7.69577 & 2.24 & -- & 0.40\\
4.1 & 322.36642 & -7.68674 & 2.24 & -- & 0.29\\
4.2 & 322.36693 & -7.68831 & 2.24 & -- & 0.35\\
4.3 & 322.36679 & -7.69497 & 2.24 & -- & 0.41\\
5.1 & 322.36422 & -7.69387 & -- & 1.67 $\pm$ 0.03 & 0.16\\
5.2 & 322.36460 & -7.69137 & -- & 1.67 $\pm$ 0.03 & 0.57\\
5.3 & 322.36243 & -7.68493 & -- & 1.67 $\pm$ 0.03 & 0.27\\
6.1 & 322.35094 & -7.69333 & 6.85 & -- & 0.69\\
6.2 & 322.35324 & -7.69744 & 6.85 & -- & 0.87\\
6.3 & 322.35394 & -7.68164 & 6.85 & -- & 1.04\\
7.1 & 322.35714 & -7.69425 & 1.357 & -- & 0.62\\
7.2 & 322.35625 & -7.69172 & 1.357 & -- & 0.84\\
7.3 & 322.35670 & -7.68554 & 1.357 & -- & 0.52\\
8.1 & 322.35698 & -7.68924 & 4.41 & -- & 0.54\\
8.2 & 322.36167 & -7.68808 & 4.41 & -- & 1.61\\
8.3 & 322.35860 & -7.68491 & 4.41 & -- & 1.86\\
8.4 & 322.36035 & -7.70094 & 4.41 & -- & 1.01\\
8.5 & 322.35419 & -7.68876 & 4.41 & -- & 1.27\\
9.1 & 322.36651 & -7.68689 & 2.24 & -- & 0.69\\
9.2 & 322.36695 & -7.68820 & 2.24 & -- & 0.43\\
9.3 & 322.36666 & -7.69525 & 2.24 & -- & 0.79\\
10.1 & 322.35762 & -7.68471 & 4.41 & -- & 0.23\\
10.2 & 322.35499 & -7.68896 & 4.41 & -- & 1.39\\
11.1 & 322.36334 & -7.69707 & 3.108 & -- & 0.79\\
11.2 & 322.36491 & -7.69010 & 3.108 & -- & 0.48\\
11.3 & 322.36167 & -7.68362 & 3.108 & -- & 0.88\\
12.1 & 322.35455 & -7.68518 & 3.897 & -- & 0.22\\
12.2 & 322.35278 & -7.68841 & 3.897 & -- & 0.45\\
$^\ast$12.3 & 322.35736 & -7.69977 & 3.897 & -- & -- \\
13.1 & 322.35330 & -7.69113 & 1.359 & -- & 0.45\\
13.2 & 322.35391 & -7.68758 & 1.359 & -- & 0.45\\
13.3 & 322.35443 & -7.69441 & 1.359 & -- & 0.13\\
$^\ast$14.1 & 322.36131 & -7.68590 & 1.452 & -- & -- \\
$^\ast$14.2 & 322.36248 & -7.69142 & 1.452 & -- & -- \\
$^\ast$14.3 & 322.36259 & -7.69360 & 1.452 & -- & -- \\
\hline
\hline
\end{tabular}
\end{center}
\end{table}

\subsection{Mass Modeling Method}
With \textsc{Lenstool} \citep{jullo2007}, we decompose the cluster gravitational potential into large-scale halos to model the main dark matter component(s) of the clusters, $\Phi_{c_{i}}$, and subhalos to model the cluster galaxies, $\Phi_{p_{j}}$, according to
\begin{eqnarray}
\Phi_{tot} = \sum\limits_{i} \Phi_{c_{i}} + \sum\limits_{j} \Phi_{p_{j}}.
\end{eqnarray}
Large-scale halos and subhalos are described with Pseudo Isothermal Elliptical Mass Distribution profiles \citep[PIEMD;][]{piemd,limousin2005,piemd2}, which are parametrized with a core radius, $r_{c}$, and a truncation radius, $r_{t}$, to calculate the projected mass density :
\begin{eqnarray}
\Sigma(R) = \frac{\sigma^{2}}{2G} \frac{r_{t}}{r_{t}-r_{c}} \left( \frac{1}{\sqrt[]{R^{2} + r_{c}^{2}}} - \frac{1}{\sqrt[]{R^{2} + r_{t}^{2}}} \right),
\end{eqnarray}
where $G$ is the gravitational constant. The projected radius, $R^{2} = x^{2}/(1+e)^{2} + y^{2}/(1-e)^{2}$, is defined with the module of the complex ellipticity, \textbf{e} from \citet{ellip}, $e = (a^{2}-b^{2})/(a^{2}+b^{2})$. In practice, $\textbf{e} = e \times e^{2i \theta}$, where $\theta$ is the orientation angle of the ellipse seen in the cluster from the observer point of view. $a$ and $b$ are respectively the semi-major and the semi-minor axes of the mass distribution, and $\sigma$ is the 1D-velocity dispersion. The position of the center, defined by ($x$,$y$), the module of the ellipticity, $e$, the orientation angle, $\theta$, the truncation and core radii, $r_{c}$ and $r_{t}$, and the velocity dispersion, $\sigma$, are the seven parameters needed to describe a PIEMD.

As pointed out by \citet{jullo2007}, the optimization of seven parameters per subhalo would lead to an under-constrained mass model. We thus consider that the luminosity of cluster galaxies is a good tracer of their mass \citep[see the discussion in][]{harvey2016}. Following such assumption, the position and ellipticity of each subhalo are fixed to their luminous counterpart, measured with \textsc{Sextractor} \citep{sextractor}. The total mass of the subhalo is then measured by rescaling the remaining PIEMD parameters for each cluster galaxy, $\sigma$, $r_{c}$, and $r_{t}$, to the ones of a reference galaxy with a luminosity $L^{\ast}$, following the \citet{faber1976} relation:

\begin{eqnarray}
\left \{
\begin{array}{c  @{=}  c}
            \sigma & \sigma^{*} \left( \frac{L}{L^{\ast}} \right)^{1/4}\\
            r_{c} & r_{c}^{*} \left( \frac{L}{L^{\ast}} \right)^{1/2}\\
            r_{t} & r_{t}^{*} \left( \frac{L}{L^{\ast}} \right)^{1/2}\\
\end{array}
\right.,
\label{scaling}
\end{eqnarray}
from which the total mass of each subhalo is derived following:
\begin{eqnarray}
M = \frac{\pi}{G} (\sigma^{*})^{2} r_{c}^{*} \left( \frac{L}{L^{\ast}} \right),
\end{eqnarray}
where $\sigma^{*}$, $r_{t}^{*}$, and $r_{c}^{*}$, are the reference velocity dispersion, truncation and core radii respectively. It was shown in previous models that $r_{c}^{*}$ is small in galaxy-scale halos and thus plays a minor role in the mass models \citep[e.g.][]{covone2006,limousin2007,piemd2}. We thus adopt a conservative value of $r_{c}^{*} \sim 0.15$\,kpc for all three clusters. 

For each model, we start by optimizing one large-scale halo per cluster. The brightest cluster galaxy (BCG), and cluster members located in the vicinity of multiple images are individually optimized as they act as small-scale perturbers.
We then add a second large-scale potential in the optimization process when a set of multiple images concentrated in a given region of the cluster core cannot be reproduced with a simple one-halo mass model. 

\textsc{Lenstool} uses a Markov Chain Monte Carlo (MCMC) process to sample the posterior density of the model, expressed as a function of the likelihood of the model, defined in \citet{jullo2007}. In practice, we minimize
\begin{eqnarray}
\chi^{2} = \sum\limits_{i} \chi_{i}^{2}\ ,
\end{eqnarray}
where
\begin{eqnarray}
\chi_{i}^{2} = \sum_{j=1}^{n_{i}} \frac{(\theta^{j}_{obs} - \theta^{j}(\textbf{p}))^{2}}{\sigma_{ij}^{2}}\ .
\end{eqnarray}
$\theta^{j}_{obs}$ is the vector position of the observed multiple image $j$, $\theta^{j}$ is the predicted vector position of image $j$, $n_{i}$ is the number of images in System $i$, and $\sigma_{ij}$ is the error on the position of image $j$ \citep[fixed at $\sim$ 0.5\arcsec\ for multiple images to account for both errors on image positions between MUSE and \emph{HST} images and line of sight effects as described in][]{jullo2007,jullo2010}. 
As a consequence, the most likely model minimizes the distance between the observed positions of the multiple images and their predicted position by the model, the rms.

In what follows, we describe the set of multiple image systems used to constrain the new mass models for each cluster, and then detail the selection of independently-optimized halos.

\begin{figure}
\centering
\includegraphics[width=0.47\textwidth]{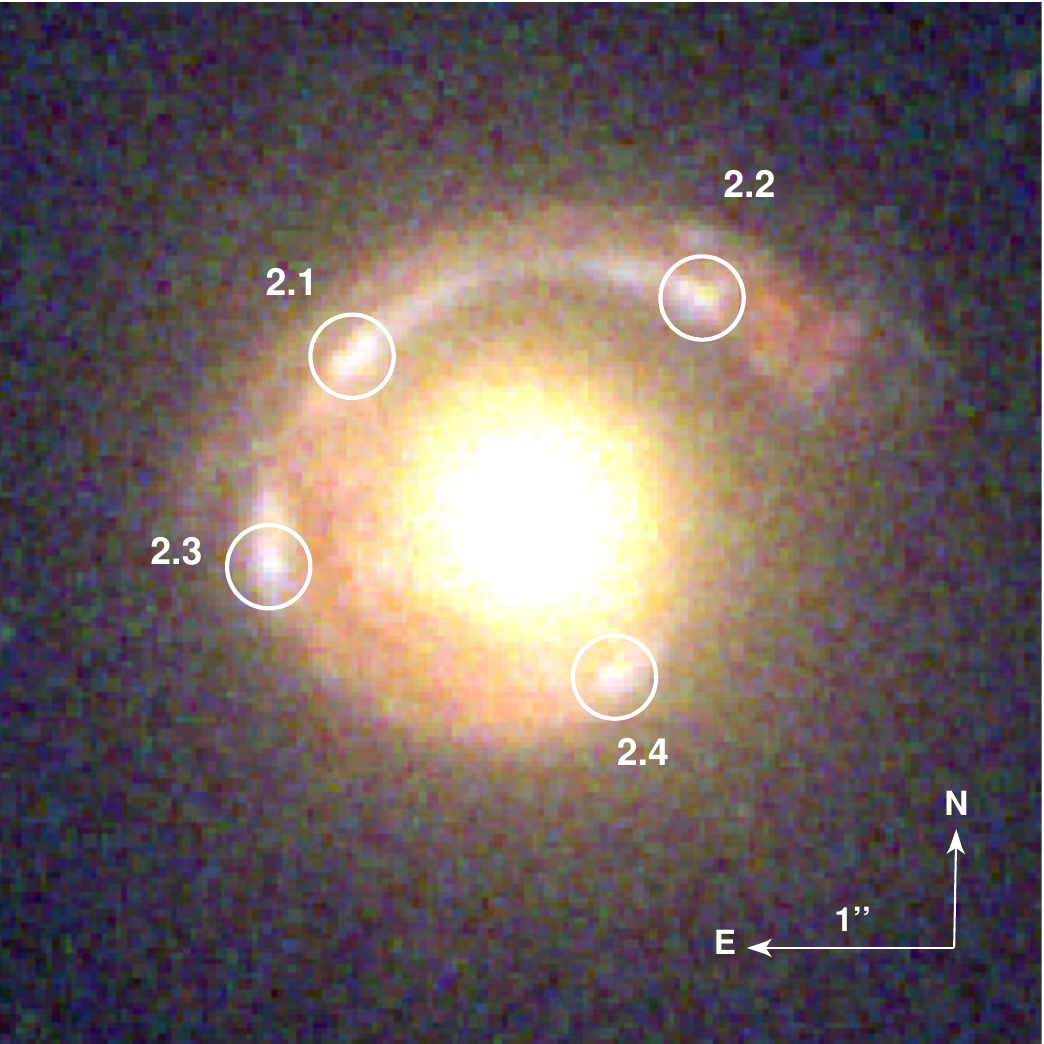}
\caption{\emph{HST} composite colour image of System\ 2 in RX\,J2129 using F475W, F606W, and F814W pass-bands. Multiple images used as constraints are highlighted by white circles.}
\label{rxj2129_sys2}
\end{figure}

\begin{figure*}
\centering
\includegraphics[width=0.99\textwidth]{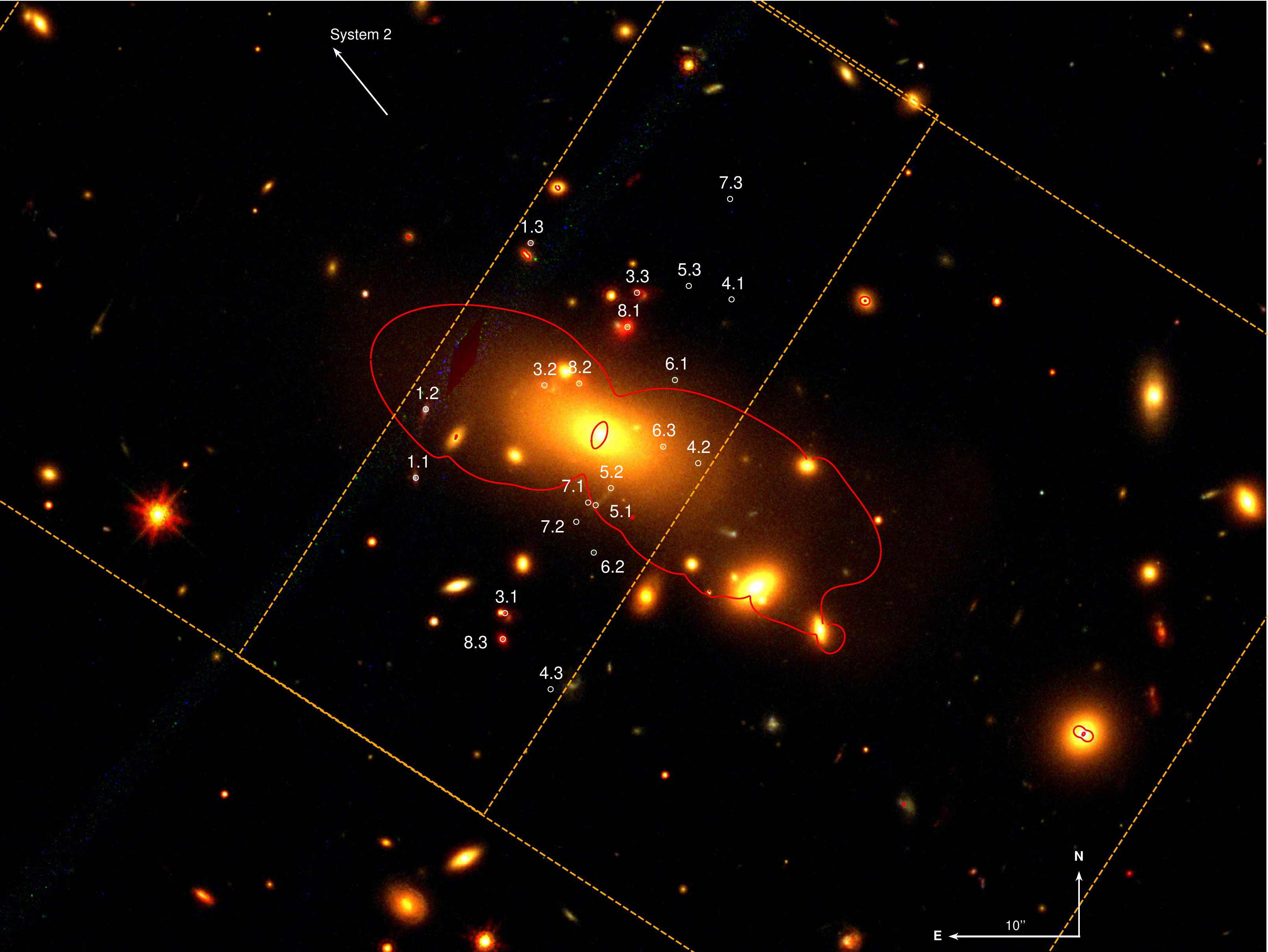}
\caption{\emph{HST} composite colour image of RX\,J2129 using F475W, F814W, and F160W pass-bands. Multiple images used as constraints are highlighted by white circles. Critical lines are displayed for a source at redshift $z=1.52$ in red. Orange dashed squares show the MUSE fields of view. System\ 2 is outside the field of view, and can be seen in Fig.\ref{rxj2129_sys2}.}
\label{rxj2129_big}
\end{figure*}

\begin{figure*}
\centering
\includegraphics[width=0.99\textwidth]{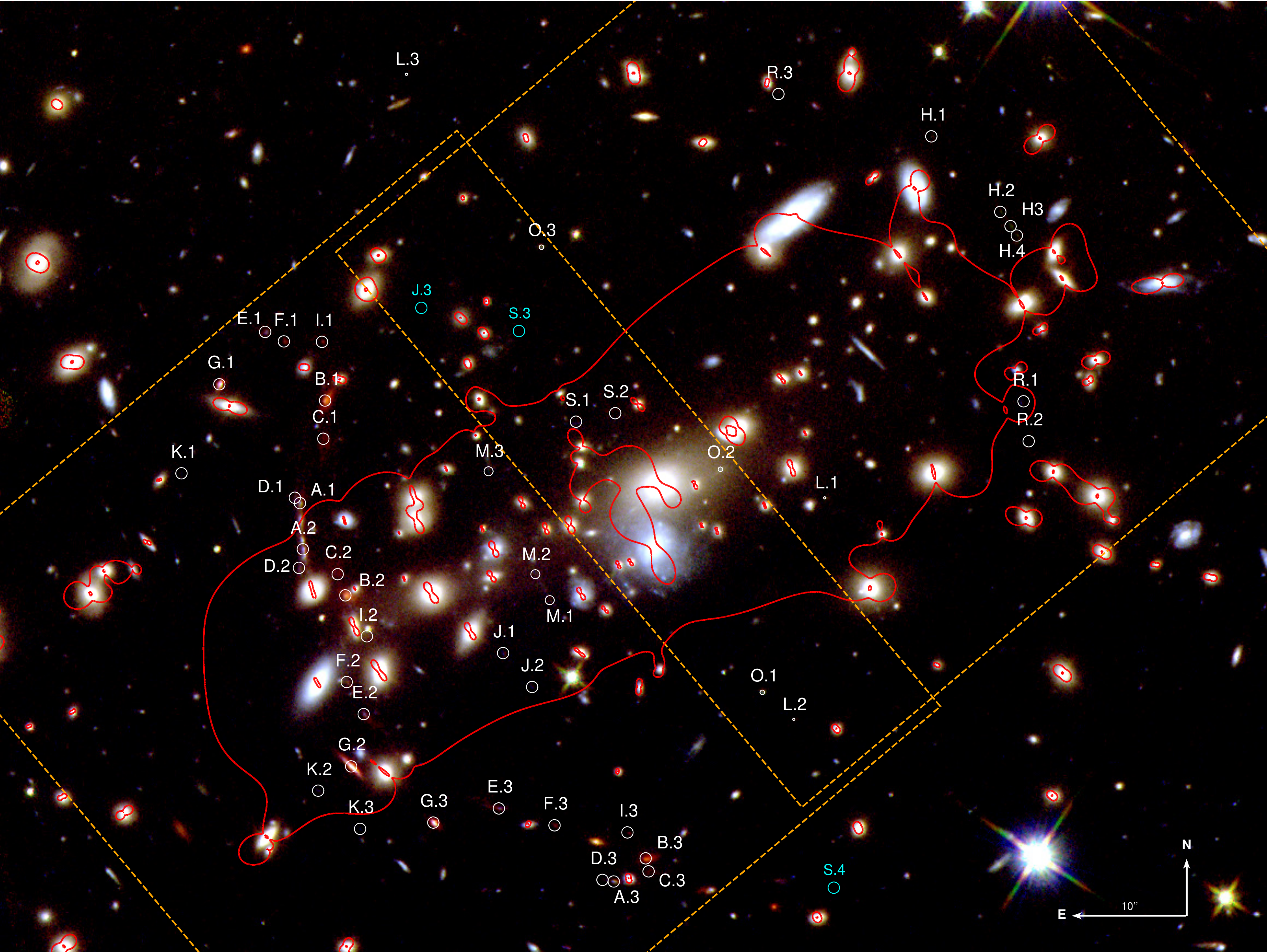}
\caption{\emph{HST} composite colour image of MS\,0451 using F814W, F110W, and F160W pass-bands. Multiple images used as constraints are highlighted by white circles. Cyan circles highlight the predicted positions of the counter-images of System\ J and System\ S. Critical lines for a source at redshift $z=2.9$ (redshift of System\ A) are shown as red lines. Orange dashed squares show the MUSE fields of view.}
\label{ms0451_big}
\end{figure*}

\begin{figure*}
\centering
\includegraphics[width=0.99\textwidth]{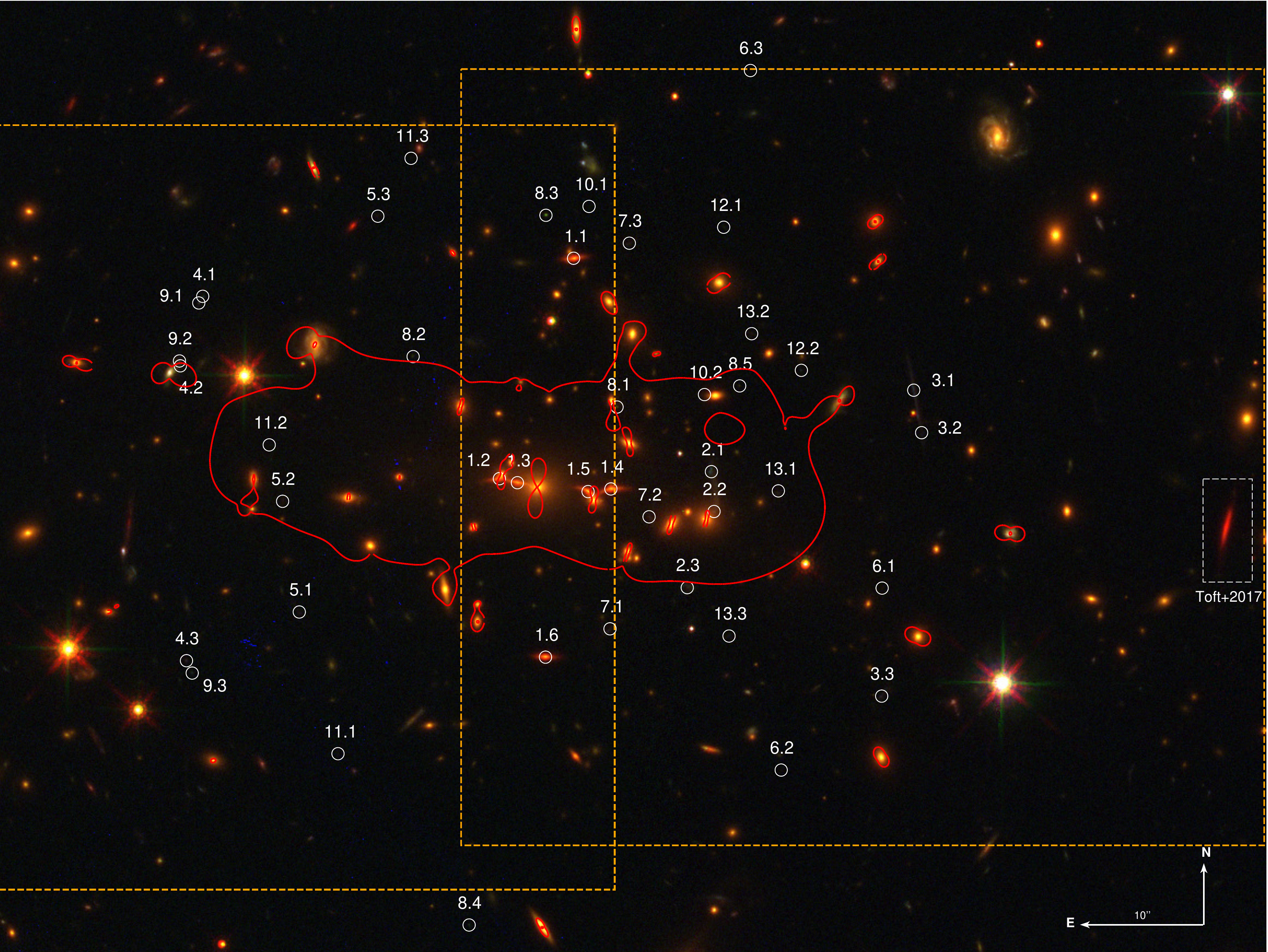}
\caption{\emph{HST} composite colour image of MACS\,J2129 using F475W, F814W, and F160W pass-bands. Multiple images used as constraints are highlighted by white circles. Critical lines are displayed as red lines for a source at redshift $z = 1.36$. Orange dashed squares show the MUSE fields of view. The white dashed box highlight the galaxy at $z=2.1478$ detected by \citet{toft2017}.}
\label{macs2129_big}
\end{figure*}

\subsection{Multiple images}
We use the catalogues of sources described in Sect.\,\ref{section3}, to carry out the search for multiple images. We start by comparing our data to the lists of multiple images used in the \emph{fiducial models}, and thus add the spectroscopic redshift information when available. 
For each sources identified as background in our catalogues, and not already identified by previous strong-lensing analyses, we use the \emph{fiducial models} to predict their multiplicity. 

If a background source is confirmed as multiple, the narrow-band datacube at the wavelength corresponding to its maximum emission is used, and combined with composite colour images made from different combinations of \emph{HST} filters, to identify all the multiple images of the system.
When a multiple image system is then confirmed, it is added as a constraint to the new mass model. 
The lists of multiple image systems used in RX\,J2129, MS\,0451, and MACS\,J2129, are given in Table\,\ref{mul_rxj2129}, Table\,\ref{mul_ms0451}, and Table\,\ref{mul_macs2129}, respectively. They are also highlighted with white circles in Fig.\,\ref{rxj2129_big}, Fig.\,\ref{ms0451_big}, and Fig.\,\ref{macs2129_big} respectively. Further details about the redshift measurements of multiple image systems are presented in Appendix\,\ref{A_A} for MS\,0451 (e.g. the strongest spectral lines, together with \emph{HST} stamps of the multiple images), and we refer the reader to \cite{caminha2019} for similar information regarding RX\,J2129 and MACS\,J2129.

\subsubsection{RX\,J2129}
Our \emph{fiducial model} was built from the parametric model presented in \cite{zitrin2015}, which includes 4 multiple images, and adapted within the \textsc{Lenstool} framework. We here present the results out of our redshift extractions from the MUSE datacube, and compare them with the results presented in \cite{caminha2019} who analyzed the same MUSE observations.

Our spectroscopic redshift measurements for System\ 1 ($z=1.522$, QF$=$3, Mg/Fe absorber), System\ 5 ($z=0.916$, QF$=$2, OII emitter), and System\ 3 ($z=1.52$, QF$=$3, Mg absorber) are in excellent agreement with the previous measurements presented by \cite{caminha2019}.
One will note that the redshift of System\ 1 was initially measured by \cite{belli2013}.
Moreover, we confirm the initial identification of 4 new multiple image Systems reported in \cite{caminha2019}. System\ 4 ($z=3.4270$, QF$=$3, [OII] emitter), System\ 6 ($z = 0.6786$, QF$=$3, [OII] emitter), System\ 7 ($z = 3.08$, QF$=$3, Ly-$\alpha$ emitter), and System\ 8 ($z = 1.52$, QF$=$3, Mg and Fe absorbers) are confirmed triply-imaged systems.

The main difference between the constraints used by \cite{caminha2019} and our analysis is the inclusion of System\ 2 here, a galaxy-galaxy lensing system. It is located relatively far from the cluster center, 81\arcsec\ from the BCG, in the vicinity of a massive isolated galaxy outside the field of view covered by MUSE (i.e. R.\,A.$=322.429$, Decl.$=0.1082$), and can be seen in Fig.\,\ref{rxj2129_sys2}. \cite{desprez2018} studied this System in detail, and its impact on the overall mass reconstruction of the cluster. 
For the analysis presented here, we assume the photometric redshift measured by \cite{desprez2018}, $z=1.61$.

\subsubsection{MS\,0451}
Our \emph{fiducial model} is based on the \textsc{Lenstool} model from \cite{mackenzie}, then revised by the identification of 8 new multiple image systems, including a $z=6.7$ quintuple image in the North of the cluster \citep[][and will be presented in detail in Richard et al. \emph{in prep.}]{knudsen2016}. We now present the new measurements for MS\,0451, and thus the constraints added to our new mass model.

\paragraph*{New multiple image systems --} 
We report the identification of two new systems of multiple images at high redshifts. System R ($z=3.7645$, QF = 3) is composed of three multiple images, and is located in the poorly-constrained Northern region of the cluster. System S ($z=4.4514$, QF = 3) is predicted to be quadruply-imaged but only two multiple images could be identified. 
These sources are Ly-$\alpha$ emitters identified thanks to the blind-search carried out directly in the datacube with \textsc{muselet}. These two systems are located within two poorly-constrained regions of the cluster, and are playing a substantial role in the improvement of the accuracy of the model as described in Sect.\,\ref{section5}.
The strongest MUSE spectral lines of Systems\ R and S are presented in Fig.~\ref{detail_spec_RS}.

\paragraph*{Confirmation \& measurement of known systems --}
We report the measurement of a spectroscopic redshift for all three multiple images of System G, $z = 2.93$ (QF$=$3, Ly-$\alpha$ emitter). We also measure a spectroscopic redshift for System A, $z = 2.92$, which is in agreement with the previous measurement from \citet{borys2004} and \cite{berciano_alba2010}. The strongest MUSE spectral lines of System\ G are presented in Fig.~\ref{detail_spec_G}.

\paragraph*{Other systems --} 
The redshift of System H is measured at $z = 6.7$ by \cite{knudsen2016}, and Richard et al.\ (in prep.). 
Image\ P.2 is located in a bright region surrounding the BCG which makes it difficult to identify. Moreover Image\ P.1 is located in the vicinity of a cluster member which increases the uncertainty on the location of the system. Therefore, System\ P was flagged as insecure and is not used in the model. For the remaining 11 systems without spectroscopic confirmation, their redshift is being optimized by the model.

\begin{figure*}
\begin{center}
\includegraphics[width=0.19\linewidth]{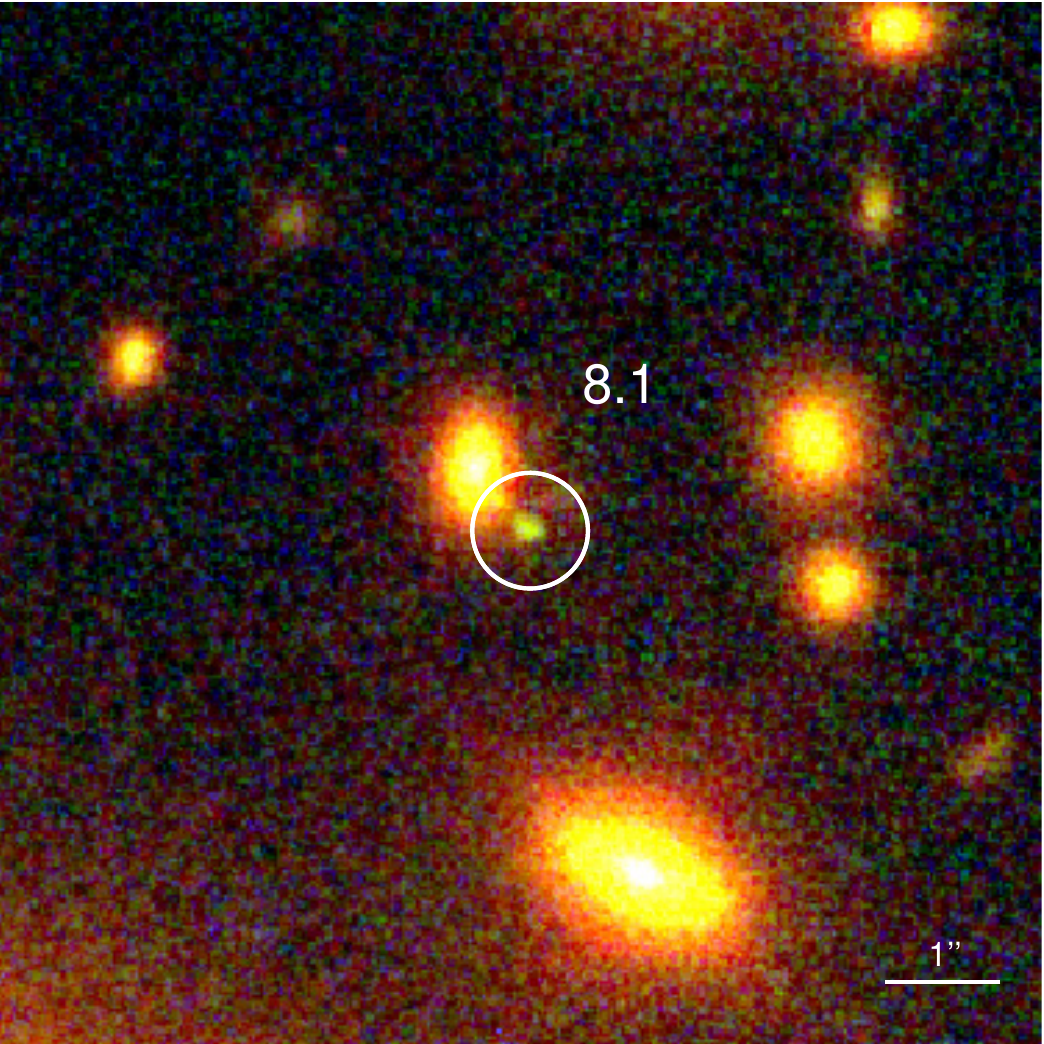}\ 
\includegraphics[width=0.19\linewidth]{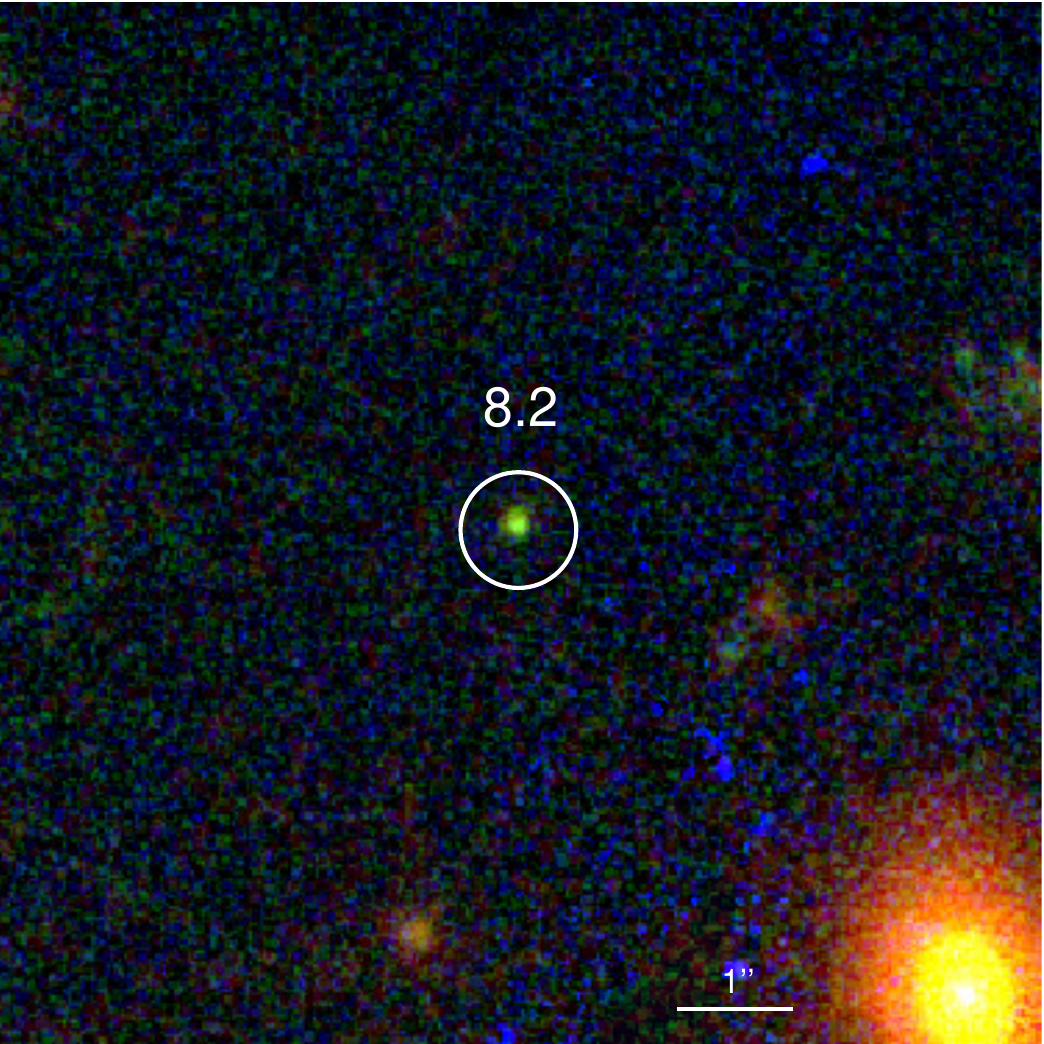}\ 
\includegraphics[width=0.19\linewidth]{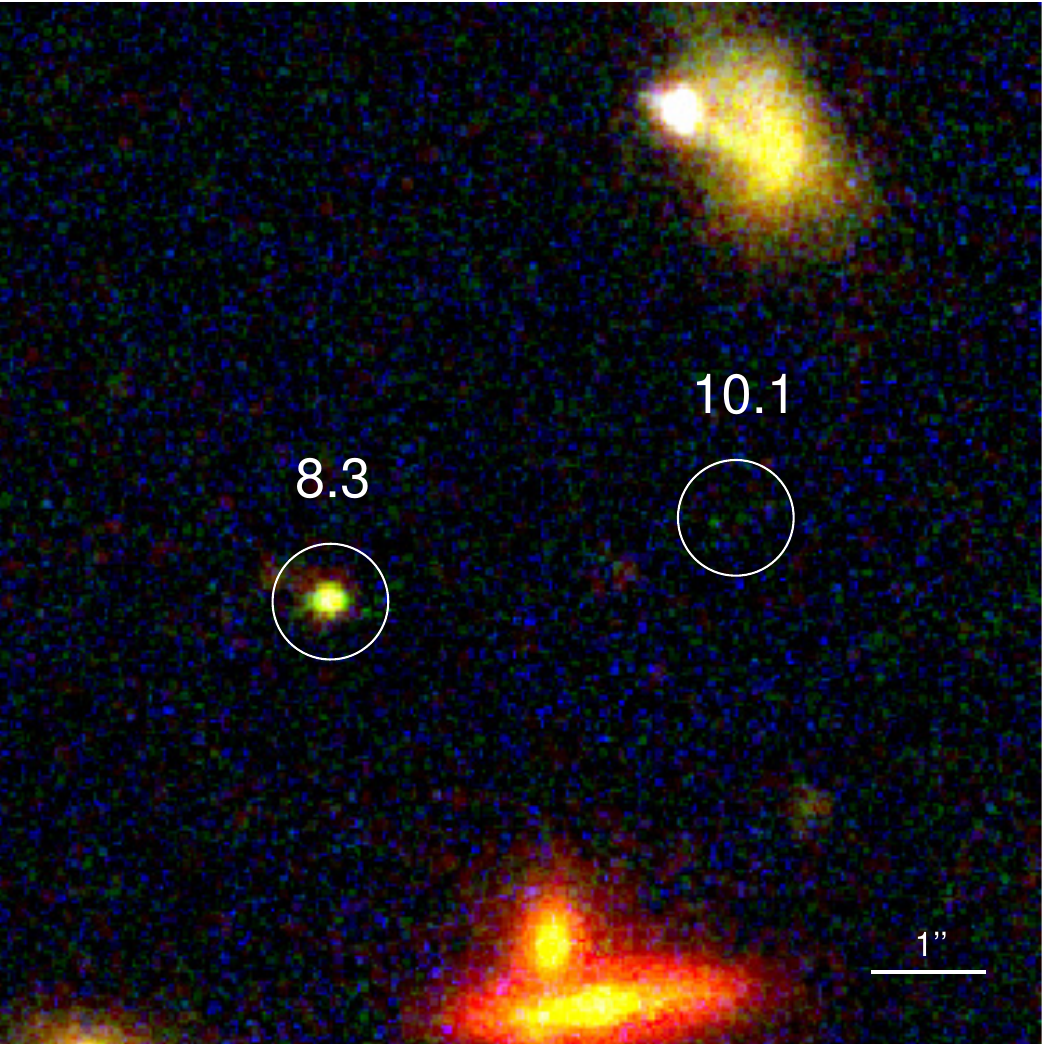}\ 
\includegraphics[width=0.19\linewidth]{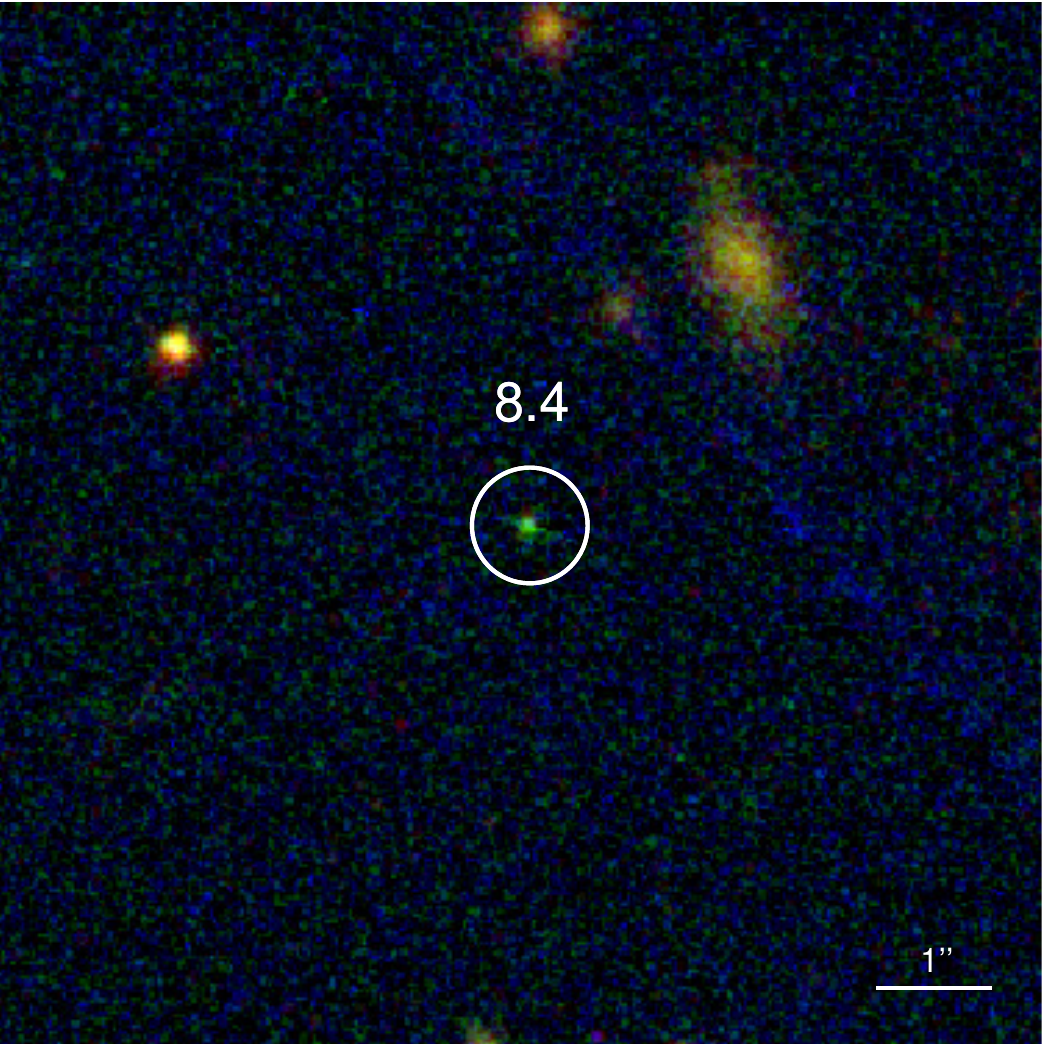}\ 
\includegraphics[width=0.19\linewidth]{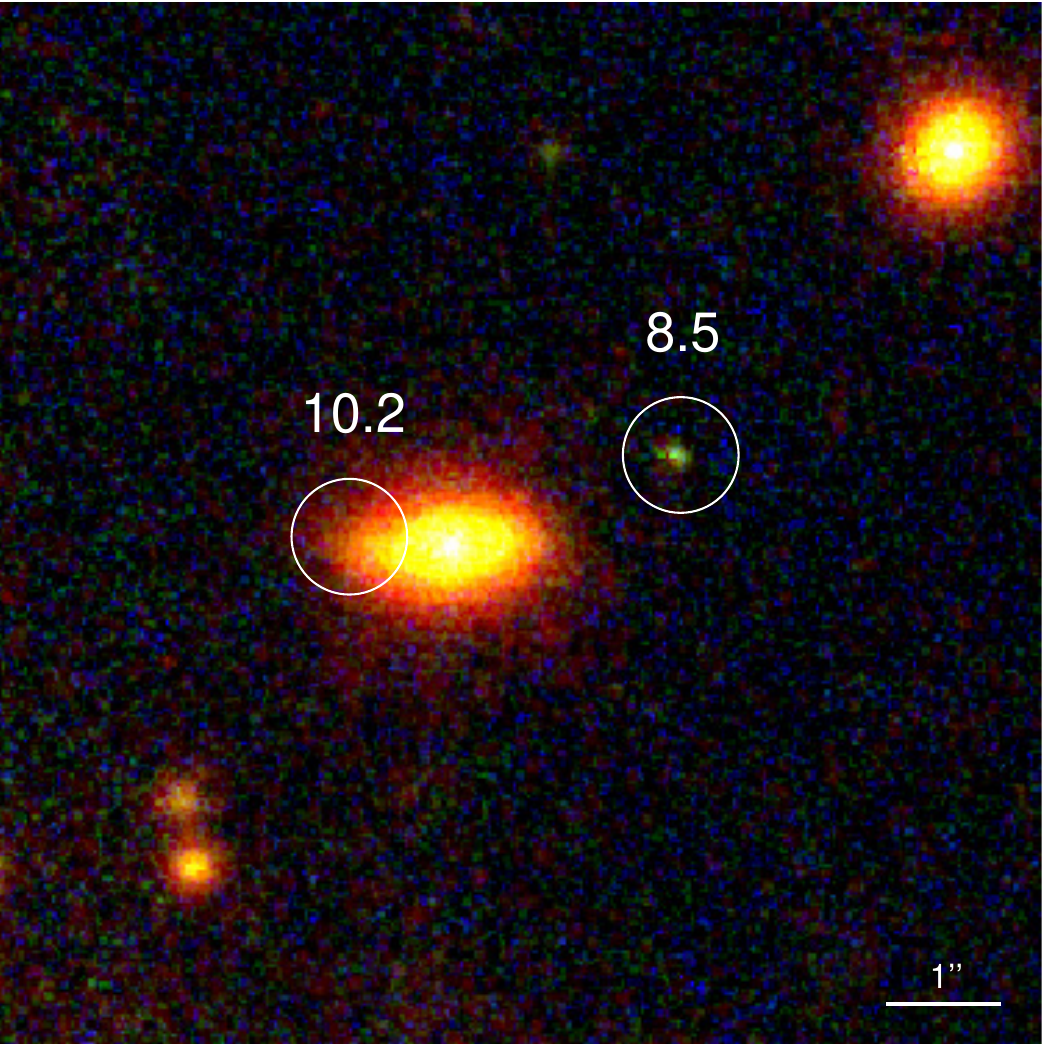}\\
\includegraphics[width=0.19\linewidth]{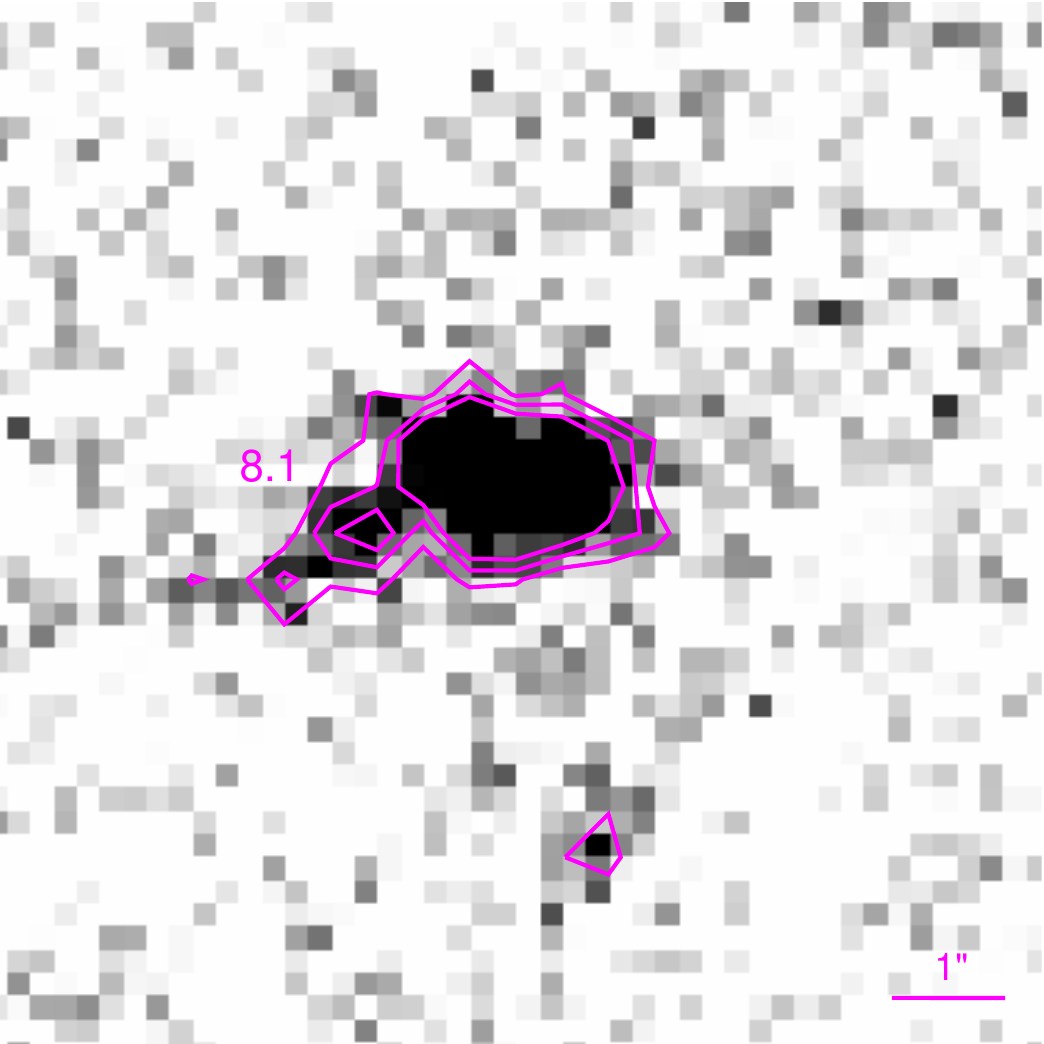}\ 
\includegraphics[width=0.19\linewidth]{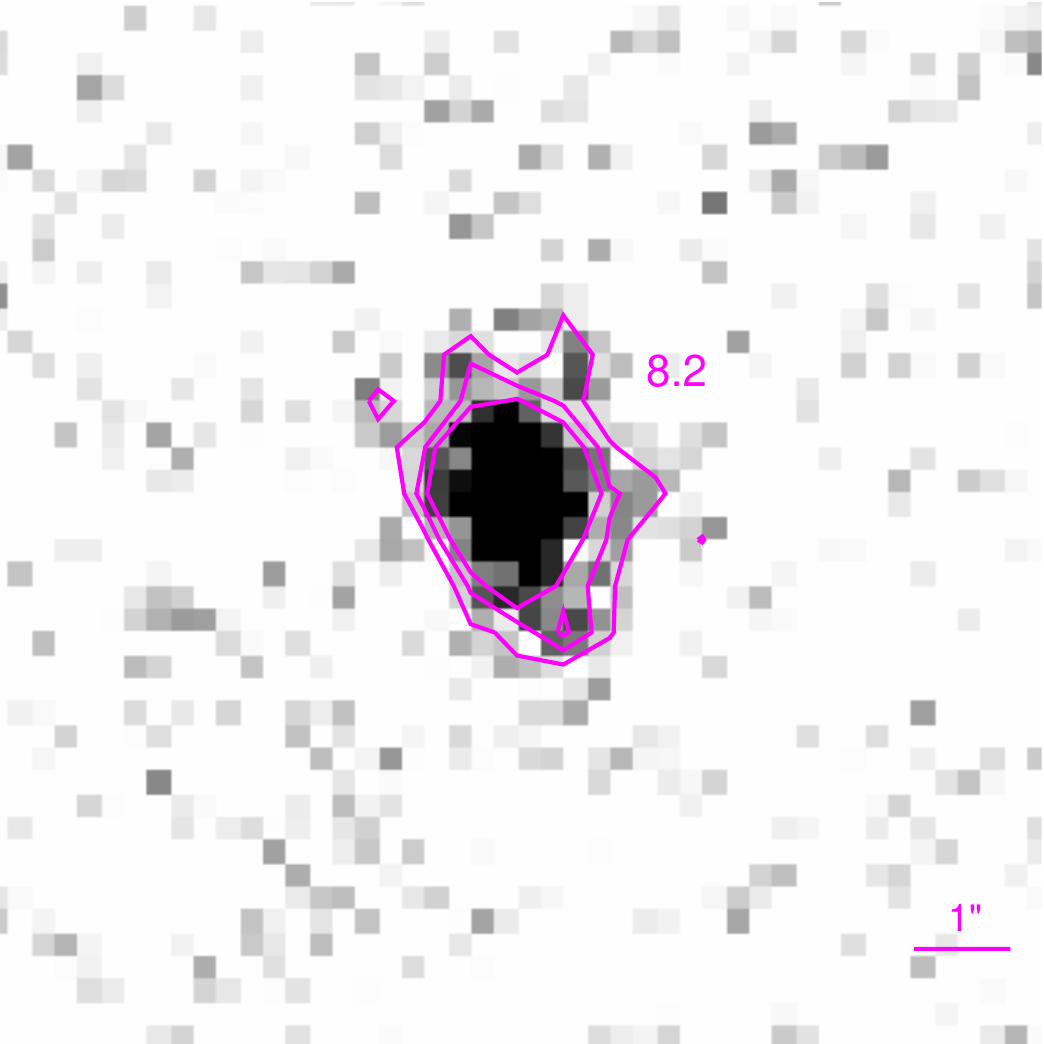}\ 
\includegraphics[width=0.19\linewidth]{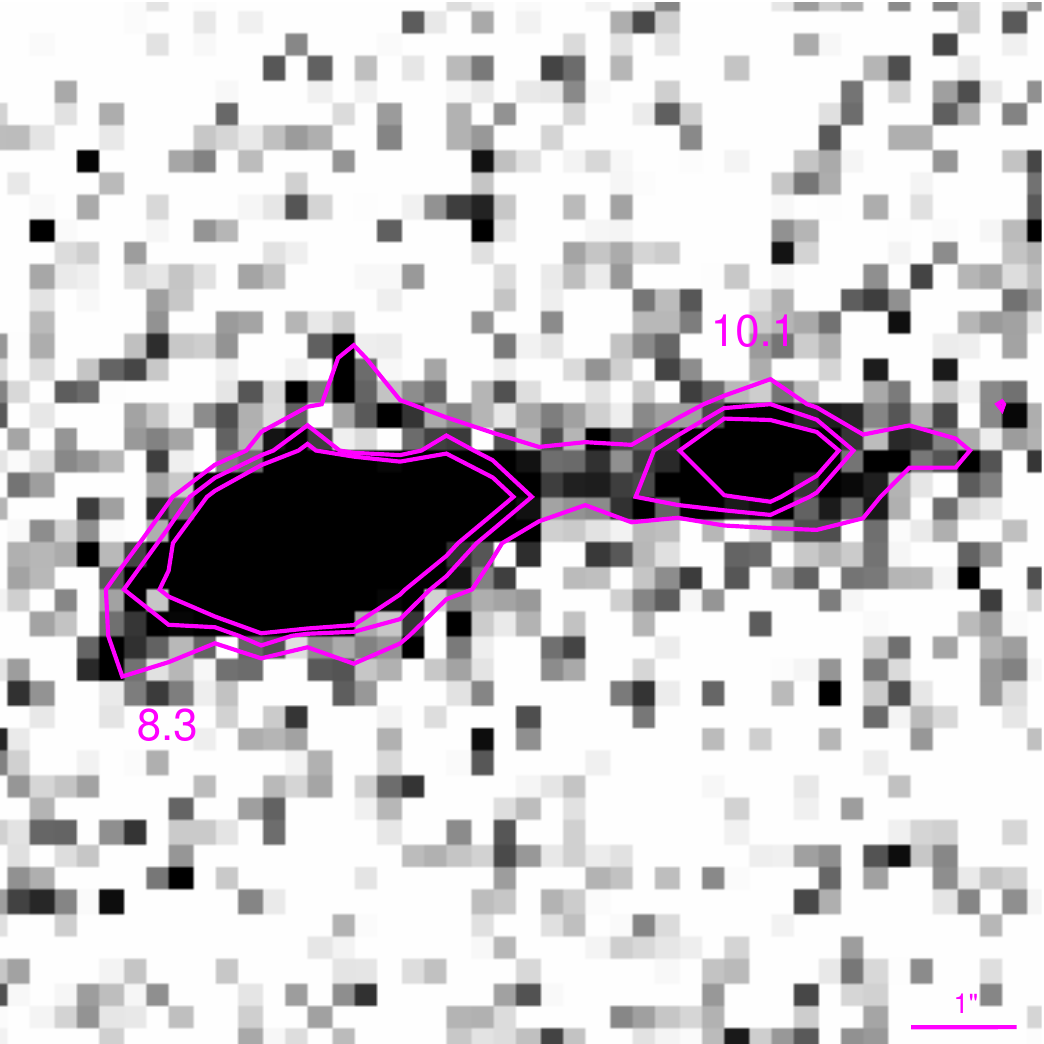}\ 
\includegraphics[width=0.19\linewidth]{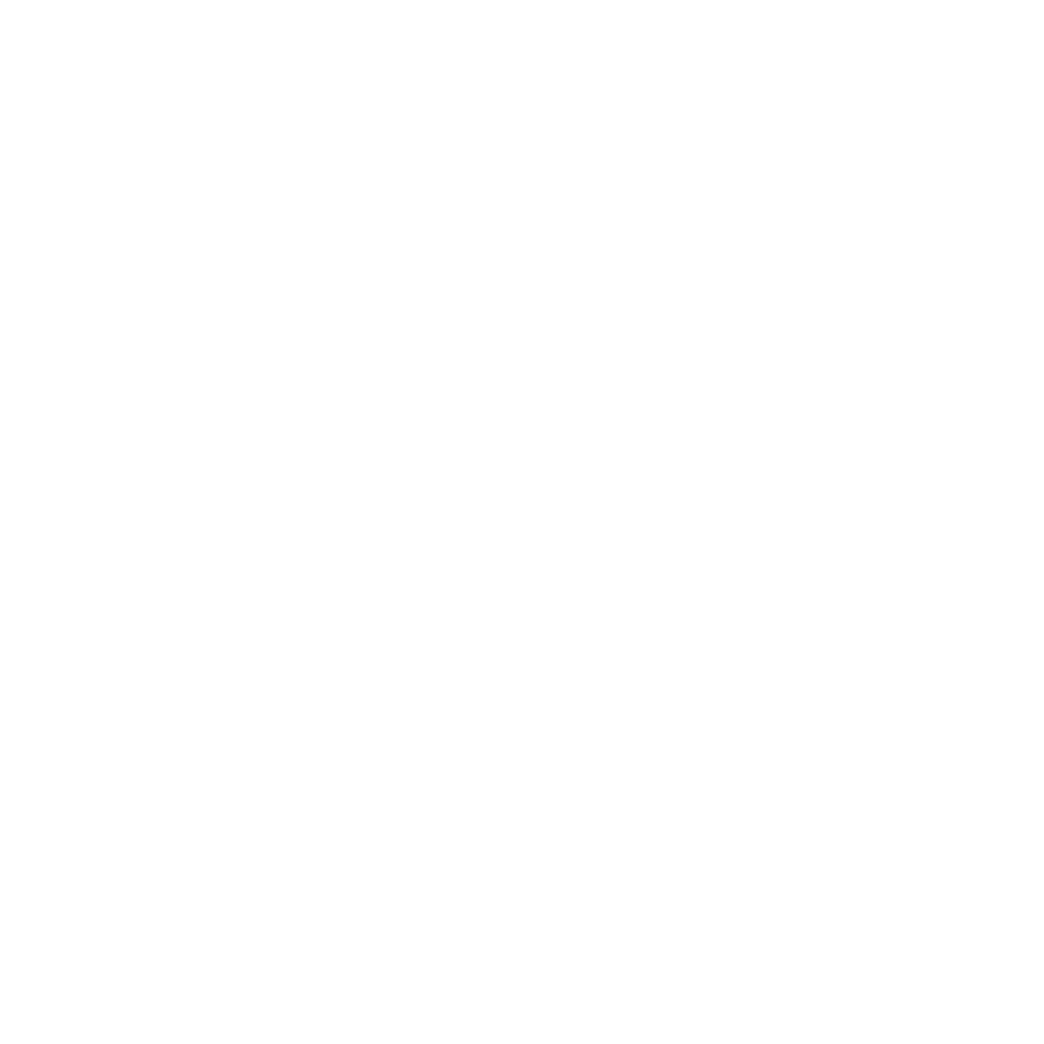}\ 
\includegraphics[width=0.19\linewidth]{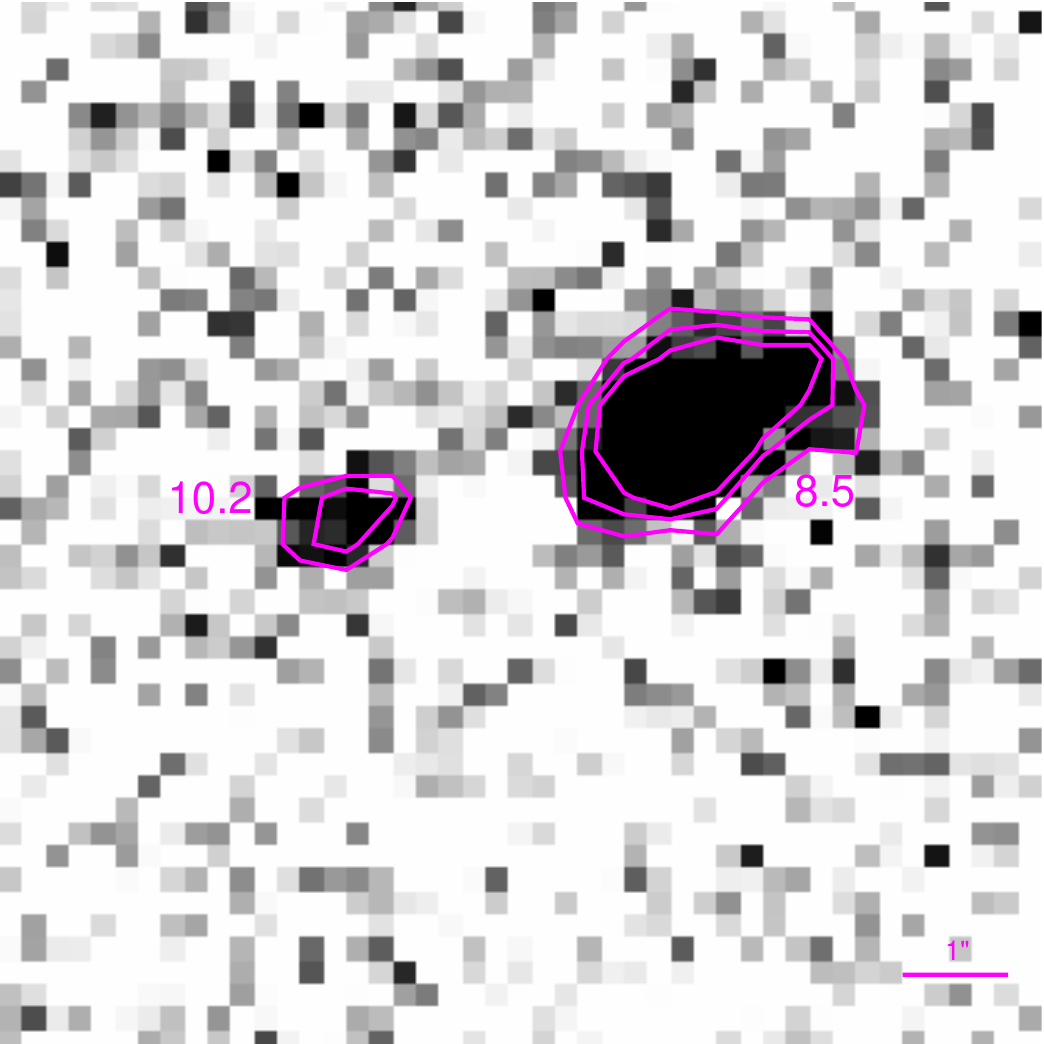}\\
\includegraphics[width=0.48\textwidth]{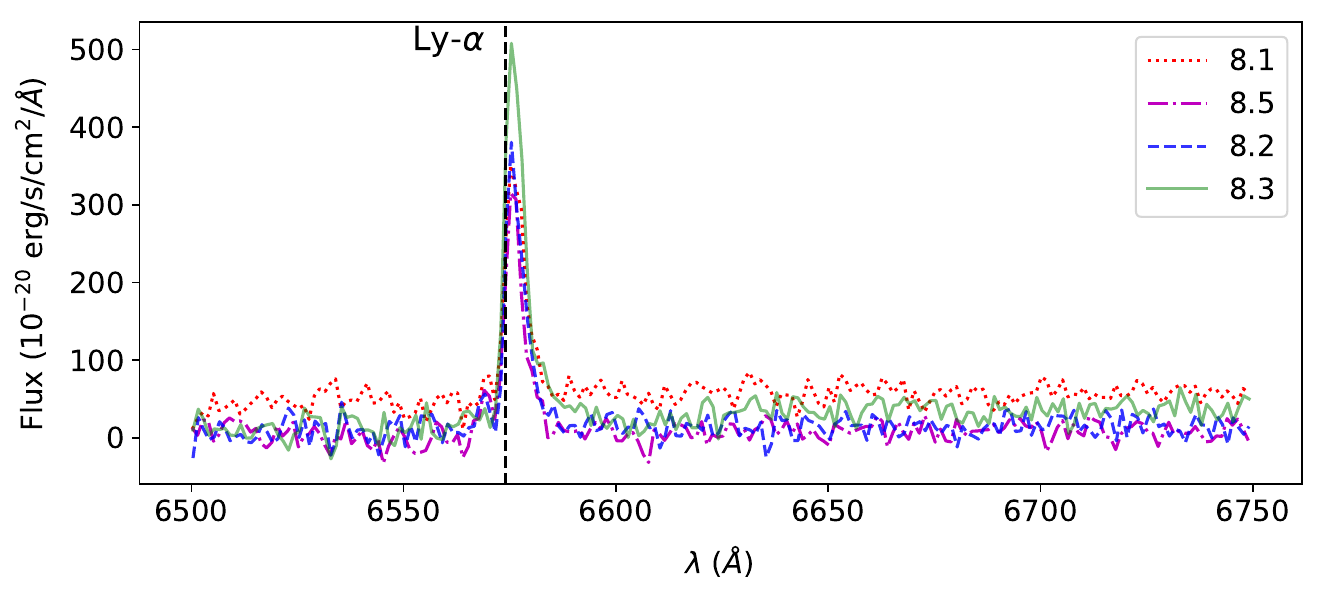}\ 
\includegraphics[width=0.48\textwidth]{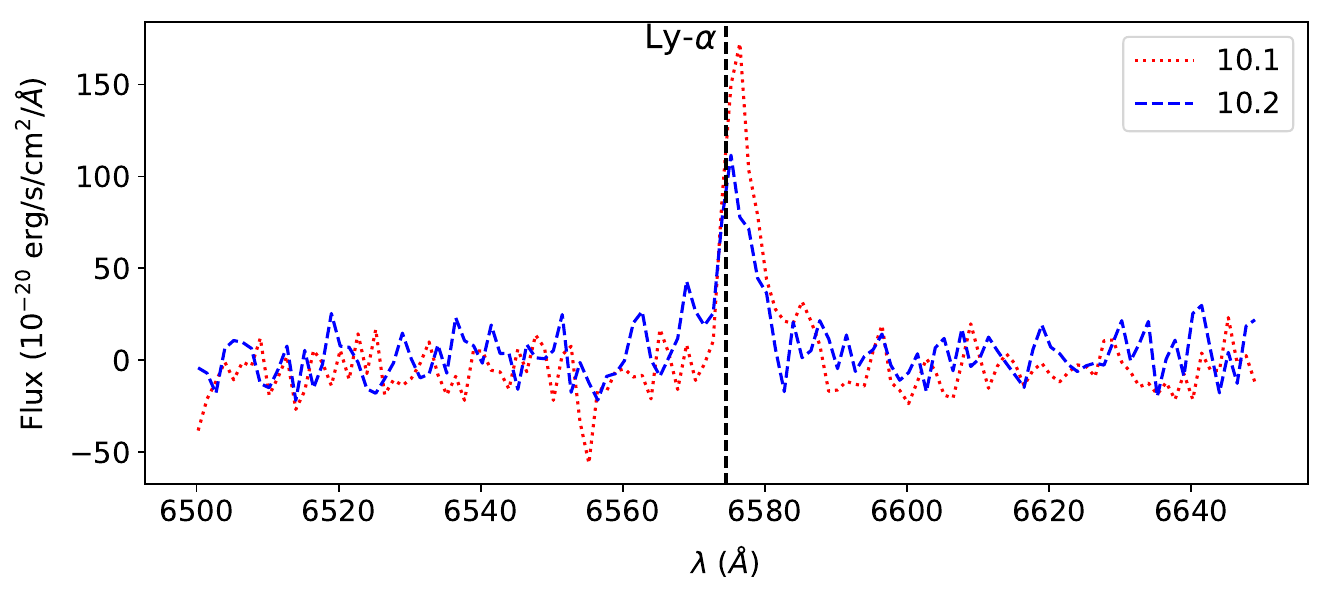}
\caption{\textbf{MACS\,J2129 --} System\ 8 \citep[System\ 10 in][]{caminha2019} and System\ 10 with a measured spectroscopic redshift, $z=4.41$. 
\emph{Top:} Composite colour \emph{HST} stamps of the multiple images 8.1 to 8.5, including 10.1 and 10.2 close to 8.3 and 8.5.
\emph{Middle:} \textsc{muselet} narrow-band datacube stamps at the wavelength of the maximum emission of the source, i.e. $\lambda=6577{\AA}$ (right panel). The contours are displayed in magenta for 2, 3 and 4\,$\times10^{-20}$\, erg\,s$^{-1}$\,arcsec$^{-2}$.
\emph{Bottom:} 
Extracted spectra from the MUSE datacube 
centered on the most prominent lines 
for images of System\ 8 (left) and System\ 10 (right).
}
\label{detail_spec_810}
\end{center}
\end{figure*}

\subsubsection{MACS\,J2129}
Our \emph{fiducial model} was built from the model presented in \citet{monna} combining CLASH photometry (see Table~\ref{hst_infos}) and VLT/VIMOS spectroscopy (PI: Rosati, ID: 186.A-0798). It relies on 9 multiple image systems, 7 of which were spectroscopically confirmed back then. 
We then extracted spectroscopic redshifts from the MUSE datacubes, and compared our results with \cite{caminha2019} who analyzed the same MUSE observations. 

Our analysis measures similar redshifts to \cite{caminha2019} for Systems\ 1, 2, 3, 7 and 8, in good agreement with \cite{monna} measurements. 
We measure a redshift of $z=4.41$ (QF$=$3, Ly-$\alpha$ emitter) for Images\ 8.1, 8.2 and 8.3, in agreement with the redshift of 8.1 measured by \citet{monna}, and confirmed by \cite{caminha2019}. The fourth image of System\ 8, Image 8.4, is outside the MUSE field of view. \cite{caminha2019} identified the counter-image\ 8.5 which we confirm as well. We show in Fig.\,\ref{detail_spec_810} composite colour \emph{HST} stamps of the five images, narrow-band images, and their spectra extracted from the MUSE datacubes.
We cannot obtain reliable redshift measurements from the extracted spectra of Systems\ 4, 5 and 9. We thus use the spectroscopic redshifts measured with VIMOS in \citet{monna} for Systems\ 4 and 9 (QF$=$2), and optimize the redshift of System\ 5 in our mass model. 
The redshift of System\ 6 ($z = 6.85$) was spectroscopically confirmed by \citet{huang2016}, and could not be measured with MUSE since the wavelength corresponding to the maximum emission is greater than the upper limit of the MUSE wavelength range ($\sim$9350\,{\AA}), as in the case of the $z=6.7$ quintuply-imaged system in MS\,0451 \citep[][Richard et al.\ \emph{in prep.}]{knudsen2016}.

With this work, we confirm the 4 new spectroscopic identifications of multiple image systems identified by \cite{caminha2019}, and present the new identification of one system: Systems\ 10 ($z = 4.41$, QF$=$3, Ly-$\alpha$ emitter), 11 ($z = 3.1081$, QF = 3, Ly-$\alpha$ emitter), 12 ($z = 3.897$, QF$=$3, Ly-$\alpha$ emitter), 13 ($z = 1.3585$, QF$=$2, [OII] emitter), and 14 ($z=1.4519$, QF$=$1, [OII] emitter).
System\ 10 is not identified by \cite{caminha2019}, however they detect Image\,10.1 which is listed in the public release of their catalogue\footnote{\url{http://cdsarc.u-strasbg.fr/viz-bin/Cat?J/A+A/632/A36}}. Our model predicts it as quadruply-imaged, although only two multiple images could be confirmed with MUSE.
This system seems to be linked to System\ 8, composed of 5 multiple images. Image\ 10.1 appears to be located close to the Ly-$\alpha$ emission of Image\ 8.3, although the two emission regions are well separated. This could imply that the two systems highlight two physically connected lensed galaxies at $z=4.41$, System\ 8 being more UV bright and detected in the \emph{HST} images.
Image\ 10.2 is located relatively close to Image\ 8.5, separated by a cluster galaxy (G3 in our model discussed in Sect.\,\ref{section:clumps} and visible in Fig.\,\ref{rms}), however the feature is similar to what is visible in the case of Images\ 8.3/10.1.
We find a possible candidate for another image of system 10 located close to Image\ 8.2, as a faint Ly-$\alpha$ tail. However the detection has a low signal-to-noise ratio, we therefore do not include that image in our model.
System\ 10 is extremely faint in the \emph{HST} image, it is thus difficult to securely identify its fourth counter-image which is predicted outside the MUSE field of view, relatively close to Image\ 8.4.
Figure\,\ref{detail_spec_810} shows composite \emph{HST} stamps of Images\ 10.1 and 10.2, their narrow-band images, and the spectra extracted from the MUSE datacubes, along with the five multiple images of System\ 8.
We therefore include the 5 images for System\ 8 in our model, and the 2 images of System\ 10 detected in the MUSE datacubes. 
Regarding System\ 14 \citep[System\ 5 in][]{caminha2019}, our spectroscopic measurement is flagged as unsecure, we therefore decide to not include this system as a constraint in our strong-lensing mass model, but list it in Table\,\ref{mul_macs2129} for consistency with \cite{caminha2019}.
\cite{caminha2019} also lists a counter-image for System\ 12, Image\ 12.3 in Table\,\ref{mul_macs2129}, System\ 9 and Image\ 9c in their analysis. This counter-image is not spectroscopically confirmed, and we are not convinced by its colour and morphology in the \emph{HST} imaging. We thus only consider it as a candidate.

\subsection{Cluster- and galaxy-scale Components}
\label{section:clumps}
For each cluster, we compare the MUSE spectroscopically confirmed cluster members to the list of cluster galaxies used in the \emph{fiducial models}. 
MUSE observations allowed us to identify 43, 112, and 89 cluster galaxies for RX\,J2129, MS\,0451, and MACS\,J2129 respectively. Among those, 15 in RX\,J2129 and 4 in MACS\,J2129 are new detections, i.e. not reported by \cite{caminha2019}. We combine those with the cluster member galaxies identified by previous works, using standard color-magnitude red-sequence selections as well as spectroscopic identifications with different instruments. We use cluster identifications from \cite{desprez2018}, \cite{monna} and Richard et al.\ (in prep.) for RX\,J2129, MS\,0451 and MACS\,J2129 respectively.
In total, we used 70, 144, and 151 cluster members for the modeling of RX\,J2129, MS\,0451, and MACS\,J2129 respectively. We used the method presented in Sect.\,\ref{section3} to optimize the parameters of each subhalo. 
As explained before, we optimize a selection of large-scale and galaxy-scale halos for each cluster. The best-fit parameters obtained are listed in Table\,\ref{tab:res_pot}, and discussed in this Section. For scaling relations, the reference magnitudes are $mag_{\rm{F814W}}=17.49$, $mag_{\rm{F814W}}=18.69$, and $mag_{\rm{F814W}}=19.19$ for RX\,J2129, MS\,0451, and MACS\,J2129 respectively.
The shapes of the individually optimized potentials from the best-fit are shown in the right panel of Fig.~\ref{rms}.

\subsubsection{RX\,J2129}
Our model contains one large-scale halo described by a PIEMD mass component. 
We model individually the BCG, and include an isolated cluster galaxy (G1 in Table\,\ref{tab:res_pot}), which acts as the lens for System\ 2 as detailed in \cite{desprez2018}. The four multiple images of System\ 2 are attributed to galaxy-galaxy lensing and are used as constraints in our model as shown in Table\,\ref{mul_rxj2129}). 
One can see a zoom in on this lensing configuration in Fig.\ref{rxj2129_big}.

Our best-fit mass model has an rms$=$0.29\arcsec. 
The best-fit parameters of the model are given in Table~\ref{tab:res_pot}, and the center of the cluster is chosen as (R.\,A.$=322.41651$,Decl.$=0.08923$).
In order to check the impact of the addition of free paramters by the inclusion of G1 as an individual potential, we run a model which treats G1 as a standard cluster member, i.e. following the \cite{faber1976} relation. While the rms ($\chi^2$, $d.o.f.$) of this model is not significantly different from our best-fit mass model, 0.28\arcsec (8,22) vs 0.29\arcsec (9,24), it has a local impact. Indeed, the rms of System\,2 is 0.15\arcsec, while we obtain an rms of 0.09\arcsec when G1 is included. We therefore conclude that G1 improves the model and is necessary to precisely recover the geometry of System\,2.

\begin{table*}
\caption{\label{tab:res_pot} Best-fit parameters of the mass models for all three clusters. 
Column\ (1) gives the cluster name. In brackets, we list the rms, the number of degrees of freedom, $d.o.f.$, and the number of multiple images respectively for each best-fit model. Column (2) is the ID of the optimized potential. The IDs of the large-scale halos are highlighted in bold. Columns (3) to (9) are respectively the R.\,A. and Decl. of the center for each halo and subhalo in arcseconds relative to the chosen center as given in Sect.\ref{section4}, the ellipticity, the orientation angle, the core radius, the truncation radius, and the velocity dispersion for each optimized potentials. The position angle, $\theta$, is given in degrees and is defined as the direction of the semi-major axis of the iso-potential, counted counterclockwise from the horizontal axis (being the R.\,A. axis). For each free parameter of the models, we indicate the $1\sigma$ error bars computed from the posterior distribution of the MCMC samples.
For scaling relations, the reference magnitudes are $mag_{\rm{F814W}}=17.49$, $mag_{\rm{F814W}}=18.69$, and $mag_{\rm{F814W}}=19.19$ for RX\,J2129, MS\,0451, and MACS\,J2129, and respectively.
Values in brackets are fixed in the model.}
\begin{center}
\begin{tabular}{ccccccccc}
\hline
\hline
Cluster & ID & R.\,A. &  Decl. & e & $\theta$ & $r_{c}$ & $r_{t}$ & $\sigma$ \\
 & & [\arcsec] & [\arcsec] &  & [deg] & [kpc] & [kpc] & [km/s] \\
\hline 
\hline
\multirow{3}{*}{RX\,J2129}
 & \textbf{C1} & $3.13 \pm 0.77$ & $-1.94\pm0.41$ & $0.69\pm0.04$ & $-23.4\pm0.8$ & $54\pm6$ & [1000] & $920\pm21$\\
 \multirow{3}{*}{(0.29\arcsec,22,25)}
 & BCG & [0.0] & [0.0] & [0.49] & [-35.4] & [0.3] & $64\pm37$ & $317 \pm 25$ \\
 & G1 & [-44.2] & [68.0] & [0.11] & [-50.6] & [0.15] & $56^{+5}_{-40}$ & $193^{+2}_{-13}$\\
   & L$^{\ast}$ galaxy & -- & -- & -- & -- & [0.15] & $10\pm5$ & $151\pm25$ \\
 \hline
\multirow{9}{*}{MS\,0451} & \textbf{C1} & $-7.5_{-1.2}^{+0.9}$ & $-2.6_{-0.7}^{+0.6}$ & $0.63_{-0.03}^{+0.04}$ & $32.2 \pm 0.5$ & $121_{-7}^{+10}$ & [1000] & $1101_{-34}^{+25}$ \\
\multirow{9}{*}{(0.60\arcsec,18,47)}
  & \textbf{C2} &  $22.3 \pm 1.6$  & $19.5 \pm 2.2$ & $0.18_{-0.06}^{+0.12}$ & $146.6_{-15.8}^{+9.3}$ & $332_{-30}^{+94}$ & $685_{-75}^{+196}$ & $811_{-68}^{+213}$ \\ 
 & BCG & [0.0] & [0.0] & [0.6] & [23.7] & [0.19] & $17 \pm 36$ & $275_{-42}^{+50}$ \\
 & G1 & [-0.49] & [-5.35] & [0.1] & [-28.7] & [0.15] & $141 \pm 31$ & $341_{-33}^{+36}$ \\ 
 & G2 & [22.42] & [26.05] & [0.51] & [-77.2] & [0.15] & $18 \pm 46$ & $105_{-23}^{+38}$ \\
 & G3 & [-29.98] & [-17.33] & [0.62] & [51.7] & [0.15] & $147 \pm 37$ &  $140_{-30}^{+19}$ \\
 & G4 & [11.68] & [23.47] & [0.79] & [34.4] & [0.15] & $102 \pm 44$ &  $45_{-22}^{+45}$ \\
 & G5 & $30.1_{-0.50}^{+0.05}$ & $6.6_{-0.4}^{+0.2}$ & $0.02_{-0.01}^{+0.6}$ & $138_{-13}^{+94}$ & [0.15] & $128_{-44}^{+55}$ & $132_{-17}^{+23}$ \\
 & G6 & [-43.84] & [0.47] & [0.44] & [53.1] & [0.15] & $29 \pm 46$ & $55 \pm 14$ \\
 & L$^{\ast}$ galaxy & -- & -- & -- & -- & [0.15] & $68.0\pm11.5$ & $157\pm11$ \\
 \hline
 \multirow{6}{*}{MACS\,J2129} 
 & \textbf{C1} & $-2.9^{+2.4}_{-0.8}$ & $1.2^{+1.1}_{-0.2}$ & $0.67 \pm 0.01$ & $171.6\pm0.5$ & $96\pm5$ & [1000] & $1125\pm20$\\    
  \multirow{6}{*}{(0.74\arcsec,33,42)}
 & \textbf{C2} & $44.6^{+0.01}_{-1.8}$ & $17.6^{+0.8}_{-2.5}$ & $0.79\pm0.10$ & $59\pm8$ & $135\pm43$ & [1000] & $565\pm4863$\\ 
 & BCG & [-0.07] & [-0.21] & [0.32] & [5.3] & [0.15] & $55^{+35}_{-5}$ & $294\pm42$ \\
 & G1 & [-2.93] & [0.81] & [0.44] & [70.0] & [0.15] & $44\pm29$ & $200\pm33$ \\
 & G2 & [4.53] & [-1.28] & [0.27] & [-31.6] & [0.15] & $43\pm29$ & $184\pm36$ \\
 & G3 & [14.42] & [7.24] & [0.47] & [4.6] & [0.15] & $7\pm2$ & $356\pm48$ \\
 & G4 & [5.92] & [6.84] & [0.39] & [88.5] & [0.15] & $85^{+25}_{-18}$ & $201\pm24$ \\ 
  & L$^{\ast}$ galaxy & -- & -- & -- & -- & [0.15] & $50^{+25}_{-19}$ & $164\pm16$ \\
 \hline
 \hline
\end{tabular}
\end{center}
\end{table*}

\subsubsection{MS\,0451}
Our initial model only included one large-scale halo centered on the BCG of MS\,0451. However, such model could not reproduce precisely the multiple images located in the North of the cluster, with an ${\rm rms}>1.0$\arcsec\ for all systems.
With a single cluster-scale halo model, the critical line at redshift $z=6.7$, corresponding to the redshift of System\ H, would not pass between Images\ H.2 and H.3, nor Images\ H.3 and H.4. 
The same applies for System\ R which, in this context, has an ${\rm rms}>1$\arcsec. 
We thus looked at the distribution of cluster members, and identified two groups of galaxies, at the cluster redshift, located North of the cluster BCG. We thus run a new model which included a second large-scale halo, also modeled with a PIEMD, centered between these two groups. 
We note that we also run a model including three large-scale halos (one for the cluster and the other two for the galaxy groups), however the best-fit model was not statistically better than the 2-halo one. We therefore decided to consider the simplest of the two models.

The BCG was modeled separately using a PIEMD profile where $\sigma$ and $r_{cut}$ are being optimized independently. 
We also include four independent subhalos to model galaxies located in the foreground of the cluster ($z = 0.0623$ - IDs\ G1, G2, G3, and G4 in Table\,\ref{tab:res_pot}). 
As the multi-plane optimization is not yet finalized in \textsc{Lenstool}, the impact of a given foreground galaxy is being assessed by projecting its mass component in the cluster plane. 

Moreover, the blind search in the \textsc{muselet} narrow-band datacube revealed an unidentified cluster member ($z=0.531$) located in the vicinity of Image\ R.1, and too faint to be seen in the \emph{HST} images.
We thus chose to include this cluster member as an individually-optimized potential in our mass model as it acts as a local small-scale perturber for System\ R (potential G5 in Table\,\ref{tab:res_pot}, OII emitter, QF\,$=$\,3). 
Finally, we add one more galaxy-scale halo, the cluster member identified as G6, located closely to Image\ K.1.

Our best-fit mass model has an rms$=$0.60\arcsec. The best-fit parameters of the model are given in Table~\ref{tab:res_pot}, and the center of the cluster is chosen as (R.\,A.$=73.54520$,Decl.$=-3.01439$). This mass model is used for the combined strong and weak-lensing analysis of MS\,0451 presented in \cite{tam2020}.

As was done for RX\,J2129, a model was run excluding all individual galaxy-scale halos, G1, G2, G3, G4, G5 and G6. The resulting model has an rms ($\chi^2$, $d.o.f.$) of 1.59\arcsec (478,34), more than a factor 2 from our best-fit model which has an rms ($\chi^2$, $d.o.f.$) of 0.6\arcsec (68, 18), demonstrating the need for these galaxies to be modeled individually.
We also assess specifically the impact of including galaxies G1, G2, G3 and G4 which are known to be foreground objects as mentioned before. We thus run a model which excludes these 4 galaxy-scale halos. The resulting best-fit model has an rms ($\chi^2$, $d.o.f.$) of 0.92\arcsec (159, 26). This again shows the necessity to include those as individual potentials, ignoring them results in a degradation of the goodness of the fit. To go even further, with this one model the rms of System\,H, the quadruply imaged galaxy at $z=6.7$, is degraded to 0.5\arcsec, to be compared with 0.16\arcsec with our best-fit mass model, and the model cannot reproduce properly the lensing configuration.
We then assess the impact of G6 in our mass model. We run a model which excludes this one individual galaxy-scale halo. The resulting model gives an rms of 0.59\arcsec. Compared to our best-fit model presented before which has an rms of 0.6\arcsec, the improvement is not significant. However, the presence of G6 impacts the rms of several systems, with the most significant one being System\,H with an rms degraded to 0.48\arcsec. This System being a stringent constraints to our model, we also consider G6 as necessary to our model.

\subsubsection{MACS\,J2129}
We use a similar approach to MS\,0451 by starting with a model as simple as possible, and, thus, including only one large-scale halo.
However, the two most Easterly systems, Systems\ 3 and 6, were poorly reproduced with this model, with an rms greater than 1\arcsec.
We thus decided to add a second large-scale halo centered on a galaxy group located North-East of the BCG. 
All large-scale halos are modeled using PIEMDs, and all their parameters are being optimized except for the truncation radius which is set to 1000\,kpc.

We individually optimize the BCG of the cluster. We then add two galaxy-scale halos to model galaxies identified as small-scale perturbers by \citet{monna}, due to their proximity to the sextuply-imaged System\ 1 (G1 and G2 in Table\,\ref{tab:res_pot}). Moreover, Images\ 10.2 and 8.1 are both located near cluster galaxies, encouraging us to include those as individual potentials in our mass model (G3 and G4 in Table\,\ref{tab:res_pot}).  

Our best-fit mass model has an rms$=$0.74\arcsec. The best-fit parameters of the model are given in Table~\ref{tab:res_pot}, and the center of the cluster is chosen as (R.\,A.$=322.35878$,Decl.$=-7.69100$).
As for RX\,J2129 and MS\,0451, we test the necessity of the individual galaxy-scale potentials in our best model. When excluding all individual galaxy-scale halos, we obtain an rms ($\chi^2$, $d.o.f.$) of 1.05\arcsec (184, 41), to be compared to 0.74\arcsec (90, 31) with our best-fit model. Such a difference comforts us in our choice of modeling G1, G2, G3 and G4 individually due to their proximity to some of the multiple images in the cluster.

\begin{figure*}
\begin{center}
 \includegraphics[width=0.5\textwidth]{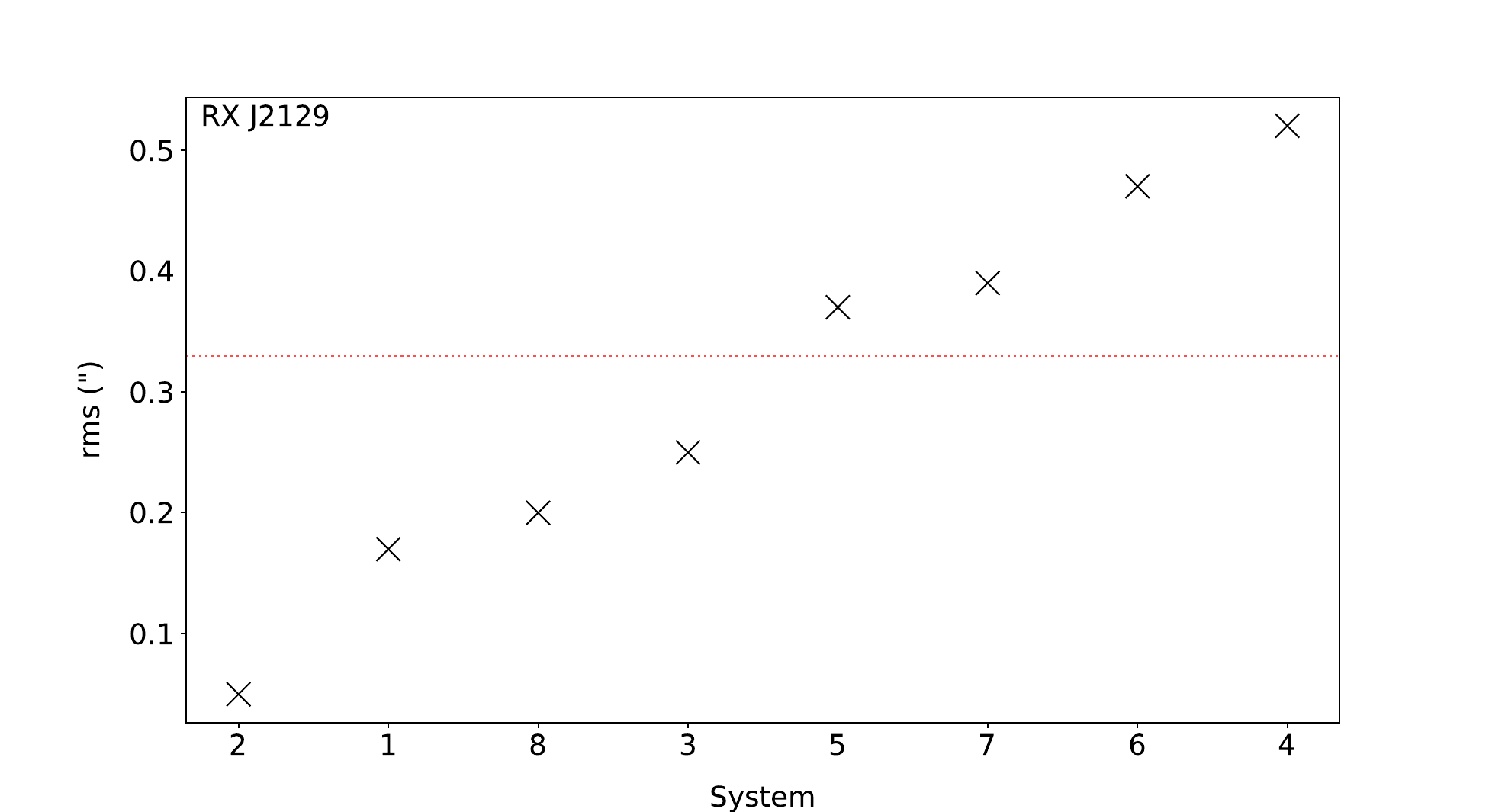}\ 
         \includegraphics[width=0.49\textwidth]{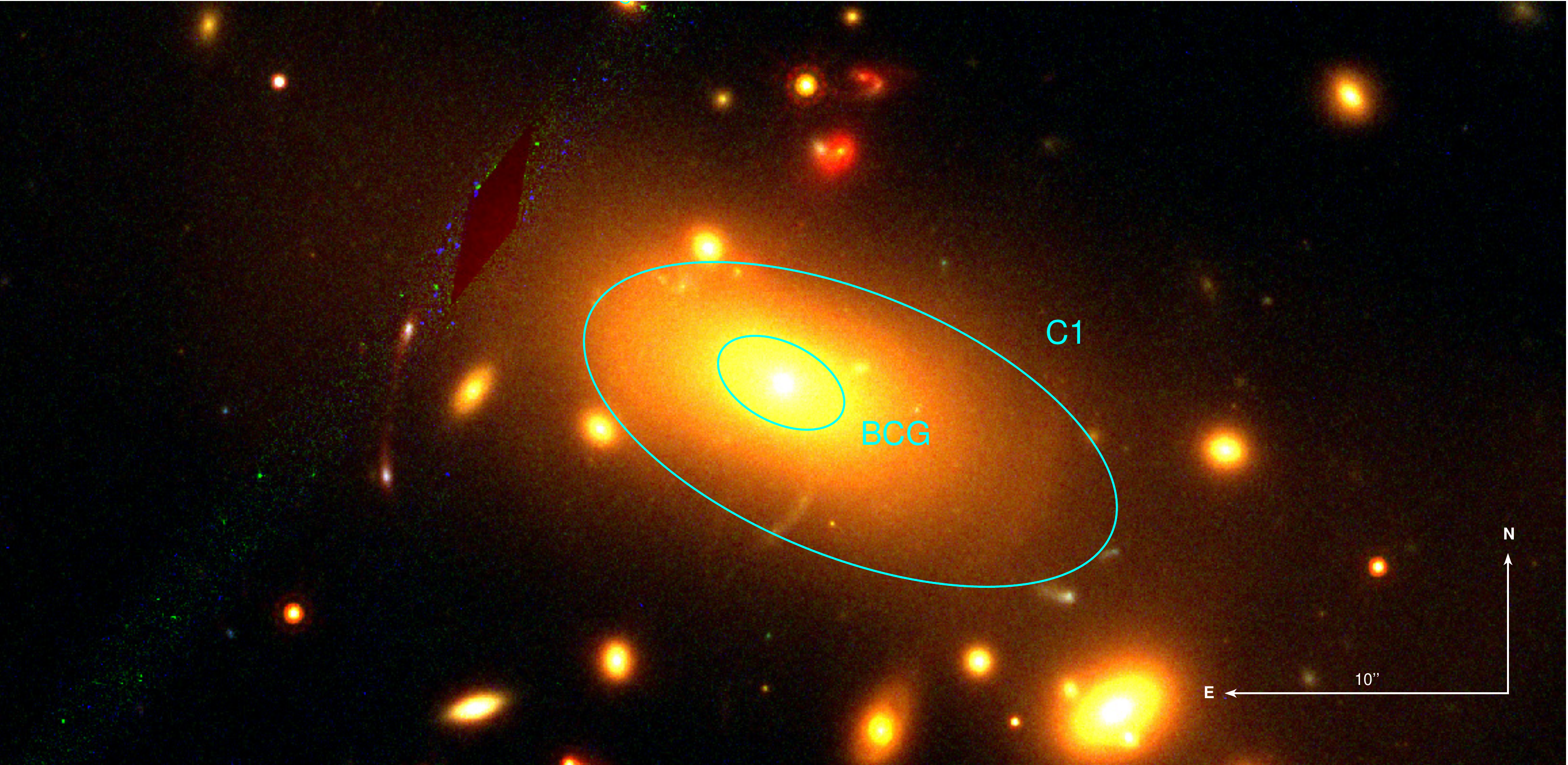}\\
         \includegraphics[width=0.5\textwidth]{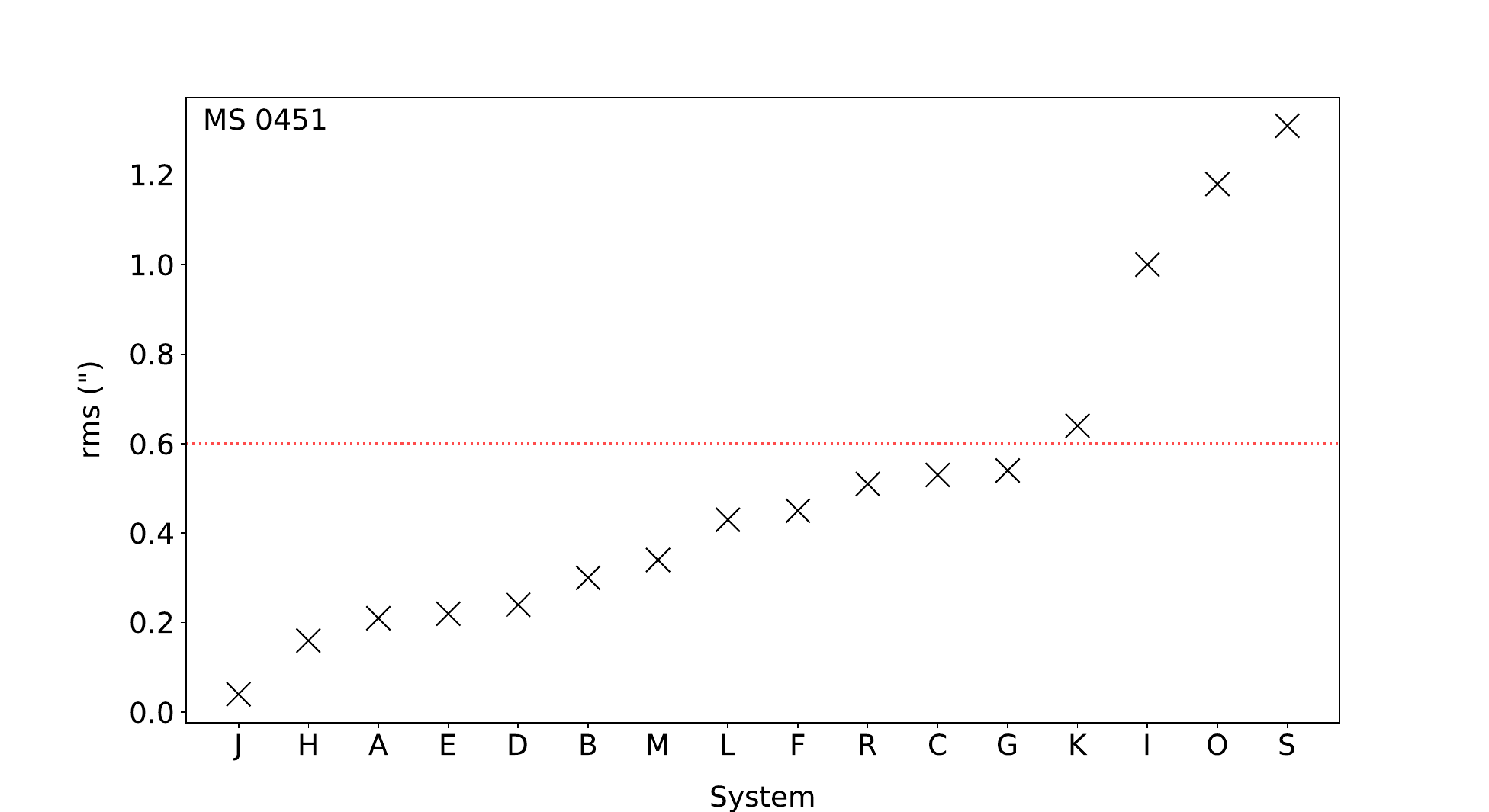}\ 
         \includegraphics[width=0.49\textwidth]{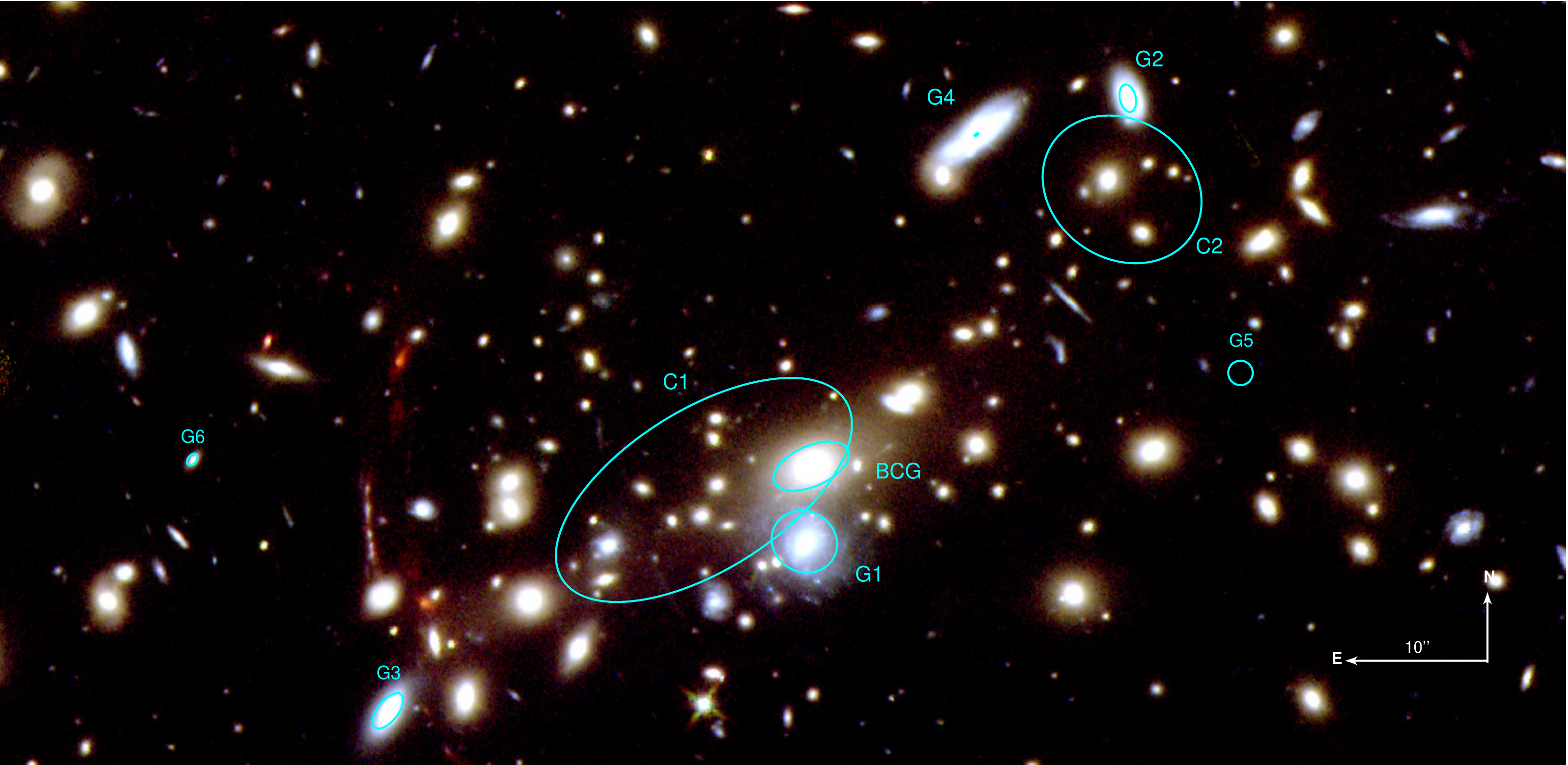}\\
         \includegraphics[width=0.5\textwidth]{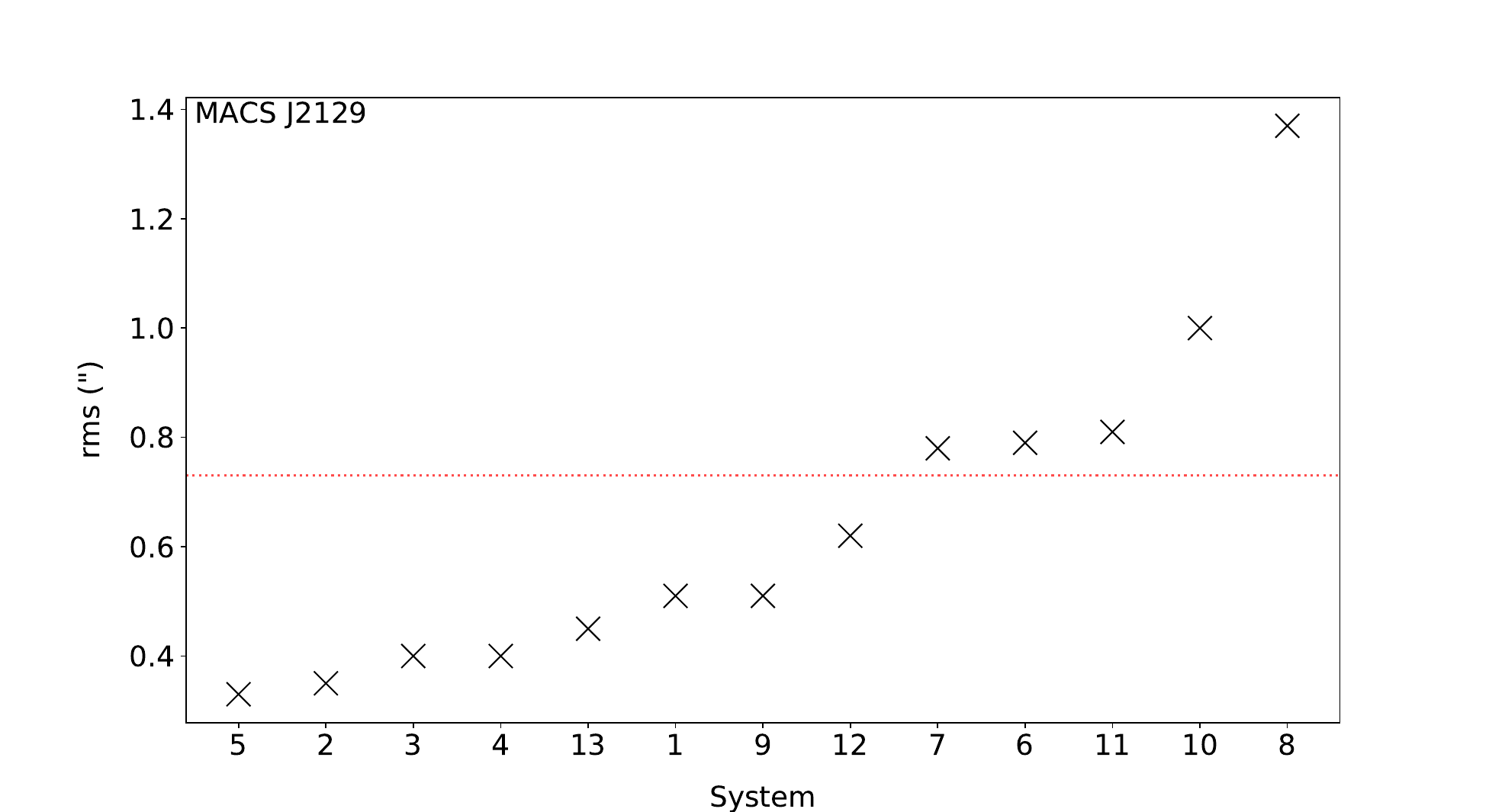}\ 
          \includegraphics[width=0.49\textwidth]{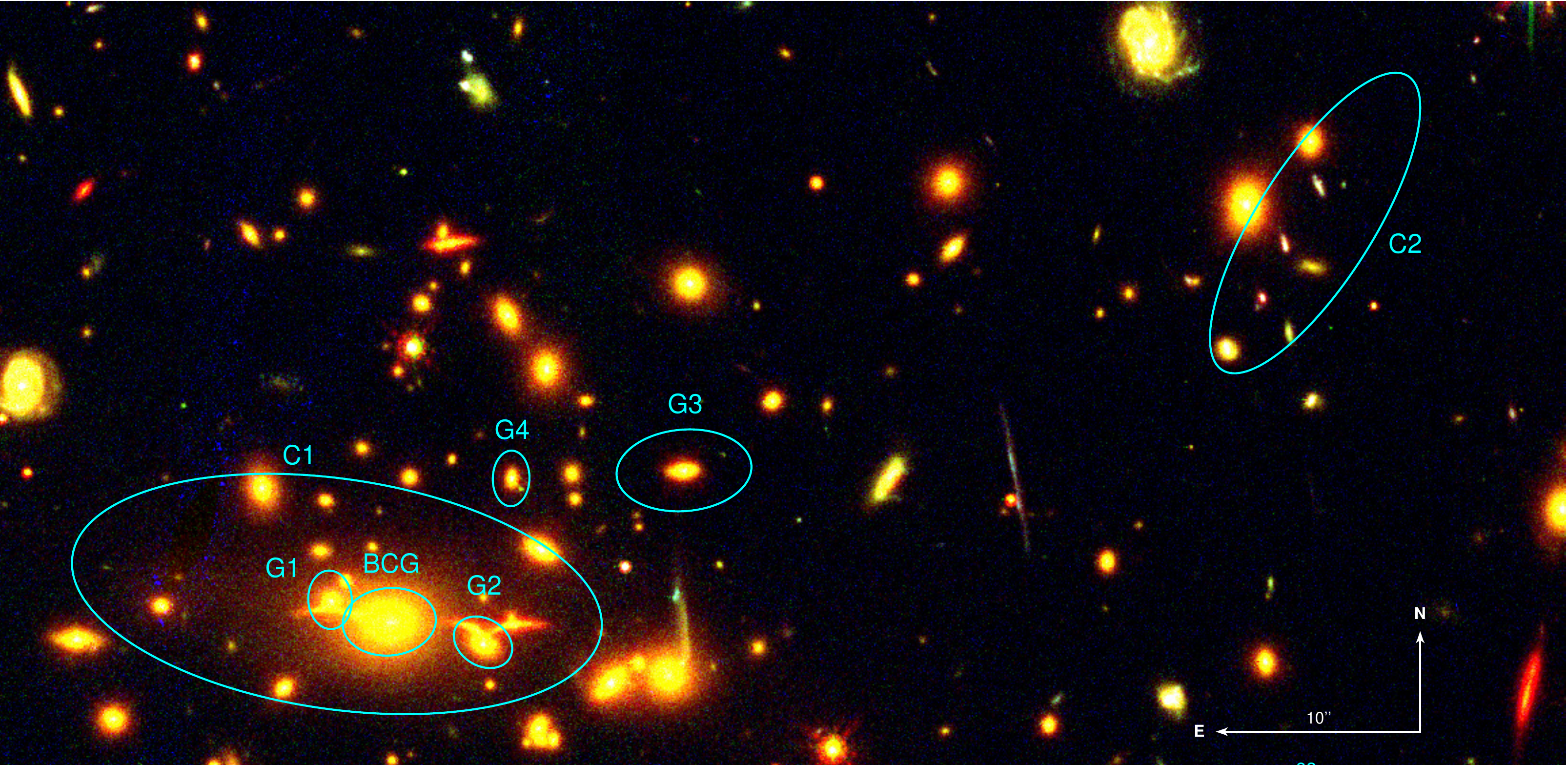}
 \caption{\textit{Left:} Rms for each multiple image system in RX\,J2129 (top), MS\,0451 (middle), and MACS\,J2129 (bottom). The red dahed line shows the rms of the best-fit mass model. \textit{Right:} Optimized potentials displayed on a composite colour image using \emph{HST} images in the F475W, F814W, F160W pass-bands for RX\,J2129 (top) and MACS\,J2129 (bottom), and in the F814W, F110W, F160W pass-bands for MS\,0451 (middle).}
\label{rms}
\end{center}
\end{figure*}

\section{Results and discussions}
\label{section5}

As presented in Sect.\,\ref{section4}, the optimization of RX\,J2129, MS\,0451, and MACS\,J2129 mass models were done using \textsc{Lenstool} \citep{jullo2007}, and the best-fit parameters for each model are given in Table\,\ref{tab:res_pot}. The list of multiple images used as constraints, together with their spectroscopic redshift or the optimized one from the models when included as a free parameter, and the rms obtained for each multiple image are provided in Table\,\ref{mul_rxj2129}, Table\,\ref{mul_ms0451}, and Table\,\ref{mul_macs2129}. 
In this Section, we discuss the improvements on the mass models brought by the MUSE data for each cluster, considering their mass distributions, density profiles, rms, and compare our results with previously published works.

\begin{figure*}
    \centering
    \includegraphics[width=\linewidth]{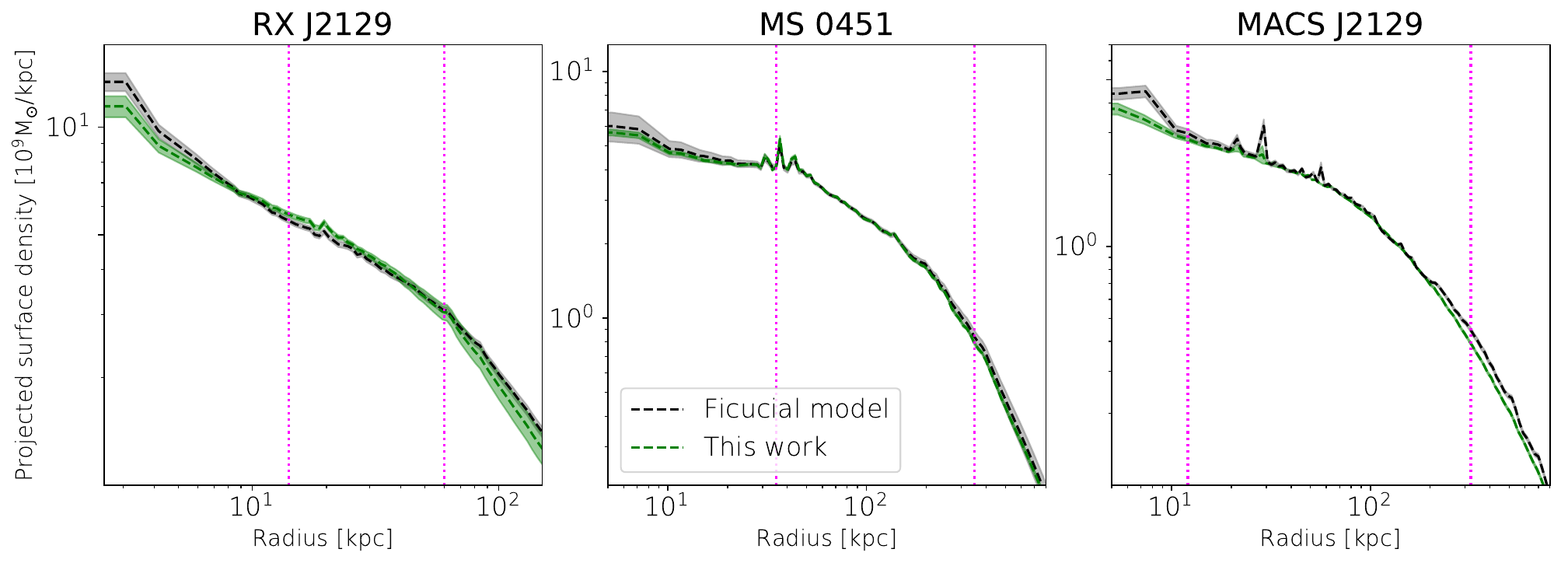}
    \caption{Radial surface density profiles for RX\,J2129 (left), MS\,0451 (middle), and MACS\,J2129 (right) derived from the strong-lensing mass models described in this work. Measurements from the \emph{fiducial model} are shown in grey, and measurements with our new mass models are shown in green. Fiducial models were built based on the analyses of \citet{caminha2019} and \citet{monna} for RX\,J2129 and MACS\,J2129. In each plot, the dashed line stands for the mean surface density, and the shaded areas indicate the 1$\sigma$ error bars. The two magenta dashed lines highlight the positions, from the cluster center, of the closest and most distant multiple image used as constraints for each model.}
    \label{fig:int_dens}
\end{figure*}

\subsection{RX\,J2129}
Our best mass model is constrained by 8 multiple image systems, 7 of them spectroscopically confirmed, and is composed of 71 potentials including one large-scale halo and 2 galaxy-scale halos independently optimized. 
The best-fit model has an rms of 0.29\arcsec.
The left panel of Fig.\,\ref{fig:int_dens} shows the density profiles for both the \emph{fiducial} and our best models. One can see that the new model predicts a significantly higher density in the very inner core of RX\,J2129, R$>$10\,kpc, compare to the \emph{fiducial model}, which then takes over at larger radii.
In terms of total mass within the multiple image region, R$<$70\,kpc, we measure $M({\rm R}<70\,\rm{kpc})=(0.30\pm0.01)\times$10$^{14}\,\msun$, in perfect agreement with $M_{\rm{fiducial}}({\rm R}<70\,\rm{kpc})=(0.30\pm0.01)\times$10$^{14}\,\msun$.

We measure an Einstein radius of $\theta_{\rm{E}}=(19\pm2)$\arcsec\ for a source redshift $z=2.0$. This is higher than previous measurements presented by \cite{rxj2129_model} and \cite{zitrin2015}, $\theta_{\rm{E,\,Richard+10}}=(9.0\pm1.4)$\arcsec, and $\theta_{\rm{E,\,Zitrin+15}}=13$\arcsec.
In terms of total mass, we measure $M({\rm R}<250\,\rm{kpc})=(1.49\pm0.04)\times$10$^{14}\,\msun$, which is in good agreement with the measurement given by \cite{rxj2129_model} of $M_{\rm{Richard+10}}({\rm R}<250\,\rm{kpc})=(1.37\pm0.37)\times$10$^{14}\,\msun$. \cite{zitrin2015} quote a mass within 13\arcsec, which corresponds to $\sim$50\,kpc at the cluster redshift, $M_{\rm{Zitrin+15}}({\rm R}<50\,\rm{kpc})=0.19\times$10$^{14}\,\msun$, which is of the same order as what we obtain with our model, $M({\rm R}<50\,\rm{kpc})=(0.17\pm0.01)\times$10$^{14}\,\msun$. 
\cite{caminha2019} quote an integrated mass of $M_{\rm{Caminha+19}}({\rm R}<200\,\rm{kpc})=(1.19\pm0.01)\times$10$^{14}\,\msun$, in excellent agreement with our value of $M({\rm R}<200\,\rm{kpc})=(1.16\pm0.03)\times$10$^{14}\,\msun$.
Thanks to discussions with \cite{desprez2018}, we could get the integrated total mass measured with their model within the multiple image region, $M_{\rm{Desprez+18}}({\rm R}<70\,\rm{kpc})=(0.29\pm0.01)\times$10$^{14}\,\msun$, value which is in good agreement with our measurement presented before.
It is important to note that at the time \cite{rxj2129_model} and \cite{zitrin2015} published their strong-lensing mass models, the spectroscopic coverage of RX\,J2129 was poor. Indeed, only one system, System\ 1 in this work, had a spectroscopic measurement. \cite{rxj2129_model} model was only constrained using that system (3 multiple images in total), and \cite{zitrin2015} model included 6 systems (18 multiple images total) as constraints. 
In our case, we have 8 systems, 23 multiple images in total as can be seen in Table\,\ref{mul_rxj2129}, all spectroscopically confirmed \citep[except System\ 2 for which we assume the photometric redshift measured by][]{desprez2018}. This could explain the discrepancies in the measured Einstein radii between these analyses and our model.

\subsection{MS\,0451}
Our best model is constrained by 16 multiple image systems, 5 spectroscopically confirmed, and is composed of 146 halos including 2 large-scale halos and 7 galaxy-scale halos independently-optimized. 
The best-fit model has an rms of 0.6\arcsec. In the middle panel of Fig.\,\ref{fig:int_dens}, we compare the density profiles obtained with the \emph{fiducial model} and the new model presented here. The two profiles are in excellent agreement, with a slightly higher density predicted in the core now, but still within the error bars of the \emph{fiducial model}, which we attribute to the stronger constraints used to optimize our model thanks to the identification of new systems in the North of MS\,0451. We measure a total mass within the multiple image region which extends up to $\sim350$\,kpc, $M({\rm R}<350\,\rm{kpc})=(3.75\pm0.11)\times$10$^{14}\,\msun$, in good agreement with the value obtained with the \emph{fiducial model}, $M_{fiducial}({\rm R}<350\,\rm{kpc})=(3.72\pm0.03)\times$10$^{14}\,\msun$.

\begin{table}
\centering
\caption{Predicted positions of the unidentified counter images from the best model of MS\,0451. Column (1) stands for the identifier of the image with respect to the system it belongs to. Columns (2) and (3) are respectively the R.\,A. and Decl. in degrees (J2000) of the images. Column (4) gives the measured magnification of the images at their predicted positions. The images are highlighted as cyan circles in Fig.\,\ref{ms0451_big}.}
\begin{center}
\begin{tabular}{cccc}
\hline
\hline
Image ID & R.\,A. & Decl. & $\mu$ \\
 & [J2000] & [J2000] & \\
\hline
\hline
J.3 & 73.55100 & -3.01006 & $3.5\pm0.3$ \\ 
S.3 & 73.54862 & -3.01063 & $13.0\pm3.0$ \\
S.4 & 73.54093 & -3.02420 & $2.2\pm0.2$ \\
\hline
\hline
\end{tabular}
\label{pred_table}
\end{center}
\end{table}

We measure an Einstein radius, $\theta_{\rm{E}} = (19 \pm 1)$\arcsec, for a source redshift of $z = 2.9$ (i.e. redshift of System\ A). 
This value is consistent with the one reported in \citet{zitrin2011}, $\theta_{\rm{E,\, Zitrin+11}} = (19\pm 2)$\arcsec. 
We measure an integrated mass within $\theta_{\rm{E}}=120\,\rm{kpc}$ of $M(<\theta_{\rm{E}})=(9.15\pm 0.08)\times$10$^{13}\,\msun$. This is slightly higher than what was measured by \cite{zitrin2011}, $M_{\rm{Zitrin+11}}(<\theta_{\rm{E}})=8.82^{+0.3}_{0.1}\times$10$^{13}\,\msun$, however their mass model only includes 4 systems of multiple images, compared to our analysis which has 16 systems as constraints. 
\cite{berciano_alba2010} measured an integrated mass within a radius of 30\arcsec, 188\,kpc at the redshift of MS\,0451, $M_{\rm{BA+09}}({\rm R}<188\,\rm{kpc})=1.73\times$10$^{14}\,\msun$, which is in excellent agreement with both our measurement, $M({\rm R}<188\,\rm{kpc})=(1.73\pm0.02)\times$10$^{14}\,\msun$, and the one from \cite{zitrin2011} at the same radius, $M_{\rm{Zitrin+11}}({\rm R}<188\,\rm{kpc})=1.80\times$10$^{14}\,\msun$. 
This strengthens our argument that the difference of mass values between \cite{zitrin2011} and our work at smaller radii might be due to the improvement of the mass model due to both an increased number of multiple images to constrain the inner mass distribution of MS\,0451, and the spectroscopic information.

The mass obtained for galaxy G5 appears relatively high (see Fig.~\ref{rms} and Table\,\ref{tab:res_pot}) considering it is not visible in any of the \emph{HST} images and only detected in the MUSE datacube. It is thus likely to be a low-mass galaxy with a velocity dispersion lower than the best-fit value measured by our mass model, $\sigma=132$\,km/s. Moreover, as it is not detected in the \emph{HST} imaging, we do not have any shape measurements for G5. The model thus optimizes all its parameters apart from its core radius, $r_{c}$. The best-fit parameters for this galaxy might be strongly degenerated. It is thus difficult to conclude.

\begin{table}
\begin{center}
\caption{Measured magnification for the lensed submillimetric galaxies identified in \citet{borys2004} and \citet{mackenzie}. We list the IDs of the multiple images following Table\,\ref{mul_ms0451}, and add the ID from \citet{mackenzie} in brackets. We here give the magnifications measured from our best model, as well as the ones reported in \citet{mackenzie}.}
\begin{tabular}{ccccc}
\hline
\hline
ID & R.\,A. & Decl. & $\mu$ & $\mu_{\rm{MK+14}}$ \\
 & [J2000] & [J2000] &  & \\
\hline
\hline
A.1/D.1 (7a) & 73.55396 & -3.01482 & $>50$ & $33\pm2$\\
A.2/D.2 (7b) & 73.55389 & -3.01595 & $34.2\pm7.9$ & $45\pm3$\\
A.3/D.3 (7c) & 73.54630 & -3.02404 & $3.0\pm0.2$ & $2.9\pm0.1$ \\
B.1 (5a) & 73.55335 & -3.01232 & $5.4\pm0.6$ & $5.3\pm0.1$\\
B.2 (5b) & 73.55285 & -3.01707 & $4.9\pm0.6$ & $6.4\pm0.1$ \\
B.3 (5c) & 73.54553 & -3.02348 & $2.9\pm0.2$ & $2.9\pm0.1$ \\
C.1 (6a) & 73.55339 & -3.01325 & $7.4\pm0.9$ & $8.2\pm0.2$ \\
C.2 (6b) & 73.55304 & -3.01656 & $5.3\pm0.6$ & $5.0\pm0.1$ \\
C.3 (6c) & 73.54545 & -3.02380 & $2.7\pm0.2$ & $2.8\pm0.1$ \\
E.1 (2a) & 73.55481 & -3.01065 & $2.9\pm0.2$ & $2.9\pm0.1$\\
E.2 (2b) & 73.55241 & -3.01996 & $7.2\pm1.4$ & $8.1\pm0.4$ \\
E.3 (2c) & 73.54911 & -3.02226 & $4.9\pm0.5$ & $6.1\pm0.1$ \\
F.1 (3a) & 73.55435 & -3.01088 & $3.0\pm0.2$ & $3.2\pm0.1$ \\
F.2 (3b) & 73.55282 & -3.01918 & $3.3\pm0.3$ & $3.0\pm0.1$ \\
F.3 (3c) & 73.54775 & -3.02268 & $3.3\pm0.3$ & $4.3\pm0.1$\\
G.1 (1a) & 73.55593 & -3.01193 & $3.8\pm0.3$ & $3.8\pm0.1$\\
G.2 (1b) & 73.55271 & -3.02124 & $10.9\pm2.1$ & $20\pm1$ \\
G.3 (1c) & 73.55071 & -3.02261 & $6.1\pm0.6$ & $7.3\pm0.1$\\
\hline
\hline
\end{tabular}
\label{ms0451_amp}
\end{center}
\end{table}

Oppositely, the cluster galaxy G4 is predicted with a velocity dispersion $\sigma=45$\,km/s, which is likely under-estimated given the size of the galaxy (see Fig.~\ref{rms}). This can be explained by the addition of the second large-scale halo in its vicinity, which might induce a non-physical 'mass transfer' between the two halos. However, as explained in Sect.\,\ref{section4}, this second large-scale halo is necessary to account for the impact of the two galaxy groups in this region, and without which we cannot recover the quintuply-imaged system at $z=6.7$ \citep[][Richard et al.\ \emph{in prep.}]{knudsen2016}.

The rms of each multiple image is given in Table\,\ref{mul_ms0451}, and the predicted position together with their magnification, $\mu$, of the unidentified counter-images of the systems used as constraints are given in Table\,\ref{pred_table}, namely System\ J and System\ S. Images\ J.3, S.3 and S.4 are highlighted as cyan circles in Fig.\,\ref{ms0451_big}.
The two systems in the North of the cluster, Systems\ H and R, are well reproduced, with an rms of 0.16\arcsec\ and 0.51\arcsec\ respectively. The same goes for the group of systems located in the vicinity of the giant arc A, South of the cluster, i.e. Systems\ A, B, C, D, E, F, G and K). 
Systems\ I and O are not as well reproduced as the others, with an rms of 1.0\arcsec\ and 1.18\arcsec\ respectively. We explain this by the location of one of their multiple images, Images\ I.2 and O.2. 
Image\ I.2 is located close to a faint cluster galaxy which is not individually optimized, but might act as a small-scale perturber. A mass model which includes the cluster galaxy close to Image I.2 does not show any significant improvements compare to our best model when considering the number of added free parameters.
Image\ O.2 is located in a region which is highly contaminated by the light of the BCG, meaning we could be missing local perturbers.

Moreover, \cite{mackenzie} discussed the group of multiply-lensed submilimetric galaxies at $z=2.9$, corresponding to Systems\ A, B, C, D, E, F, and G in this work.
They provide magnification estimates for each image, in excellent agreement with the magnifications measured with our best model. Table\,\ref{ms0451_amp} gives the measured magnifications by this analysis as well as the ones measured in \cite{mackenzie}. The largest differences are seen for highly magnified images such as Images\ A.1, A.2, G.2 and G.3. This was to be expected as in high magnification regions such measurements have large uncertainties. Indeed magnification is supposedly infinite on critical lines, making these measurements difficult to believe, apart from the fact that lensed galaxies located in these regions are highly magnified.
One more thing to consider is the unambiguous measurement of the redshift of System\ G with MUSE of $z=2.93$. Indeed this value differs from the one used in \cite{mackenzie}, $z=3.11\pm0.03$, which was initially measured by \cite{takata2003}.

\paragraph*{System S -- }Although spectroscopically confirmed, System\ S is not well reproduced by the model, with an rms of 1.3\arcsec. Both Images\ S.1 and S.2 are in the high magnification region of the cluster, close to the radial critical line at the redshift of the source, $z = 4.45$, with a measured magnification for S.1 of $\mu > 300$. 
System S is predicted to be quadruply-imaged. Images\ S.1 and S.2 are only detected in the MUSE datacube, as their proximity to the BCG makes them undetectable in the \emph{HST} images. 
Counter-images are predicted further away from the BCG, with relatively low magnifications as quoted in Table\,\ref{pred_table}. 
That could explain why Image\ S.3 is not detected in the MUSE datacube, while Image\ S.4 is predicted outside the MUSE field of view as can be seen in Fig.\,\ref{ms0451_big} (cyan circles). However the lack of photometric identification of Images\ S.1 and S.2 in the \emph{HST} images makes them impossible to identify as we have no idea of their colour nor morphology.

\subsection{MACS\,J2129}
Our best model is constrained by 13 multiple image systems, 12 spectroscopically confirmed, and is composed of 151 halos including 2 large-scale halos and 4 galaxy-scale halos independently optimized. The constraints used in this model are the same as the ones presented in \cite{caminha2019}, except for a newly identified system, System\ 10, and the removal of System\ 14, as discussed in Sect.\,\ref{section4}.
The best-fit model has an rms of 0.74\arcsec. 
The right panel of Fig.\,\ref{fig:int_dens} shows the density profiles obtained with this work and our \emph{fiducial model}. One can notice differences between the two density profiles in the very inner region of the cluster, R$<10$\,kpc, however this discrepancy is difficult to interpret as within this region the density is dominated by the stellar content of the BCG, and thus we lack constraints from strong-lensing. 
At R$>$10\,kpc, the new model exhibits overdensity peaks not visible in the \emph{fiducial model} density profile. This is explained by the inclusion of new constraints in this region, both multiple images and cluster galaxies up to R$\sim$100\,kpc.
In terms of total mass within the multiple image region, which extends up to $\sim$300\,kpc, we measure $M({\rm R}<300\,\rm{kpc})=(2.94\pm0.02)\times$10$^{14}\,\msun$, in good agreement with the value from the \emph{fiducial model}, $M_{\rm{fiducial}}({\rm R}<300\,\rm{kpc})=(2.97\pm0.02)\times$10$^{14}\,\msun$.

The properties of the main dark matter halo (namely C1 in Table\,\ref{tab:res_pot}) are consistent with those reported in \citet{monna} and \cite{caminha2019}. Our best-fit model also favours a relatively large core radius, $r_{c}=(96\pm5)$\,kpc, in excellent agreement with \cite{monna} measurement of $r_{c,\,\rm{Monna+17}}=(101\pm13)$\,kpc.
\cite{caminha2019}, who are also using the \textsc{Lenstool} software, quote a core radius for the main cluster-scale halo of $r_{c,\,\rm{Caminha+19}}=(74\pm4)$\,kpc. This is slightly lower than what we obtain. The second cluster-scale halo they include has similar properties to this work (C2 in Table\,\ref{tab:res_pot}), however their model favors it closer to the BCG, $\sim$250\,kpc \emph{vs} $\sim$320\,kpc in this work. This might explain the lower value of $r_{c}$ for C1 obtained by \cite{caminha2019} compared to our work.
We find that velocity dispersions of the galaxy-scale halos, BCG, G1 and G2, are overestimated compared to their luminous counterparts. This is explained by the correlation between the baryonic mass distribution within central galaxies, and the size of the core of the dark matter halo \citep{newman2013a,newman2013b}, which was already evidenced by \citet{monna}. 

We measure an Einstein radius for a source redshift $z=2.0$ of $\theta_{\rm{E}}=(23\pm2)$\arcsec, which is within the error bars of what \cite{zitrin2011,zitrin2015} measured for a similar source redshift, $\theta_{\rm{E,\,Zitrin+11,15}}=(19\pm2)$\arcsec. 
\cite{monna} quote an Einstein radius for a source redshift $z=1.36$, $\theta_{\rm{E,\,Monna+17}}=(14\pm2)$\arcsec, of similar order to what we measure at the same source redshift, $\theta_{\rm{E}}=(18\pm1)$\arcsec.
In order to compare our model with \cite{monna}, \cite{zitrin2011} and \cite{zitrin2015}, we now quote masses within a radius of 19\arcsec, which corresponds to $\sim$130\,kpc at the redshift of MACS\,J2129. We measure a total mass of $M({\rm R}<130\,\rm{kpc})=(1.04\pm0.01)\times$10$^{14}\,\msun$, slightly higher than what is measured by \cite{monna}, $M_{\rm{Monna+17}}({\rm R}<130\,\rm{kpc})=(0.89\pm0.01)\times$10$^{14}\,\msun$, but within the error bars of the measurements given by \cite{zitrin2011} and \cite{zitrin2015}, $M_{\rm{Zitrin+11,15}}({\rm R}<130\,\rm{kpc})=0.92\pm0.09)\times$10$^{14}\,\msun$.
\cite{caminha2019} quote a mass of $M_{\rm{Caminha+19}}({\rm R}<200\,\rm{kpc})=(1.84\pm0.01)\times$10$^{14}\,\msun$, in excellent agreement with our measurement of $M({\rm R}<200\,\rm{kpc})=(1.81\pm0.02)\times$10$^{14}\,\msun$.

Globally, multiple images are well recovered by our model, except for System\ 8 which has an rms$=$1.37\arcsec. 
The new multiple image system reported in this work, i.e. Systems\ 10, is well-recovered, with an rms$=$0.61\arcsec. 
In particular, we note that the inclusion of the cluster galaxy G3 (see Table\,\ref{tab:res_pot}) is critical for the recovery of this system.
The addition of the second cluster-scale halo in the North-East of the cluster significantly improves the recovery of Systems\ 3 and 6 compared to the model presented by \citet{monna}. 
We measure an rms of 0.40\arcsec, to be compared to $\sim$0.8\arcsec\ for System\ 3 when G3 is not included, and 0.79\arcsec\ to be compared to 1.7\arcsec\ for System\ 6. 
While 4 over the 5 multiple images of System 8 are spectroscopically confirmed, it has an rms of 1.37\arcsec. We note that the inclusion of galaxy G4 in the model seems to be responsible for that, as it degrades the accuracy of the reconstruction of System\ 8 (Image\ 8.3 is poorly-reproduced). 
However, G4 is necessary to recover precisely Systems\ 1 and 7, decreasing the rms of the overall model from 0.95\arcsec\ to 0.73\arcsec.

Finally, MACS\,J2129 also hosts a particularly red and bright single strongly-lensed galaxy, West of the cluster center as shown in Fig.\,\ref{macs2129_big} ($\alpha=322.34311$, $\delta=-7.69203$). \cite{toft2017} presented a detailed analysis of this $z=2.1472$ compact galaxy, spectroscopically confirmed thanks to VLT/X-Shooter observations, and which revealed to be a fast-spinning, rotationally supported disk galaxy. Thanks to their lensing mass model, \cite{toft2017} measured a magnification of $\mu_{\rm{Toft+17}}=4.2\pm0.6$. While this galaxy is not used in the mass model presented here as it is singly-imaged, we can measure its predicted magnification. We measure $\mu=3.7\pm0.2$, a slightly lower value than the one from \cite{toft2017}, but in good agreement within error bars. One should note that their strong-lensing mass model was only constrained by two multiple image systems, namely Systems\ 1, and 3 in Table\,\ref{mul_macs2129}.

\section{Summary and conclusions}
\label{section6}

In this paper, we present new strong-lensing mass models for three galaxy clusters, MS\,0451, MACS\,J2129, and RX\,J2129, which include new VLT/MUSE observations.
We combine the MUSE datacubes with high resolution imaging from \emph{HST} available for each cluster to maximize the number of extracted spectra. 
We measure the redshift of each source with a dedicated software, \textsc{ifs-redex} \citep[][]{ifs}, allowing for a wavelet-based filtration of the spectra. 
Our conclusions are as follows:

\textbullet \ We measure 158, 171, and 189 secured or likely spectroscopic redshifts in the RX\,J2129, MS\,0451, and MACS\,J2129 MUSE datacubes respectively. 
For MS\,0451, we identify 2 new systems of multiple images located in the North of the cluster core, the least constrained region, confirm the redshift of System\ A measured by \cite{borys2004}, and measure the redshift of the three multiple images of System\ G. For RX\,J2129 and MACS\,J2129, we obtain measurements in excellent agreement with \cite{caminha2019}. We report a new multiple image system detection, System\ 10 at $z=4.41$, in MACS\,J2129.
Finally, the MUSE datacubes allowed us to spectroscopically confirm 43, 112, and 89 cluster members in RX\,J2129, MS\,0451, and MACS\,J2129 respectively. Among those, in RX\,J2129 and MACS\,J2129, 15 and 4 cluster members respectively are new identifications, i.e. not reported by \cite{caminha2019}.

\textbullet \ We carried out a fruitful blind search while combining \emph{HST} imaging with MUSE datacubes using \textsc{muselet}. It played a decisive role in the identification of multiple image systems since it highlighted strong emission lines invisible in the \emph{HST} images due to either the faintness of the sources at wavelengths not corresponding to maximum emissions, or their proximity to luminous emitters in the foreground. This was particularly interesting for MS\,0451 where the blind search revealed Systems\ R and S, both located in the North of the cluster, which before this analysis was lacking strong-lensing constraints.

\textbullet \ Our models are optimized using the parametric version of the \textsc{Lenstool} algorithm \citep{jullo2007}. The multiple image systems in the three clusters were reproduced with an rms of 0.28\arcsec, 0.6\arcsec, and 0.74\arcsec\ in RX\,J2129, MS\,0451, and MACS\,J2129 respectively. We measure integrated aperture masses in good agreement or within the error bars of the ones published in previous analyses \cite{berciano_alba2010,rxj2129_model,zitrin2011,zitrin2015,monna,desprez2018,caminha2019}.

\textbullet \ The addition of a second cluster-scale halo in MS\,0451 and MACS\,J2129 mass models are necessary to minimize the rms of the two models, and to recover the multiple image systems geometry in the North and North-East of the two clusters respectively. \cite{caminha2019} presented a similar mass model for MACS\,J2129, with their second cluster-scale halo located in the same region as us, slightly closer to the BCG than in our case, $\sim$250\,kpc away from the BCG compared to $\sim$320\,kpc in our case.
Similarly, the addition of some cluster galaxies located in the vicinity of multiple images played a decisive role in the reconstruction of the mass distribution, e.g. the rms of MACS\,J2129 improved from 0.95\arcsec\ to 0.74\arcsec\ just by the addition of G4.

\textbullet \ We compare our magnification measurements of the submillimetric galaxy group multiply-lensed by MS\,0451, at $z=2.91$, with the results published by \cite{mackenzie}. The two analyses show excellent agreement, with some differences for the most highly magnified images. This was to be expected as magnification measurements close to the critical line, mathematically a region where magnification is supposed to be infinite, have high uncertainties.

\textbullet \ Our mass model of MACS\,J2129 allowed us to measure the magnification of the singly imaged $z=2.1472$ galaxy identified by \cite{toft2017}, $\mu=3.7\pm0.2$. This values is within the error bars of the initial measurement from \cite{toft2017}, $\mu_{\rm{Toft+17}}=4.2\pm0.6$, which was derived using a mass model only constrained by two systems of multiple images, namely Systems\ 1 and 3 from our analysis.

\textbullet \ Further investigations have to be carried out to identify the missing counter-images presented in Table\,\ref{pred_table}, and confirm System\ P in MS\,0451. Moreover, we identify a group of galaxies at $z=0.06$ in MS\,0451 which impacts the multiple image geometry, strengthening the need to properly implement multi-plane optimization in \textsc{Lenstool}. 

More generally, our analysis highlights again the power of MUSE to secure and identify strong-lensing features in cluster cores. Such observations are mandatory to recover precisely and accurately the mass distribution in cluster cores. Without such lensing mass models, cluster lenses cannot be used efficiently as gravitational telescopes, as the mass distribution needs to be known in order to recover the intrinsic physical properties of the lensed galaxies \citep{vanzella2015,patricio2016,toft2017,johnson2017,claeyssens2019}. 
Moreover, the physics in place in clusters themselves is highly dependent on how well we can recover their mass distribution. Indeed, while a multi-wavelength analysis is needed to observe all their components (stars and gas), only gravitational lensing gives us an estimate of their total mass, and thus indirectly of their dark matter content and distribution. With such knowledge, we can hope to use galaxy clusters as probes of the nature of dark matter \citep[][]{jauzac2016b,jauzac2018b,robertson2019}.

\section*{Acknowledgements}

MJ is supported by the United Kingdom Research and Innovation (UKRI) Future Leaders Fellowship `Using Cosmic Beasts to uncover the Nature of Dark Matter' (grant number MR/S017216/1).
BK, JPK  and  MJ acknowledge  support from the ERC advanced grant LIDA.
\section*{Data availability}

The data underlying this article will be shared on reasonable request to the corresponding author.


\bibliographystyle{mnras}
\bibliography{Bibliography} 

\begin{thebibliography}{}
\makeatletter
\relax
\def\mn@urlcharsother{\let\do\@makeother \do\$\do\&\do\#\do\^\do\_\do\%\do\~}
\def\mn@doi{\begingroup\mn@urlcharsother \@ifnextchar [ {\mn@doi@}
  {\mn@doi@[]}}
\def\mn@doi@[#1]#2{\def\@tempa{#1}\ifx\@tempa\@empty \href
  {http://dx.doi.org/#2} {doi:#2}\else \href {http://dx.doi.org/#2} {#1}\fi
  \endgroup}
\def\mn@eprint#1#2{\mn@eprint@#1:#2::\@nil}
\def\mn@eprint@arXiv#1{\href {http://arxiv.org/abs/#1} {{\tt arXiv:#1}}}
\def\mn@eprint@dblp#1{\href {http://dblp.uni-trier.de/rec/bibtex/#1.xml}
  {dblp:#1}}
\def\mn@eprint@#1:#2:#3:#4\@nil{\def\@tempa {#1}\def\@tempb {#2}\def\@tempc
  {#3}\ifx \@tempc \@empty \let \@tempc \@tempb \let \@tempb \@tempa \fi \ifx
  \@tempb \@empty \def\@tempb {arXiv}\fi \@ifundefined
  {mn@eprint@\@tempb}{\@tempb:\@tempc}{\expandafter \expandafter \csname
  mn@eprint@\@tempb\endcsname \expandafter{\@tempc}}}

\bibitem[\protect\citeauthoryear{{Acebron}, {Jullo}, {Limousin}, {Tilquin},
  {Giocoli}, {Jauzac}, {Mahler}  \& {Richard}}{{Acebron}
  et~al.}{2017}]{acebron}
{Acebron} A.,  {Jullo} E.,  {Limousin} M.,  {Tilquin} A.,  {Giocoli} C.,
  {Jauzac} M.,  {Mahler} G.,   {Richard} J.,  2017, \mn@doi [\mnras]
  {10.1093/mnras/stx1330}, \href
  {http://adsabs.harvard.edu/abs/2017MNRAS.470.1809A} {470, 1809}

\bibitem[\protect\citeauthoryear{{Atek} et~al.,}{{Atek}
  et~al.}{2015}]{atek2015}
{Atek} H.,  et~al., 2015, \mn@doi [\apj] {10.1088/0004-637X/814/1/69}, \href
  {http://adsabs.harvard.edu/abs/2015ApJ...814...69A} {814, 69}

\bibitem[\protect\citeauthoryear{{Atek}, {Richard}, {Kneib}  \&
  {Schaerer}}{{Atek} et~al.}{2018}]{atek2018}
{Atek} H.,  {Richard} J.,  {Kneib} J.-P.,   {Schaerer} D.,  2018, \mn@doi
  [\mnras] {10.1093/mnras/sty1820}, \href
  {https://ui.adsabs.harvard.edu/abs/2018MNRAS.479.5184A} {479, 5184}

\bibitem[\protect\citeauthoryear{{Bacon} et~al.,}{{Bacon}
  et~al.}{2010}]{bacon2010}
{Bacon} R.,  et~al., 2010, in Ground-based and Airborne Instrumentation for
  Astronomy III. p. 773508, \mn@doi{10.1117/12.856027}

\bibitem[\protect\citeauthoryear{{Bacon} et~al.,}{{Bacon}
  et~al.}{2015}]{bacon2015}
{Bacon} R.,  et~al., 2015, \mn@doi [\aap] {10.1051/0004-6361/201425419}, \href
  {http://adsabs.harvard.edu/abs/2015A%26A...575A..75B} {575, A75}

\bibitem[\protect\citeauthoryear{{Bacon}, {Piqueras}, {Conseil}, {Richard}  \&
  {Shepherd}}{{Bacon} et~al.}{2016}]{mpdaf}
{Bacon} R.,  {Piqueras} L.,  {Conseil} S.,  {Richard} J.,   {Shepherd} M.,
  2016, {MPDAF: MUSE Python Data Analysis Framework}, Astrophysics Source Code
  Library (\mn@eprint {ascl} {1611.003})

\bibitem[\protect\citeauthoryear{{Bacon} et~al.,}{{Bacon}
  et~al.}{2017}]{bacon2017}
{Bacon} R.,  et~al., 2017, \mn@doi [\aap] {10.1051/0004-6361/201730833}, \href
  {https://ui.adsabs.harvard.edu/abs/2017A&A...608A...1B} {608, A1}

\bibitem[\protect\citeauthoryear{{Bartelmann} \& {Maturi}}{{Bartelmann} \&
  {Maturi}}{2017}]{bartelmann2017}
{Bartelmann} M.,  {Maturi} M.,  2017, \mn@doi [Scholarpedia]
  {10.4249/scholarpedia.32440}, \href
  {https://ui.adsabs.harvard.edu/abs/2017SchpJ..1232440B} {12, 32440}

\bibitem[\protect\citeauthoryear{{Belli}, {Jones}, {Ellis}  \&
  {Richard}}{{Belli} et~al.}{2013}]{belli2013}
{Belli} S.,  {Jones} T.,  {Ellis} R.~S.,   {Richard} J.,  2013, \mn@doi [\apj]
  {10.1088/0004-637X/772/2/141}, \href
  {https://ui.adsabs.harvard.edu/abs/2013ApJ...772..141B} {772, 141}

\bibitem[\protect\citeauthoryear{{Berciano Alba}, {Garrett}, {Koopmans}  \&
  {Wucknitz}}{{Berciano Alba} et~al.}{2007}]{berciano_alba2007}
{Berciano Alba} A.,  {Garrett} M.~A.,  {Koopmans} L.~V.~E.,   {Wucknitz} O.,
  2007, \mn@doi [\aap] {10.1051/0004-6361:20065223}, \href
  {https://ui.adsabs.harvard.edu/abs/2007A&A...462..903B} {462, 903}

\bibitem[\protect\citeauthoryear{{Berciano Alba}, {Koopmans}, {Garrett},
  {Wucknitz}  \& {Limousin}}{{Berciano Alba} et~al.}{2010}]{berciano_alba2010}
{Berciano Alba} A.,  {Koopmans} L.~V.~E.,  {Garrett} M.~A.,  {Wucknitz} O.,
  {Limousin} M.,  2010, \mn@doi [\aap] {10.1051/0004-6361/200912903}, \href
  {https://ui.adsabs.harvard.edu/abs/2010A&A...509A..54B} {509, A54}

\bibitem[\protect\citeauthoryear{{Bertin} \& {Arnouts}}{{Bertin} \&
  {Arnouts}}{1996}]{sextractor}
{Bertin} E.,  {Arnouts} S.,  1996, \mn@doi [\aaps] {10.1051/aas:1996164}, \href
  {http://adsabs.harvard.edu/abs/1996A%26AS..117..393B} {117, 393}

\bibitem[\protect\citeauthoryear{{Borys} et~al.,}{{Borys}
  et~al.}{2004}]{borys2004}
{Borys} C.,  et~al., 2004, \mn@doi [\mnras] {10.1111/j.1365-2966.2004.07982.x},
  \href {http://adsabs.harvard.edu/abs/2004MNRAS.352..759B} {352, 759}

\bibitem[\protect\citeauthoryear{{Bouwens}, {Oesch}, {Illingworth}, {Ellis}  \&
  {Stefanon}}{{Bouwens} et~al.}{2017}]{earlyuni2}
{Bouwens} R.~J.,  {Oesch} P.~A.,  {Illingworth} G.~D.,  {Ellis} R.~S.,
  {Stefanon} M.,  2017, \mn@doi [\apj] {10.3847/1538-4357/aa70a4}, \href
  {http://adsabs.harvard.edu/abs/2017ApJ...843..129B} {843, 129}

\bibitem[\protect\citeauthoryear{{Caminha} et~al.,}{{Caminha}
  et~al.}{2017a}]{caminha2017b}
{Caminha} G.~B.,  et~al., 2017a, \mn@doi [\aap] {10.1051/0004-6361/201629297},
  \href {https://ui.adsabs.harvard.edu/abs/2017A&A...600A..90C} {600, A90}

\bibitem[\protect\citeauthoryear{{Caminha} et~al.,}{{Caminha}
  et~al.}{2017b}]{caminha2017a}
{Caminha} G.~B.,  et~al., 2017b, \mn@doi [\aap] {10.1051/0004-6361/201731498},
  \href {https://ui.adsabs.harvard.edu/abs/2017A&A...607A..93C} {607, A93}

\bibitem[\protect\citeauthoryear{{Caminha} et~al.,}{{Caminha}
  et~al.}{2019}]{caminha2019}
{Caminha} G.~B.,  et~al., 2019, \mn@doi [\aap] {10.1051/0004-6361/201935454},
  \href {https://ui.adsabs.harvard.edu/abs/2019A&A...632A..36C} {632, A36}

\bibitem[\protect\citeauthoryear{{Cerny} et~al.,}{{Cerny}
  et~al.}{2018}]{cerny2018}
{Cerny} C.,  et~al., 2018, \mn@doi [\apj] {10.3847/1538-4357/aabe7b}, \href
  {https://ui.adsabs.harvard.edu/abs/2018ApJ...859..159C} {859, 159}

\bibitem[\protect\citeauthoryear{{Chiriv{\`\i}}, {Suyu}, {Grillo}, {Halkola},
  {Balestra}, {Caminha}, {Mercurio}  \& {Rosati}}{{Chiriv{\`\i}}
  et~al.}{2018}]{chirivi2018}
{Chiriv{\`\i}} G.,  {Suyu} S.~H.,  {Grillo} C.,  {Halkola} A.,  {Balestra} I.,
  {Caminha} G.~B.,  {Mercurio} A.,   {Rosati} P.,  2018, \mn@doi [\aap]
  {10.1051/0004-6361/201731433}, \href
  {https://ui.adsabs.harvard.edu/abs/2018A&A...614A...8C} {614, A8}

\bibitem[\protect\citeauthoryear{{Claeyssens} et~al.,}{{Claeyssens}
  et~al.}{2019}]{claeyssens2019}
{Claeyssens} A.,  et~al., 2019, \mn@doi [\mnras] {10.1093/mnras/stz2492}, \href
  {https://ui.adsabs.harvard.edu/abs/2019MNRAS.489.5022C} {489, 5022}

\bibitem[\protect\citeauthoryear{{Coe}, {Bradley}  \& {Zitrin}}{{Coe}
  et~al.}{2015}]{coe14}
{Coe} D.,  {Bradley} L.,   {Zitrin} A.,  2015, \mn@doi [\apj]
  {10.1088/0004-637X/800/2/84}, \href
  {https://ui.adsabs.harvard.edu/abs/2015ApJ...800...84C} {800, 84}

\bibitem[\protect\citeauthoryear{{Covone}, {Kneib}, {Soucail}, {Richard},
  {Jullo}  \& {Ebeling}}{{Covone} et~al.}{2006}]{covone2006}
{Covone} G.,  {Kneib} J.-P.,  {Soucail} G.,  {Richard} J.,  {Jullo} E.,
  {Ebeling} H.,  2006, \mn@doi [\aap] {10.1051/0004-6361:20053384}, \href
  {http://adsabs.harvard.edu/abs/2006A%26A...456..409C} {456, 409}

\bibitem[\protect\citeauthoryear{{Desprez}, {Richard}, {Jauzac}, {Martinez},
  {Siana}  \& {Cl{\'e}ment}}{{Desprez} et~al.}{2018}]{desprez2018}
{Desprez} G.,  {Richard} J.,  {Jauzac} M.,  {Martinez} J.,  {Siana} B.,
  {Cl{\'e}ment} B.,  2018, \mn@doi [\mnras] {10.1093/mnras/sty1666}, \href
  {https://ui.adsabs.harvard.edu/abs/2018MNRAS.479.2630D} {479, 2630}

\bibitem[\protect\citeauthoryear{{Diego}, {Broadhurst}, {Wong}, {Silk}, {Lim},
  {Zheng}, {Lam}  \& {Ford}}{{Diego} et~al.}{2016}]{diego2017}
{Diego} J.~M.,  {Broadhurst} T.,  {Wong} J.,  {Silk} J.,  {Lim} J.,  {Zheng}
  W.,  {Lam} D.,   {Ford} H.,  2016, \mn@doi [\mnras] {10.1093/mnras/stw865},
  \href {https://ui.adsabs.harvard.edu/abs/2016MNRAS.459.3447D} {459, 3447}

\bibitem[\protect\citeauthoryear{{Diego} et~al.,}{{Diego}
  et~al.}{2018}]{diego2018}
{Diego} J.~M.,  et~al., 2018, \mn@doi [\apj] {10.3847/1538-4357/aab617}, \href
  {https://ui.adsabs.harvard.edu/abs/2018ApJ...857...25D} {857, 25}

\bibitem[\protect\citeauthoryear{{Donahue}, {Gaskin}, {Patel}, {Joy}, {Clowe}
  \& {Hughes}}{{Donahue} et~al.}{2003}]{donahue2003}
{Donahue} M.,  {Gaskin} J.~A.,  {Patel} S.~K.,  {Joy} M.,  {Clowe} D.,
  {Hughes} J.~P.,  2003, \mn@doi [\apj] {10.1086/378688}, \href
  {https://ui.adsabs.harvard.edu/abs/2003ApJ...598..190D} {598, 190}

\bibitem[\protect\citeauthoryear{{Ebeling}, {Barrett}, {Donovan}, {Ma}, {Edge}
  \& {van Speybroeck}}{{Ebeling} et~al.}{2007}]{ebeling2007}
{Ebeling} H.,  {Barrett} E.,  {Donovan} D.,  {Ma} C.-J.,  {Edge} A.~C.,   {van
  Speybroeck} L.,  2007, \mn@doi [\apjl] {10.1086/518603}, \href
  {http://adsabs.harvard.edu/abs/2007ApJ...661L..33E} {661, L33}

\bibitem[\protect\citeauthoryear{{El{\'{\i}}asd{\'o}ttir}
  et~al.,}{{El{\'{\i}}asd{\'o}ttir} et~al.}{2007}]{piemd2}
{El{\'{\i}}asd{\'o}ttir} {\'A}.,  et~al., 2007, preprint, \href
  {http://adsabs.harvard.edu/abs/2007arXiv0710.5636E} {} (\mn@eprint {arXiv}
  {0710.5636})

\bibitem[\protect\citeauthoryear{{Ellingson}, {Yee}, {Abraham}, {Morris}  \&
  {Carlberg}}{{Ellingson} et~al.}{1998}]{ellingson1998}
{Ellingson} E.,  {Yee} H.~K.~C.,  {Abraham} R.~G.,  {Morris} S.~L.,
  {Carlberg} R.~G.,  1998, \mn@doi [\apjs] {10.1086/313106}, \href
  {https://ui.adsabs.harvard.edu/abs/1998ApJS..116..247E} {116, 247}

\bibitem[\protect\citeauthoryear{{Faber} \& {Jackson}}{{Faber} \&
  {Jackson}}{1976}]{faber1976}
{Faber} S.~M.,  {Jackson} R.~E.,  1976, \mn@doi [\apj] {10.1086/154215}, \href
  {http://adsabs.harvard.edu/abs/1976ApJ...204..668F} {204, 668}

\bibitem[\protect\citeauthoryear{{Ford} et~al.,}{{Ford} et~al.}{2003}]{acs}
{Ford} H.~C.,  et~al., 2003, in {Blades} J.~C.,  {Siegmund} O.~H.~W.,  eds,
  \procspie Vol. 4854, Future EUV/UV and Visible Space Astrophysics Missions
  and Instrumentation.. pp 81--94, \mn@doi{10.1117/12.460040}

\bibitem[\protect\citeauthoryear{{Geach} et~al.,}{{Geach}
  et~al.}{2006}]{geach2006}
{Geach} J.~E.,  et~al., 2006, \mn@doi [\apj] {10.1086/506469}, \href
  {https://ui.adsabs.harvard.edu/abs/2006ApJ...649..661G} {649, 661}

\bibitem[\protect\citeauthoryear{{Gioia}, {Maccacaro}, {Schild}, {Wolter},
  {Stocke}, {Morris}  \& {Henry}}{{Gioia} et~al.}{1990}]{gioia}
{Gioia} I.~M.,  {Maccacaro} T.,  {Schild} R.~E.,  {Wolter} A.,  {Stocke} J.~T.,
   {Morris} S.~L.,   {Henry} J.~P.,  1990, \mn@doi [\apjs] {10.1086/191426},
  \href {http://adsabs.harvard.edu/abs/1990ApJS...72..567G} {72, 567}

\bibitem[\protect\citeauthoryear{{Grillo} et~al.,}{{Grillo}
  et~al.}{2016}]{grillo2016}
{Grillo} C.,  et~al., 2016, \mn@doi [\apj] {10.3847/0004-637X/822/2/78}, \href
  {https://ui.adsabs.harvard.edu/abs/2016ApJ...822...78G} {822, 78}

\bibitem[\protect\citeauthoryear{{Harvey} et~al.,}{{Harvey}
  et~al.}{2014}]{harvey2014}
{Harvey} D.,  et~al., 2014, \mn@doi [\mnras] {10.1093/mnras/stu337}, \href
  {http://adsabs.harvard.edu/abs/2014MNRAS.441..404H} {441, 404}

\bibitem[\protect\citeauthoryear{{Harvey}, {Massey}, {Kitching}, {Taylor}  \&
  {Tittley}}{{Harvey} et~al.}{2015}]{harvey2015}
{Harvey} D.,  {Massey} R.,  {Kitching} T.,  {Taylor} A.,   {Tittley} E.,  2015,
  \mn@doi [Science] {10.1126/science.1261381}, \href
  {http://adsabs.harvard.edu/abs/2015Sci...347.1462H} {347, 1462}

\bibitem[\protect\citeauthoryear{{Harvey}, {Kneib}  \& {Jauzac}}{{Harvey}
  et~al.}{2016}]{harvey2016}
{Harvey} D.,  {Kneib} J.~P.,   {Jauzac} M.,  2016, \mn@doi [\mnras]
  {10.1093/mnras/stw295}, \href
  {http://adsabs.harvard.edu/abs/2016MNRAS.458..660H} {458, 660}

\bibitem[\protect\citeauthoryear{{Hoekstra}, {Bartelmann}, {Dahle}, {Israel},
  {Limousin}  \& {Meneghetti}}{{Hoekstra} et~al.}{2013}]{hoekstra2013}
{Hoekstra} H.,  {Bartelmann} M.,  {Dahle} H.,  {Israel} H.,  {Limousin} M.,
  {Meneghetti} M.,  2013, \mn@doi [\ssr] {10.1007/s11214-013-9978-5}, \href
  {https://ui.adsabs.harvard.edu/abs/2013SSRv..177...75H} {177, 75}

\bibitem[\protect\citeauthoryear{{Huang} et~al.,}{{Huang}
  et~al.}{2016}]{huang2016}
{Huang} K.-H.,  et~al., 2016, \mn@doi [\apjl] {10.3847/2041-8205/823/1/L14},
  \href {http://adsabs.harvard.edu/abs/2016ApJ...823L..14H} {823, L14}

\bibitem[\protect\citeauthoryear{{Ishigaki}, {Kawamata}, {Ouchi}, {Oguri},
  {Shimasaku}  \& {Ono}}{{Ishigaki} et~al.}{2018}]{ishigaki2018}
{Ishigaki} M.,  {Kawamata} R.,  {Ouchi} M.,  {Oguri} M.,  {Shimasaku} K.,
  {Ono} Y.,  2018, \mn@doi [\apj] {10.3847/1538-4357/aaa544}, \href
  {https://ui.adsabs.harvard.edu/abs/2018ApJ...854...73I} {854, 73}

\bibitem[\protect\citeauthoryear{{Jauzac} et~al.,}{{Jauzac}
  et~al.}{2014}]{jauzac2014}
{Jauzac} M.,  et~al., 2014, \mn@doi [\mnras] {10.1093/mnras/stu1355}, \href
  {https://ui.adsabs.harvard.edu/abs/2014MNRAS.443.1549J} {443, 1549}

\bibitem[\protect\citeauthoryear{{Jauzac} et~al.,}{{Jauzac}
  et~al.}{2015}]{jauzacmacsj0416}
{Jauzac} M.,  et~al., 2015, \mn@doi [\mnras] {10.1093/mnras/stu2425}, \href
  {http://adsabs.harvard.edu/abs/2015MNRAS.446.4132J} {446, 4132}

\bibitem[\protect\citeauthoryear{{Jauzac} et~al.,}{{Jauzac}
  et~al.}{2016a}]{jauzac2016a}
{Jauzac} M.,  et~al., 2016a, \mn@doi [\mnras] {10.1093/mnras/stw069}, \href
  {http://adsabs.harvard.edu/abs/2016MNRAS.457.2029J} {457, 2029}

\bibitem[\protect\citeauthoryear{{Jauzac} et~al.,}{{Jauzac}
  et~al.}{2016b}]{jauzac2016b}
{Jauzac} M.,  et~al., 2016b, \mn@doi [\mnras] {10.1093/mnras/stw2251}, \href
  {https://ui.adsabs.harvard.edu/abs/2016MNRAS.463.3876J} {463, 3876}

\bibitem[\protect\citeauthoryear{{Jauzac} et~al.,}{{Jauzac}
  et~al.}{2018}]{jauzac2018b}
{Jauzac} M.,  et~al., 2018, \mn@doi [\mnras] {10.1093/mnras/sty2366}, \href
  {https://ui.adsabs.harvard.edu/abs/2018MNRAS.481.2901J} {481, 2901}

\bibitem[\protect\citeauthoryear{{Jauzac} et~al.,}{{Jauzac}
  et~al.}{2019}]{jauzac2019}
{Jauzac} M.,  et~al., 2019, \mn@doi [\mnras] {10.1093/mnras/sty3312}, \href
  {https://ui.adsabs.harvard.edu/abs/2019MNRAS.483.3082J} {483, 3082}

\bibitem[\protect\citeauthoryear{{Johnson} \& {Sharon}}{{Johnson} \&
  {Sharon}}{2016}]{johnson2016}
{Johnson} T.~L.,  {Sharon} K.,  2016, \mn@doi [\apj]
  {10.3847/0004-637X/832/1/82}, \href
  {https://ui.adsabs.harvard.edu/abs/2016ApJ...832...82J} {832, 82}

\bibitem[\protect\citeauthoryear{{Johnson}, {Sharon}, {Bayliss}, {Gladders},
  {Coe}  \& {Ebeling}}{{Johnson} et~al.}{2014}]{johnson2014}
{Johnson} T.~L.,  {Sharon} K.,  {Bayliss} M.~B.,  {Gladders} M.~D.,  {Coe} D.,
   {Ebeling} H.,  2014, \mn@doi [\apj] {10.1088/0004-637X/797/1/48}, \href
  {https://ui.adsabs.harvard.edu/abs/2014ApJ...797...48J} {797, 48}

\bibitem[\protect\citeauthoryear{{Johnson} et~al.,}{{Johnson}
  et~al.}{2017}]{johnson2017}
{Johnson} T.~L.,  et~al., 2017, \mn@doi [\apjl] {10.3847/2041-8213/aa7516},
  \href {https://ui.adsabs.harvard.edu/abs/2017ApJ...843L..21J} {843, L21}

\bibitem[\protect\citeauthoryear{{Joye} \& {Mandel}}{{Joye} \&
  {Mandel}}{2003}]{ds9}
{Joye} W.~A.,  {Mandel} E.,  2003, in {Payne} H.~E.,  {Jedrzejewski} R.~I.,
  {Hook} R.~N.,  eds,  Astronomical Society of the Pacific Conference Series
  Vol. 295, Astronomical Data Analysis Software and Systems XII. p.~489

\bibitem[\protect\citeauthoryear{{Jullo}, {Kneib}, {Limousin},
  {El{\'{\i}}asd{\'o}ttir}, {Marshall}  \& {Verdugo}}{{Jullo}
  et~al.}{2007}]{jullo2007}
{Jullo} E.,  {Kneib} J.-P.,  {Limousin} M.,  {El{\'{\i}}asd{\'o}ttir} {\'A}.,
  {Marshall} P.~J.,   {Verdugo} T.,  2007, \mn@doi [New Journal of Physics]
  {10.1088/1367-2630/9/12/447}, \href
  {http://adsabs.harvard.edu/abs/2007NJPh....9..447J} {9, 447}

\bibitem[\protect\citeauthoryear{{Jullo}, {Natarajan}, {Kneib}, {D'Aloisio},
  {Limousin}, {Richard}  \& {Schimd}}{{Jullo} et~al.}{2010a}]{jullo2010}
{Jullo} E.,  {Natarajan} P.,  {Kneib} J.-P.,  {D'Aloisio} A.,  {Limousin} M.,
  {Richard} J.,   {Schimd} C.,  2010a, \mn@doi [Science]
  {10.1126/science.1185759}, \href
  {http://adsabs.harvard.edu/abs/2010Sci...329..924J} {329, 924}

\bibitem[\protect\citeauthoryear{{Jullo}, {Natarajan}, {Kneib}, {D'Aloisio},
  {Limousin}, {Richard}  \& {Schimd}}{{Jullo} et~al.}{2010b}]{jullo2011}
{Jullo} E.,  {Natarajan} P.,  {Kneib} J.~P.,  {D'Aloisio} A.,  {Limousin} M.,
  {Richard} J.,   {Schimd} C.,  2010b, \mn@doi [Science]
  {10.1126/science.1185759}, \href
  {https://ui.adsabs.harvard.edu/abs/2010Sci...329..924J} {329, 924}

\bibitem[\protect\citeauthoryear{{Kassiola} \& {Kovner}}{{Kassiola} \&
  {Kovner}}{1993}]{piemd}
{Kassiola} A.,  {Kovner} I.,  1993, \mn@doi [\apj] {10.1086/173325}, \href
  {http://adsabs.harvard.edu/abs/1993ApJ...417..450K} {417, 450}

\bibitem[\protect\citeauthoryear{{Kilbinger}}{{Kilbinger}}{2015}]{kilbinger2015}
{Kilbinger} M.,  2015, \mn@doi [Reports on Progress in Physics]
  {10.1088/0034-4885/78/8/086901}, \href
  {https://ui.adsabs.harvard.edu/abs/2015RPPh...78h6901K} {78, 086901}

\bibitem[\protect\citeauthoryear{{Kneib} \& {Natarajan}}{{Kneib} \&
  {Natarajan}}{2011}]{kneib2011}
{Kneib} J.-P.,  {Natarajan} P.,  2011, \mn@doi [\aapr]
  {10.1007/s00159-011-0047-3}, \href
  {http://adsabs.harvard.edu/abs/2011A%26ARv..19...47K} {19, 47}

\bibitem[\protect\citeauthoryear{{Kneib}, {Ellis}, {Smail}, {Couch}  \&
  {Sharples}}{{Kneib} et~al.}{1996}]{kneib1996}
{Kneib} J.-P.,  {Ellis} R.~S.,  {Smail} I.,  {Couch} W.~J.,   {Sharples} R.~M.,
   1996, \mn@doi [\apj] {10.1086/177995}, \href
  {http://adsabs.harvard.edu/abs/1996ApJ...471..643K} {471, 643}

\bibitem[\protect\citeauthoryear{{Knudsen}, {Richard}, {Kneib}, {Jauzac},
  {Cl{\'e}ment}, {Drouart}, {Egami}  \& {Lindroos}}{{Knudsen}
  et~al.}{2016}]{knudsen2016}
{Knudsen} K.~K.,  {Richard} J.,  {Kneib} J.-P.,  {Jauzac} M.,  {Cl{\'e}ment}
  B.,  {Drouart} G.,  {Egami} E.,   {Lindroos} L.,  2016, \mn@doi [\mnras]
  {10.1093/mnrasl/slw114}, \href
  {https://ui.adsabs.harvard.edu/abs/2016MNRAS.462L...6K} {462, L6}

\bibitem[\protect\citeauthoryear{{LaRoque} et~al.,}{{LaRoque}
  et~al.}{2003}]{laroque2003}
{LaRoque} S.~J.,  et~al., 2003, \mn@doi [\apj] {10.1086/345500}, \href
  {https://ui.adsabs.harvard.edu/abs/2003ApJ...583..559L} {583, 559}

\bibitem[\protect\citeauthoryear{{Lagattuta} et~al.,}{{Lagattuta}
  et~al.}{2017}]{lagatutta}
{Lagattuta} D.~J.,  et~al., 2017, \mn@doi [\mnras] {10.1093/mnras/stx1079},
  \href {http://adsabs.harvard.edu/abs/2017MNRAS.469.3946L} {469, 3946}

\bibitem[\protect\citeauthoryear{{Lagattuta} et~al.,}{{Lagattuta}
  et~al.}{2019}]{lagattuta2019}
{Lagattuta} D.~J.,  et~al., 2019, \mn@doi [\mnras] {10.1093/mnras/stz620},
  \href {https://ui.adsabs.harvard.edu/abs/2019MNRAS.485.3738L} {485, 3738}

\bibitem[\protect\citeauthoryear{{Limousin}, {Kneib}  \&
  {Natarajan}}{{Limousin} et~al.}{2005}]{limousin2005}
{Limousin} M.,  {Kneib} J.-P.,   {Natarajan} P.,  2005, \mn@doi [\mnras]
  {10.1111/j.1365-2966.2004.08449.x}, \href
  {http://adsabs.harvard.edu/abs/2005MNRAS.356..309L} {356, 309}

\bibitem[\protect\citeauthoryear{{Limousin} et~al.,}{{Limousin}
  et~al.}{2007}]{limousin2007}
{Limousin} M.,  et~al., 2007, \mn@doi [\apj] {10.1086/521293}, \href
  {http://adsabs.harvard.edu/abs/2007ApJ...668..643L} {668, 643}

\bibitem[\protect\citeauthoryear{{Livermore}, {Finkelstein}  \&
  {Lotz}}{{Livermore} et~al.}{2017}]{livermore2017}
{Livermore} R.~C.,  {Finkelstein} S.~L.,   {Lotz} J.~M.,  2017, \mn@doi [\apj]
  {10.3847/1538-4357/835/2/113}, \href
  {https://ui.adsabs.harvard.edu/abs/2017ApJ...835..113L} {835, 113}

\bibitem[\protect\citeauthoryear{{Lotz} et~al.,}{{Lotz} et~al.}{2017}]{hff}
{Lotz} J.~M.,  et~al., 2017, \mn@doi [\apj] {10.3847/1538-4357/837/1/97}, \href
  {http://adsabs.harvard.edu/abs/2017ApJ...837...97L} {837, 97}

\bibitem[\protect\citeauthoryear{{Ma}, {Ebeling}, {Donovan}  \& {Barrett}}{{Ma}
  et~al.}{2008}]{ma08}
{Ma} C.-J.,  {Ebeling} H.,  {Donovan} D.,   {Barrett} E.,  2008, \mn@doi [\apj]
  {10.1086/589991}, \href
  {https://ui.adsabs.harvard.edu/abs/2008ApJ...684..160M} {684, 160}

\bibitem[\protect\citeauthoryear{{MacKenzie} et~al.,}{{MacKenzie}
  et~al.}{2014}]{mackenzie}
{MacKenzie} T.~P.,  et~al., 2014, \mn@doi [\mnras] {10.1093/mnras/stu1623},
  \href {http://adsabs.harvard.edu/abs/2014MNRAS.445..201M} {445, 201}

\bibitem[\protect\citeauthoryear{{Mahler} et~al.,}{{Mahler}
  et~al.}{2018}]{mahler2018}
{Mahler} G.,  et~al., 2018, \mn@doi [\mnras] {10.1093/mnras/stx1971}, \href
  {https://ui.adsabs.harvard.edu/abs/2018MNRAS.473..663M} {473, 663}

\bibitem[\protect\citeauthoryear{{Mahler} et~al.,}{{Mahler}
  et~al.}{2019}]{mahler19}
{Mahler} G.,  et~al., 2019, \mn@doi [\apj] {10.3847/1538-4357/ab042b}, \href
  {https://ui.adsabs.harvard.edu/abs/2019ApJ...873...96M} {873, 96}

\bibitem[\protect\citeauthoryear{{Massey}, {Kitching}  \& {Richard}}{{Massey}
  et~al.}{2010}]{massey2010}
{Massey} R.,  {Kitching} T.,   {Richard} J.,  2010, \mn@doi [Reports on
  Progress in Physics] {10.1088/0034-4885/73/8/086901}, \href
  {https://ui.adsabs.harvard.edu/abs/2010RPPh...73h6901M} {73, 086901}

\bibitem[\protect\citeauthoryear{{Molnar}, {Hughes}, {Donahue}  \&
  {Joy}}{{Molnar} et~al.}{2002}]{molnar2002}
{Molnar} S.~M.,  {Hughes} J.~P.,  {Donahue} M.,   {Joy} M.,  2002, \mn@doi
  [\apjl] {10.1086/342086}, \href
  {https://ui.adsabs.harvard.edu/abs/2002ApJ...573L..91M} {573, L91}

\bibitem[\protect\citeauthoryear{{Monna} et~al.,}{{Monna} et~al.}{2017}]{monna}
{Monna} A.,  et~al., 2017, \mn@doi [\mnras] {10.1093/mnras/stx015}, \href
  {http://adsabs.harvard.edu/abs/2017MNRAS.466.4094M} {466, 4094}

\bibitem[\protect\citeauthoryear{{Natarajan} \& {Kneib}}{{Natarajan} \&
  {Kneib}}{1997}]{ellip}
{Natarajan} P.,  {Kneib} J.-P.,  1997, \mn@doi [\mnras]
  {10.1093/mnras/287.4.833}, \href
  {http://adsabs.harvard.edu/abs/1997MNRAS.287..833N} {287, 833}

\bibitem[\protect\citeauthoryear{{Natarajan} et~al.,}{{Natarajan}
  et~al.}{2017}]{natarajan2017}
{Natarajan} P.,  et~al., 2017, \mn@doi [\mnras] {10.1093/mnras/stw3385}, \href
  {http://adsabs.harvard.edu/abs/2017MNRAS.468.1962N} {468, 1962}

\bibitem[\protect\citeauthoryear{{Newman}, {Treu}, {Ellis}, {Sand }, {Nipoti},
  {Richard}  \& {Jullo}}{{Newman} et~al.}{2013a}]{newman2013a}
{Newman} A.~B.,  {Treu} T.,  {Ellis} R.~S.,  {Sand } D.~J.,  {Nipoti} C.,
  {Richard} J.,   {Jullo} E.,  2013a, \mn@doi [\apj]
  {10.1088/0004-637X/765/1/24}, \href
  {https://ui.adsabs.harvard.edu/abs/2013ApJ...765...24N} {765, 24}

\bibitem[\protect\citeauthoryear{{Newman}, {Treu}, {Ellis}  \& {Sand
  }}{{Newman} et~al.}{2013b}]{newman2013b}
{Newman} A.~B.,  {Treu} T.,  {Ellis} R.~S.,   {Sand } D.~J.,  2013b, \mn@doi
  [\apj] {10.1088/0004-637X/765/1/25}, \href
  {https://ui.adsabs.harvard.edu/abs/2013ApJ...765...25N} {765, 25}

\bibitem[\protect\citeauthoryear{{Niemiec}, {Jauzac}, {Jullo}, {Limousin},
  {Sharon}, {Kneib}, {Natarajan}  \& {Richard}}{{Niemiec}
  et~al.}{2020}]{niemec2020}
{Niemiec} A.,  {Jauzac} M.,  {Jullo} E.,  {Limousin} M.,  {Sharon} K.,  {Kneib}
  J.-P.,  {Natarajan} P.,   {Richard} J.,  2020, \mn@doi [\mnras]
  {10.1093/mnras/staa473}, \href
  {https://ui.adsabs.harvard.edu/abs/2020MNRAS.493.3331N} {493, 3331}

\bibitem[\protect\citeauthoryear{{Patr{\'\i}cio} et~al.,}{{Patr{\'\i}cio}
  et~al.}{2016}]{patricio2016}
{Patr{\'\i}cio} V.,  et~al., 2016, \mn@doi [\mnras] {10.1093/mnras/stv2859},
  \href {https://ui.adsabs.harvard.edu/abs/2016MNRAS.456.4191P} {456, 4191}

\bibitem[\protect\citeauthoryear{{Postman} et~al.,}{{Postman}
  et~al.}{2012}]{postmanclash}
{Postman} M.,  et~al., 2012, \mn@doi [\apjs] {10.1088/0067-0049/199/2/25},
  \href {http://adsabs.harvard.edu/abs/2012ApJS..199...25P} {199, 25}

\bibitem[\protect\citeauthoryear{{Remolina Gonz{\'a}lez}, {Sharon}  \&
  {Mahler}}{{Remolina Gonz{\'a}lez} et~al.}{2018}]{gonzalez2019}
{Remolina Gonz{\'a}lez} J.~D.,  {Sharon} K.,   {Mahler} G.,  2018, \mn@doi
  [\apj] {10.3847/1538-4357/aacf8e}, \href
  {https://ui.adsabs.harvard.edu/abs/2018ApJ...863...60R} {863, 60}

\bibitem[\protect\citeauthoryear{{Rescigno} et~al.,}{{Rescigno}
  et~al.}{2020}]{rescigno2020}
{Rescigno} U.,  et~al., 2020, \mn@doi [\aap] {10.1051/0004-6361/201936590},
  \href {https://ui.adsabs.harvard.edu/abs/2020A&A...635A..98R} {635, A98}

\bibitem[\protect\citeauthoryear{{Rexroth}, {Kneib}, {Joseph}, {Richard}  \&
  {Her}}{{Rexroth} et~al.}{2017}]{ifs}
{Rexroth} M.,  {Kneib} J.-P.,  {Joseph} R.,  {Richard} J.,   {Her} R.,  2017,
  preprint, \href {http://adsabs.harvard.edu/abs/2017arXiv170309239R} {}
  (\mn@eprint {arXiv} {1703.09239})

\bibitem[\protect\citeauthoryear{{Richard} et~al.,}{{Richard}
  et~al.}{2010}]{rxj2129_model}
{Richard} J.,  et~al., 2010, \mn@doi [\mnras]
  {10.1111/j.1365-2966.2009.16274.x}, \href
  {http://adsabs.harvard.edu/abs/2010MNRAS.404..325R} {404, 325}

\bibitem[\protect\citeauthoryear{{Richard} et~al.,}{{Richard}
  et~al.}{2014}]{richard2014}
{Richard} J.,  et~al., 2014, \mn@doi [\mnras] {10.1093/mnras/stu1395}, \href
  {http://adsabs.harvard.edu/abs/2014MNRAS.444..268R} {444, 268}

\bibitem[\protect\citeauthoryear{{Richard} et~al.,}{{Richard}
  et~al.}{2015}]{richard2015}
{Richard} J.,  et~al., 2015, \mn@doi [\mnras] {10.1093/mnrasl/slu150}, \href
  {http://adsabs.harvard.edu/abs/2015MNRAS.446L..16R} {446, L16}

\bibitem[\protect\citeauthoryear{{Richard} et~al.,}{{Richard}
  et~al.}{2020}]{richard2020}
{Richard} J.,  et~al., 2020, arXiv e-prints, \href
  {https://ui.adsabs.harvard.edu/abs/2020arXiv200909784R} {p. arXiv:2009.09784}

\bibitem[\protect\citeauthoryear{{Robertson}, {Harvey}, {Massey}, {Eke},
  {McCarthy}, {Jauzac}, {Li}  \& {Schaye}}{{Robertson}
  et~al.}{2019}]{robertson2019}
{Robertson} A.,  {Harvey} D.,  {Massey} R.,  {Eke} V.,  {McCarthy} I.~G.,
  {Jauzac} M.,  {Li} B.,   {Schaye} J.,  2019, \mn@doi [\mnras]
  {10.1093/mnras/stz1815}, \href
  {https://ui.adsabs.harvard.edu/abs/2019MNRAS.488.3646R} {488, 3646}

\bibitem[\protect\citeauthoryear{{Rosati} et~al.,}{{Rosati}
  et~al.}{2014}]{rosati2014}
{Rosati} P.,  et~al., 2014, The Messenger, \href
  {http://adsabs.harvard.edu/abs/2014Msngr.158...48R} {158, 48}

\bibitem[\protect\citeauthoryear{{Schmidt} et~al.,}{{Schmidt}
  et~al.}{2014}]{schmidt2014}
{Schmidt} K.~B.,  et~al., 2014, \mn@doi [\apjl] {10.1088/2041-8205/782/2/L36},
  \href {https://ui.adsabs.harvard.edu/abs/2014ApJ...782L..36S} {782, L36}

\bibitem[\protect\citeauthoryear{{Sharon} et~al.,}{{Sharon}
  et~al.}{2020}]{sharon2020}
{Sharon} K.,  et~al., 2020, \mn@doi [\apjs] {10.3847/1538-4365/ab5f13}, \href
  {https://ui.adsabs.harvard.edu/abs/2020ApJS..247...12S} {247, 12}

\bibitem[\protect\citeauthoryear{{Soto}, {Lilly}, {Bacon}, {Richard}  \&
  {Conseil}}{{Soto} et~al.}{2016}]{zap}
{Soto} K.~T.,  {Lilly} S.~J.,  {Bacon} R.,  {Richard} J.,   {Conseil} S.,
  2016, \mn@doi [\mnras] {10.1093/mnras/stw474}, \href
  {http://adsabs.harvard.edu/abs/2016MNRAS.458.3210S} {458, 3210}

\bibitem[\protect\citeauthoryear{{Steinhardt} et~al.,}{{Steinhardt}
  et~al.}{2020}]{steinhardt2020}
{Steinhardt} C.~L.,  et~al., 2020, \mn@doi [\apjs] {10.3847/1538-4365/ab75ed},
  \href {https://ui.adsabs.harvard.edu/abs/2020ApJS..247...64S} {247, 64}

\bibitem[\protect\citeauthoryear{{Takata} et~al.,}{{Takata}
  et~al.}{2003}]{takata2003}
{Takata} T.,  et~al., 2003, \mn@doi [\pasj] {10.1093/pasj/55.4.789}, \href
  {https://ui.adsabs.harvard.edu/abs/2003PASJ...55..789T} {55, 789}

\bibitem[\protect\citeauthoryear{{Tam} et~al.,}{{Tam} et~al.}{2020}]{tam2020}
{Tam} S.-I.,  et~al., 2020, \mn@doi [\mnras] {10.1093/mnras/staa1828}, \href
  {https://ui.adsabs.harvard.edu/abs/2020MNRAS.496.4032T} {496, 4032}

\bibitem[\protect\citeauthoryear{{Toft} et~al.,}{{Toft}
  et~al.}{2017}]{toft2017}
{Toft} S.,  et~al., 2017, \mn@doi [\nat] {10.1038/nature22388}, \href
  {https://ui.adsabs.harvard.edu/abs/2017Natur.546..510T} {546, 510}

\bibitem[\protect\citeauthoryear{{Treu} \& {Ellis}}{{Treu} \&
  {Ellis}}{2015}]{treu2015}
{Treu} T.,  {Ellis} R.~S.,  2015, \mn@doi [Contemporary Physics]
  {10.1080/00107514.2015.1006001}, \href
  {https://ui.adsabs.harvard.edu/abs/2015ConPh..56...17T} {56, 17}

\bibitem[\protect\citeauthoryear{{Treu} et~al.,}{{Treu}
  et~al.}{2015}]{treu_glass}
{Treu} T.,  et~al., 2015, \mn@doi [\apj] {10.1088/0004-637X/812/2/114}, \href
  {https://ui.adsabs.harvard.edu/abs/2015ApJ...812..114T} {812, 114}

\bibitem[\protect\citeauthoryear{{Vanzella} et~al.,}{{Vanzella}
  et~al.}{2017}]{vanzella2015}
{Vanzella} E.,  et~al., 2017, \mn@doi [\mnras] {10.1093/mnras/stw2442}, \href
  {https://ui.adsabs.harvard.edu/abs/2017MNRAS.465.3803V} {465, 3803}

\bibitem[\protect\citeauthoryear{{Weilbacher}, {Streicher}, {Urrutia}, {Jarno},
  {P{\'e}contal-Rousset}, {Bacon}  \& {B{\"o}hm}}{{Weilbacher}
  et~al.}{2012}]{pipeline2}
{Weilbacher} P.~M.,  {Streicher} O.,  {Urrutia} T.,  {Jarno} A.,
  {P{\'e}contal-Rousset} A.,  {Bacon} R.,   {B{\"o}hm} P.,  2012, in Software
  and Cyberinfrastructure for Astronomy II. p. 84510B,
  \mn@doi{10.1117/12.925114}

\bibitem[\protect\citeauthoryear{{Weilbacher}, {Streicher}, {Urrutia},
  {P{\'e}contal-Rousset}, {Jarno}  \& {Bacon}}{{Weilbacher}
  et~al.}{2014}]{pipeline}
{Weilbacher} P.~M.,  {Streicher} O.,  {Urrutia} T.,  {P{\'e}contal-Rousset} A.,
   {Jarno} A.,   {Bacon} R.,  2014, in {Manset} N.,  {Forshay} P.,  eds,
  Astronomical Society of the Pacific Conference Series Vol. 485, Astronomical
  Data Analysis Software and Systems XXIII. p.~451 (\mn@eprint {arXiv}
  {1507.00034})

\bibitem[\protect\citeauthoryear{{Williams} et~al.,}{{Williams}
  et~al.}{2018}]{williams18}
{Williams} P.~R.,  et~al., 2018, \mn@doi [\mnras] {10.1093/mnrasl/sly043},
  \href {https://ui.adsabs.harvard.edu/abs/2018MNRAS.477L..70W} {477, L70}

\bibitem[\protect\citeauthoryear{{Wittman}, {Golovich}  \& {Dawson}}{{Wittman}
  et~al.}{2017}]{counterdavid}
{Wittman} D.,  {Golovich} N.,   {Dawson} W.~A.,  2017, preprint, \href
  {http://adsabs.harvard.edu/abs/2017arXiv170105877W} {} (\mn@eprint {arXiv}
  {1701.05877})

\bibitem[\protect\citeauthoryear{{Zitrin}, {Broadhurst}, {Barkana}, {Rephaeli}
  \& {Ben{\'{\i}}tez}}{{Zitrin} et~al.}{2011a}]{zitrin2011_b}
{Zitrin} A.,  {Broadhurst} T.,  {Barkana} R.,  {Rephaeli} Y.,
  {Ben{\'{\i}}tez} N.,  2011a, \mn@doi [\mnras]
  {10.1111/j.1365-2966.2010.17574.x}, \href
  {http://adsabs.harvard.edu/abs/2011MNRAS.410.1939Z} {410, 1939}

\bibitem[\protect\citeauthoryear{{Zitrin} et~al.,}{{Zitrin}
  et~al.}{2011b}]{zitrin2011}
{Zitrin} A.,  et~al., 2011b, \mn@doi [\apj] {10.1088/0004-637X/742/2/117},
  \href {http://adsabs.harvard.edu/abs/2011ApJ...742..117Z} {742, 117}

\bibitem[\protect\citeauthoryear{{Zitrin} et~al.,}{{Zitrin}
  et~al.}{2015}]{zitrin2015}
{Zitrin} A.,  et~al., 2015, \mn@doi [\apj] {10.1088/0004-637X/801/1/44}, \href
  {http://cdsads.u-strasbg.fr/abs/2015ApJ...801...44Z} {801, 44}

\makeatother
\end{thebibliography}




\appendix

\section{List of secure and likely redshifts extracted from the MUSE datacubes}
\label{tab}
We here give the redshifts extracted with a quality flag greater than 2 (i.e., secure or likely) from the MUSE datacubes in Table\,\ref{rxj2129_z}, Table\,\ref{ms0451_z}, and Table\,\ref{macs2129_z} for RX\,J2129, MS\,0451, and MACS\,J2129 respectively.
\newpage
\begin{table}
\begin{tabular}{cc}
\, & \,\\
\end{tabular}
\caption{List of redshifts measured with a quality flag larger than 2 in RX\,J2129. Column (1) is the ID of the source ("sing" and "cont" stand for singular and continuum emission line in the \textsc{muselet} catalogue). If it exists, the index stands for the pointing's number. Columns (2) and (3) are the R.\,A. and the Dec. in degrees (J2000). Column (4) is the redshift of the source, and column (5) the systematic error on the redshift. Column (6) is the QF of the determination.}
\label{rxj2129_z}
\end{table}
\scriptsize
\begin{center}
\tablefirsthead{\hline
                                \multicolumn{1}{c}{ID} &
                                \multicolumn{1}{c}{R.\,A.} &
                                \multicolumn{1}{c}{Decl.} &
                                \multicolumn{1}{c}{$z$} &
                                \multicolumn{1}{c}{$z_{\rm err}$ [$\times 10^{-4}$]} &
                                \multicolumn{1}{c}{QF}
                                \\ \hline  }
\tablehead{\hline \multicolumn{6}{l}{\small\sl continued from previous page}\\
           \hline \multicolumn{1}{c}{ID} &
                                \multicolumn{1}{c}{R.\,A.} &
                                \multicolumn{1}{c}{Decl.} &
                                \multicolumn{1}{c}{$z$} &
                                \multicolumn{1}{c}{$z_{\rm err}$ $\times 10^{-4}$]} &
                                \multicolumn{1}{c}{QF}
                                \\ \hline  }
\tabletail{\hline\multicolumn{6}{r}{\small\sl continued on next page}\\\hline}
\tablelasttail{\hline}
\par
\begin{supertabular}{cccccc}
$964$ & 322.42145 & 0.09219  & 0.0 & 0.37 & 3\\
$971$ & 322.42834 & 0.09236 & 0.0 & 1.47 & 3\\
$927$ & 322.43222 & 0.09100 & 0.0 & 2.19 & 3\\
$937$ & 322.43228 & 0.09141 & 0.0 & 0.36 & 3\\
$829$ & 322.42813 & 0.08774 & 0.0 & 1.62 & 3\\
$919$ & 322.40698 & 0.09054 & 0.0 & 1.55 & 3\\
$711$ & 322.42581 & 0.08754 & 0.0 & 0.26 & 3\\
$1079$ & 322.41464 & 0.0970 & 0.0 & 1.11 & 3\\
$1030$ & 322.41000 & 0.0947 & 0.0 & 6.14 & 3\\
$614$ & 322.41513 & 0.08205 & 0.0 & 2.71 & 3\\
$953$ & 322.40814 & 0.09203 & 0.0 & 1.50 & 3\\
$951$ & 322.41626 & 0.09215 & 0.0 & 1.18 & 3\\
$849$ & 322.40720 & 0.08800 & 0.0 & 2.04 & 2\\
$818$ & 322.41064 & 0.08742 & 0.0 & 3.62 & 3\\
$805$ & 322.42130 & 0.08697 & 0.0 & 2.16 & 3\\
$724$ & 322.42001 & 0.08529 & 0.0 & 0.55 & 3\\
$745$ & 322.41858 & 0.08547 & 0.0 & 5.39 & 3\\
$1197$ & 322.42480 & 0.10001 & 0.0 & 2.43 & 3\\
$639$ & 322.41571 & 0.08298  & 0.1288 & 0.10 & 3\\
$783$ & 322.41132 & 0.08633 & 0.1343 & 0.10 & 3\\
$741$ & 322.41953 & 0.08604 & 0.1346 & 0.56 & 3\\
$776$ & 322.40485 & 0.09006 & 0.1375 & 0.08 & 3\\
$1068$ & 322.4202 & 0.09866 & 0.1379 & 0.08 & 3\\
$973$ & 322.40921 & 0.09290 & 0.1804 & 0.13 & 3\\
$625$ & 322.41287 & 0.08312 & 0.2145 & 0.10 & 3\\
$832$ & 322.41330 & 0.08755 & 0.2150 & 0.11 & 3\\
$806$ & 322.41379 & 0.08716 & 0.2151 & 0.10 & 3\\
$1114$ & 322.42868 & 0.09705 & 0.2255 & 0.84 & 2\\
$963$ & 322.42215 & 0.09274 & 0.2258 & 0.80 & 3\\
$794$ & 322.41827 & 0.08879 & 0.2261 & 0.74 & 3\\
$1233$ & 322.42120 & 0.1019 & 0.2281 & 1.19 & 2\\
$926$ & 322.41092 & 0.09205 & 0.2293 & 0.88 & 3\\
$980$ & 322.41806 & 0.09302 & 0.2304 & 1.88 & 3\\
$637$ & 322.41708 & 0.08393 & 0.2305 & 0.08 & 3\\
$533$ & 322.41476 & 0.07965 & 0.2316 & 1.75 & 2\\
$982$ & 322.41580 & 0.09283 & 0.2320 & 3.14 & 2\\
$931$ & 322.41611 & 0.09153 & 0.2320 & 0.13 & 3\\
$761$ & 322.41953 & 0.08919 & 0.2322 & 0.46 & 3\\
$731$ & 322.41367 & 0.08621 & 0.2328 & 1.70 & 3\\
$739$ & 322.41422 & 0.08590 & 0.2330 & 1.98 & 2\\
$816$ & 322.41214 & 0.08857 & 0.2334 & 0.81 & 3\\
$755$ & 322.41812 & 0.08650 & 0.2336 & 0.66 & 3\\
$764$ & 322.41458 & 0.08650 & 0.2337 & 0.89 & 3\\
$454$ & 322.40631 & 0.08292 & 0.2339 & 0.21 & 3\\
$642$ & 322.41647 & 0.08923 & 0.2339 & 0.08 & 3\\
$879$ & 322.41574 & 0.08937 & 0.2334 & 0.09 & 3\\
$870$ & 322.41312 & 0.08859 & 0.2340 & 4.14 & 2\\
$825$ & 322.41602 & 0.08785 & 0.2341 & 0.69 & 2\\
$766$ & 322.40286 & 0.08782 & 0.2348 & 0.37 & 3\\
$592$ & 322.41321 & 0.08601 & 0.2348 & 0.28 & 3\\
$860$ & 322.41345 & 0.08871 & 0.2351 & 0.66 & 2\\
$598$ & 322.41187 & 0.08512 & 0.2359 & 0.58 & 3\\
$sing192$ & 322.41581 & 0.08748 & 0.2360 & 8.67 & 2\\
$732$ & 322.40494 & 0.08631 & 0.2364 & 0.68 & 3\\
$1006$ & 322.41739 & 0.09443 & 0.2368 & 0.94 & 3\\
$1157$ & 322.42090 & 0.09899 & 0.2369 & 2.08 & 3\\
$989$ & 322.42053 & 0.09340 & 0.2371 & 0.56 & 2\\
$1072$ & 322.42831 & 0.09789 & 0.2374 & 0.49 & 3\\
$1093$ & 322.41595 & 0.09773 & 0.2375 & 1.10 & 3\\
$565$ & 322.41010 & 0.08144 & 0.2376 & 0.08 & 3\\
$873$ & 322.41724 & 0.09055 & 0.2376 & 0.57 & 3\\
$824$ & 322.42810 & 0.08839 & 0.2377 & 0.96 & 3\\
$908$ & 322.41174 & 0.09003 & 0.2410 & 0.21 & 3\\
$627$ & 322.41910 & 0.08249 & 0.3570 & 0.84 & 2\\
$1031$ & 322.41589 & 0.09451 & 0.3827 & 0.86 & 2\\
$cont46$ & 322.41693 & 0.08472 & 0.3857 & 0.20 & 3\\
$881$ & 322.40048 & 0.08928 & 0.4215 & 0.78 & 2\\
$833$ & 322.40549 & 0.08799 & 0.4221 & 2.08 & 2\\
$693$ & 322.42096 & 0.08438 & 0.4229 & 0.12 & 3\\
$682$ & 322.40857 & 0.08410 & 0.4243 & 0.11 & 3\\
$712$ & 322.41260 & 0.08445 & 0.4275 & 0.30 & 2\\
$615$ & 322.41312 & 0.08187 & 0.4278 & 0.35 & 3\\
$1007$ & 322.42987 & 0.09354 & 0.4340 & 1.00 & 2\\
$838$ & 322.40366 & 0.08796 & 0.4345 & 0.10 & 3\\
$1016$ & 322.41971 & 0.09376 & 0.4468 & 0.60 & 2\\
$861$ & 322.42651 & 0.08892 & 0.5509 & 0.27 & 3\\
$1073$ & 322.41763 & 0.09626 & 0.5552 & 1.98 & 3\\
$699$ & 322.42334 & 0.08441 & 0.5640 & 0.10 & 3\\
$1335$ & 322.42438 & 0.10441 & 0.5788 & 0.17 & 3\\
$1348$ & 322.42416 & 0.10436 & 0.5791 & 0.11 & 3\\
$1084$ & 322.41412 & 0.09651 & 0.5885 & 0.10 & 3\\
$1122$ & 322.41443 & 0.09732 & 0.5917 & 0.30 & 3\\
$1116$ & 322.41434 & 0.09722 & 0.5918 & 0.14 & 3\\
$1119$ & 322.41467 & 0.09737 & 0.5920 & 1.65 & 3\\
$1178$ & 322.42508 & 0.09938 & 0.5925 & 0.11 & 3\\
$cont27$ & 322.42634 & 0.09195 & 0.5925 & 0.23 & 3\\
$787$ & 322.40781 & 0.08611 & 0.6076 & 0.13 & 3\\
$1109$ & 322.41208 & 0.09708 & 0.6165 & 0.17 & 3\\
$728$ & 322.40771 & 0.08536 & 0.6165 & 0.51 & 3\\
$sing111$ & 322.42676 & 0.09212 & 0.6166 & 0.21 & 3\\
$978$ & 322.43118 & 0.09246 & 0.6170 & 0.13 & 3\\
$944$ & 322.42719 & 0.09139 & 0.6174 & 0.12 & 3\\
$1015$ & 322.42349 & 0.09445 & 0.6176 & 0.17 & 3\\
$1053$ & 322.42023 & 0.09528 & 0.6188 & 0.38 & 3\\
$673$ & 322.41553 & 0.08582 & 0.6710 & 0.28 & 3\\
$948$ & 322.41388 & 0.09157 & 0.6711 & 0.18 & 3\\
$1312$ & 322.42239 & 0.10377 & 0.6766 & 3.16 & 3\\
$1271$ & 322.42261 & 0.10200 & 0.6784 & 0.57 & 3\\
$909$ & 322.41492 & 0.09038 & 0.6785 & 0.10 & 3\\
$809$ & 322.41663 & 0.08674 & 0.6787 & 0.11 & 3\\
$sing120$ & 322.41510 & 0.08896 & 0.6787 & 0.11 & 3\\
$1218$ & 322.42322 & 0.10211 & 0.6788 & 0.16 & 3\\
$911$ & 322.41232 & 0.09019 & 0.7077 & 0.21 & 3\\
$918$ & 322.43179 & 0.09065 & 0.8354 & 0.12 & 3\\
$468$ & 322.41049 & 0.07667 & 0.8867 & 1.22 & 2\\
$1226$ & 322.42401 & 0.10125 & 0.9129 & 0.41 & 3\\
$1318$ & 322.42368 & 0.10390 & 0.9129 & 0.91 & 3\\
$521$ & 322.41177 & 0.07885 & 0.9159 & 0.46 & 3\\
$486$ & 322.41129 & 0.07855 & 0.9161 & 0.23 & 3\\
$676$ & 322.41629 & 0.08429 & 0.9356 & 0.16 & 3\\
$1282$ & 322.42450 & 0.10220 & 1.0093 & 1.53 & 3\\
$820$ & 322.41119 & 0.08696 & 1.0400 & 6.95 & 2\\
$sing92$ & 322.40498 & 0.08465 & 1.1037 & 0.85 & 2\\
$821$ & 322.42691 & 0.08723 & 1.1385 & 0.46 & 3\\
$1262$ & 322.41989 & 0.10158 & 1.1388 & 0.42 & 3\\
$733$ & 322.40469 & 0.08524 & 1.2261 & 0.60 & 2\\
$700$ & 322.40466 & 0.08507 & 1.2263 & 0.38 & 3\\
$702$ & 322.42157 & 0.08442 & 1.2580 & 0.45 & 2\\
$1018$ & 322.42865 & 0.09394 & 1.3328 & 1.18 & 2\\
$936$ & 322.41592 & 0.09150 & 1.5180 & 0.52 & 2\\
$848$ & 322.42038 & 0.08833 & 1.5189 & 4.43 & 2\\
$888$ & 322.42017 & 0.08975 & 1.5189 & 0.84 & 2\\
\textbf{$716$} & 322.41855 & 0.08491 & 1.5197 & 1.10 & 3\\
$970$ & 322.41571 & 0.09224 & 1.5199 & 2.9300 & 3\\
\textbf{$734$} & 322.41843 & 0.08537 & 1.5201 & 1.20 & 3\\
$958$ & 322.41559 & 0.09214 & 1.5203 & 1.68 & 3\\
$549$ & 322.40863 & 0.08002 & 1.8500 & 3.29 & 2\\
$sing4$ & 322.40996 & 0.07653 & 3.0132 & 0.35 & 2\\
$sing1$ & 322.41096 & 0.09547 & 3.0156 & 1.14 & 2\\
$758$ & 322.40863 & 0.08562 & 3.0477 & 1.63 & 3\\
\textbf{$sing6$} & 322.41675 & 0.08780 & 3.0800 & 0.54 & 2\\
\textbf{$sing7$} & 322.41700 & 0.08739 & 3.0800 & 0.62 & 2\\
$sing16$ & 322.41769 & 0.09553 & 3.3276 & 1.04 & 2\\
$sing9$ & 322.42956 & 0.09470 & 3.3470 & 1.09 & 2\\
$sing9_{2}$ & 322.42199 & 0.08357 & 3.3475 & 0.92 & 2\\
$sing18$ & 322.43055 & 0.09488 & 3.4463 & 0.76 & 2\\
$sing14$ & 322.40222 & 0.09064 & 3.4934 & 0.75 & 2\\
$sing15$ & 322.41505 & 0.08659 & 3.5650 & 0.55 & 2\\
$sing24$ & 322.42393 & 0.08668 & 3.7030 & 0.39 & 2\\
$736$ & 322.40628 & 0.08511 & 3.7581 & 0.65 & 3\\
$sing20$ & 322.41356 & 0.07947 & 3.7690 & 0.80 & 3\\
$sing34$ & 322.40363 & 0.08834 & 4.2716 & 0.54 & 3\\
$sing35$ & 322.40910 & 0.09220 & 4.2821 & 0.24 & 3\\
$1155$ & 322.41635 & 0.09813 & 4.2841 & 0.62 & 2\\
$sing38$ & 322.40695 & 0.08124 & 4.3374 & 0.69 & 3\\
$sing42$ & 322.41034 & 0.07776 & 4.3708 & 1.20 & 3\\
$sing44$ & 322.40929 & 0.09256 & 4.4358 & 1.72 & 3\\
$sing54$ & 322.40709 & 0.09171 & 4.6575 & 0.19 & 3\\
$586$ & 322.40720 & 0.08109 & 4.6921 & 1.70 & 3\\
$sing62$ & 322.41410 & 0.09775 & 4.8453 & 1.74 & 3\\
$544$ & 322.40845 & 0.07981 & 4.8669 & 0.93 & 2\\
$sing64$ & 322.41584 & 0.07975 & 4.8888 & 0.13 & 2\\
$1230$ & 322.42508 & 0.10080 & 5.2960 & 0.88 & 3\\
$686$ & 322.42145 & 0.08394 & 5.3507 & 1.41 & 2\\
$sing89$ & 322.40703 & 0.09010 & 5.4250 & 0.85 & 2\\
$647$ & 322.41882 & 0.08293 & 5.5346 & 0.57 & 3\\
\hline
\end{supertabular}
\end{center}

\newpage
\begin{table}
\begin{tabular}{cc}
\, & \,\\
\end{tabular}
\caption{List of spectroscopically confirmed redshifts in MS\,0451. The columns are the same as in Table\,\ref{rxj2129_z}.}
\label{ms0451_z}
\end{table}
\scriptsize
\begin{center}
\tablefirsthead{\hline
\hline 
                                \multicolumn{1}{c}{ID} &
                                \multicolumn{1}{c}{R.\,A.} &
                                \multicolumn{1}{c}{Decl.} &
                                \multicolumn{1}{c}{$z$} &
                                \multicolumn{1}{c}{$z_{\rm err}$ [$\times 10^{-4}$]} &
                                \multicolumn{1}{c}{QF}
                                \\ \hline \hline   }
\tablehead{\hline \multicolumn{6}{l}{\small\sl continued from previous page}\\
           \hline \multicolumn{1}{c}{ID} &
                                \multicolumn{1}{c}{R.\,A.} &
                                \multicolumn{1}{c}{Decl.} &
                                \multicolumn{1}{c}{$z$} &
                                \multicolumn{1}{c}{$z_{\rm err}$ [$\times 10^{-4}$]} &
                                \multicolumn{1}{c}{QF}
                                \\ \hline \hline  }
\tabletail{\hline\multicolumn{6}{r}{\small\sl continued on next page}\\\hline}
\tablelasttail{\hline}
\par
\begin{supertabular}{cccccc}
$1878$ & 73.537475 & -3.00266 & -0.0001 & 0.14 & 3\\
$1192$ & 73.547325 & -3.01903 & 0.0011 & 0.18 & 3\\
$1305$ & 73.546982 & -3.01852 & 0.0023 & 0.13 & 3\\
$1321$ & 73.552307 & -3.01785 & 0.0036 & 0.85 & 2\\
$1194$ & 73.545318 & -3.01586 & 0.0625 & 0.08 & 3\\ 
$cont15_2$ & 73.53930 & -3.00815 & 0.1297 & 0.14 & 2\\
$1609$ & 73.53894 & -3.00713 & 0.1299 & 0.03 & 3\\
$1123$ & 73.55352 & -3.01917 & 0.1564 & 0.08 & 3\\
$sing29_2$ & 73.54294 & -3.00923 & 0.1565 & 0.15 & 3\\
$1513$ & 73.54194 & -3.00785 & 0.1568 & 0.08 & 3\\
$1716$ & 73.54447 & -3.00823 & 0.1573 & 0.26 & 3\\
$1503$ & 73.53747 & -3.01237 & 0.1861 & 0.08 & 3\\
$cont31_2$ & 73.53782 & -3.01343 & 0.1863 & 0.25 & 3\\
$1583$ & 73.53293 & -3.00942 & 0.1866 & 0.08 & 3\\
$cont30_2$ & 73.53317 & -3.01044 & 0.1868 & 0.23 & 3\\
$1685$ & 73.53237 & -3.00902 & 0.1868 & 0.17 & 3\\
$sing41_2$ & 73.53141 & -3.00906 & 0.1872 & 0.28 & 3\\
$1696$ & 73.53548 & -3.00767 & 0.2814 & 0.09 & 3\\
$sing49_2$ & 73.53582 & -3.00868 & 0.2814 & 0.17 & 3\\
$1491$ & 73.55356 & -3.01292 & 0.2953 & 0.11 & 3\\
$cont21_2$ & 73.53517 & -3.00646 & 0.3548 & 0.14 & 3\\
$1722$ & 73.53260 & -3.00783 & 0.3551 & 0.16 & 3\\
$968$ & 73.55311 & -3.02863 & 0.3590 & 0.33 & 2\\
$1375$ & 73.54730 & -3.01645 & 0.4177 & 0.25 & 3\\
$1307$ & 73.54710 & -3.01700 & 0.4182 & 0.13 & 3\\
$1456$ & 73.54355 & -3.01312 & 0.4440 & 0.23 & 3\\
$1485$ & 73.54031 & -3.01211 & 0.4440 & 0.11 & 3\\
$cont25_2$ & 73.54061 & -3.01318 & 0.4441 & 0.45 & 3\\
$1514$ & 73.54678 & -3.01241 & 0.4464 & 0.32 & 3\\
$1348$ & 73.54920 & -3.01592 & 0.4471 & 0.15 & 3\\
$1827$ & 73.54581 & -3.00432 & 0.4906 & 0.94 & 3\\
$1399$ & 73.53449 & -3.01462 & 0.5255 & 1.48 & 3\\
$1313$ & 73.55365 & -3.01691 & 0.5257 & 0.86 & 3\\
$1594$ & 73.53535 & -3.00932 & 0.5270 & 1.04 & 3\\
$1553$ & 73.55235 & -3.00960 & 0.5292 & 1.68 & 3\\
$1301$ & 73.56181 & -3.01744 & 0.5297 & 0.59 & 3\\
$1459$ & 73.54328 & -3.01295 & 0.5304 & 1.38 & 3\\
$1535$ & 73.53647 & -3.01155 & 0.5307 & 0.16 & 3\\
$1065$ & 73.55477 & -3.02293 & 0.5310 & 0.78 & 3\\
$1418$ & 73.53849 & -3.01405 & 0.5310 & 0.98 & 3\\
$1290$ & 73.55906 & -3.01700 & 0.5312 & 0.83 & 3\\
$1610$ & 73.53555 & -3.00867 & 0.5314 & 1.36 & 3\\
$1329$ & 73.54651 & -3.01740 & 0.5317 & 1.32 & 3\\
$1493$ & 73.54958 & -3.01227 & 0.5317 & 1.88 & 3\\
$1557$ & 73.53633 & -3.00996 & 0.5319 & 0.98 & 3\\
$1163$ & 73.55202 & -3.01889 & 0.5319 & 0.51 & 3\\
$1479$ & 73.55567 & -3.01243 & 0.5322 & 0.68 & 3\\
$1401$ & 73.54736 & -3.01534 & 0.5324 & 1.83 & 3\\
$1346$ & 73.54618 & -3.01631 & 0.5330 & 0.1 & 3\\
$1512$ & 73.54220 & -3.01177 & 0.5330 & 0.83 & 3\\
$1377$ & 73.55113 & -3.01519 & 0.5331 & 0.58 & 3\\
$1787$ & 73.53588 & -3.00591 & 0.5332 & 1.62 & 3\\
$1383$ & 73.56202 & -3.01602 & 0.5335 & 1.85 & 3\\
$1332$ & 73.56181 & -3.01693 & 0.5343 & 1.03 & 3\\
$1406$ & 73.53975 & -3.01555 & 0.5346 & 3.25 & 2\\
$1453$ & 73.55038 & -3.01396 & 0.5351 & 1.52 & 3\\
$1409$ & 73.54795 & -3.01543 & 0.5352 & 1.4 & 3\\
$1193$ & 73.54980 & -3.01794 & 0.5352 & 0.63 & 3\\
$1098$ & 73.55824 & -3.02235 & 0.5360 & 0.7 & 3\\
$1435$ & 73.53557 & -3.01403 & 0.5367 & 1.12 & 3\\
$1433$ & 73.54195 & -3.01394 & 0.5367 & 1.54 & 3\\
$1361$ & 73.55141 & -3.01664 & 0.5368 & 2.64 & 3\\
$1126$ & 73.55975 & -3.02196 & 0.5368 & 0.74 & 3\\
$1288$ & 73.54518 & -3.01884 & 0.5369 & 2.33 & 2\\
$1419$ & 73.54950 & -3.01542 & 0.5370 & 2.19 & 2\\
$1673$ & 73.55204 & -3.00876 & 0.5371 & 1.51 & 3\\
$1174$ & 73.56169 & -3.01828 & 0.5373 & 0.62 & 3\\
$1149$ & 73.54620 & -3.02133 & 0.5374 & 1.7 & 2\\
$1590$ & 73.53869 & -3.00978 & 0.5380 & 1.85 & 3\\
$1568$ & 73.53587 & -3.01056 & 0.5384 & 2.26 & 2\\
$1518$ & 73.55385 & -3.01149 & 0.5384 & 1.66 & 3\\
$1452$ & 73.54713 & -3.01382 & 0.5386 & 2.55 & 3\\
$1428$ & 73.54431 & -3.01434 & 0.5389 & 1.3 & 3\\
$1187$ & 73.54519 & -3.01439 & 0.5391 & 0.79 & 3\\
$1121$ & 73.55189 & -3.02135 & 0.5391 & 0.41 & 3\\
$1297$ & 73.55075 & -3.01699 & 0.5391 & 0.62 & 3\\
$1425$ & 73.54261 & -3.01485 & 0.5392 & 1.3 & 3\\
$1364$ & 73.54588 & -3.01624 & 0.5393 & 3.08 & 3\\
$1429$ & 73.54853 & -3.01480 & 0.5398 & 1.89 & 3\\
$1439$ & 73.55738 & -3.01423 & 0.5406 & 3.04 & 2\\
$1785$ & 73.54843 & -3.00590 & 0.5406 & 1.12 & 3\\
$1592$ & 73.54039 & -3.00990 & 0.5410 & 2.01 & 3\\
$1295$ & 73.54714 & -3.01843 & 0.5410 & 1.66 & 3\\
$1180$ & 73.55950 & -3.01988 & 0.5412 & 3.07 & 3\\
$1475$ & 73.54966 & -3.01315 & 0.5415 & 2.6 & 3\\
$1352$ & 73.55261 & -3.01689 & 0.5418 & 2.34 & 3\\
$1840$ & 73.54255 & -3.00455 & 0.5418 & 1.54 & 3\\
$1499$ & 73.54571 & -3.01239 & 0.5422 & 1.31 & 3\\
$1870$ & 73.54051 & -3.00389 & 0.5424 & 1.15 & 2\\
$1088$ & 73.55472 & -3.02325 & 0.5426 & 1.74 & 3\\
$1359$ & 73.55804 & -3.01641 & 0.5430 & 2.05 & 2\\
$1344$ & 73.55874 & -3.01645 & 0.5431 & 0.93 & 3\\
$1386$ & 73.53622 & -3.01516 & 0.5436 & 1.21 & 3\\
$1405$ & 73.55115 & -3.01468 & 0.5439 & 0.73 & 3\\
$1403$ & 73.55153 & -3.01552 & 0.5441 & 3.58 & 3\\
$1465$ & 73.54709 & -3.01342 & 0.5441 & 2.38 & 3\\
$1597$ & 73.54262 & -3.00865 & 0.5442 & 0.15 & 3\\
$1402$ & 73.55286 & -3.01521 & 0.5444 & 0.12 & 3\\
$1152$ & 73.55135 & -3.02103 & 0.5449 & 1.79 & 3\\
$1704$ & 73.53858 & -3.00841 & 0.5450 & 0.42 & 3\\
$1416$ & 73.54414 & -3.01535 & 0.5450 & 0.59 & 3\\
$1606$ & 73.53937 & -3.00873 & 0.5451 & 0.88 & 3\\
$1807$ & 73.54054 & -3.00432 & 0.5451 & 1.56 & 3\\
$1380$ & 73.55768 & -3.01575 & 0.5452 & 0.29 & 3\\
$1281$ & 73.55263 & -3.01781 & 0.5453 & 0.92 & 3\\
$1574$ & 73.55003 & -3.01028 & 0.5453 & 0.7 & 3\\
$1964$ & 73.54366 & -3.00105 & 0.5456 & 2.45 & 3\\
$1423$ & 73.54706 & -3.01473 & 0.5456 & 0.92 & 3\\
$1511$ & 73.54173 & -3.01164 & 0.5457 & 3.19 & 3\\
$1020$ & 73.55282 & -3.02561 & 0.5463 & 0.83 & 3\\
$1661$ & 73.53979 & -3.00974 & 0.5463 & 2.37 & 2\\
$1795$ & 73.54411 & -3.00601 & 0.5465 & 2.16 & 3\\
$1294$ & 73.54004 & -3.01687 & 0.5477 & 0.6 & 3\\
$1504$ & 73.53470 & -3.01182 & 0.5484 & 0.72 & 2\\
$1517$ & 73.54986 & -3.01139 & 0.5487 & 2.71 & 3\\
$1404$ & 73.54378 & -3.01547 & 0.5488 & 0.17 & 3\\
$1356$ & 73.54927 & -3.01659 & 0.5520 & 1.04 & 3\\
$1534$ & 73.53453 & -3.01131 & 0.5540 & 2.09 & 3\\
$1571$ & 73.54946 & -3.01065 & 0.5543 & 1.39 & 3\\
$1523$ & 73.55298 & -3.01178 & 0.5551 & 1.59 & 3\\
$1151$ & 73.55249 & -3.02087 & 0.5649 & 0.11 & 3\\
$1168$ & 73.55403 & -3.02082 & 0.5652 & 0.13 & 3\\
$sing44_2$ & 73.54777 & -3.01296 & 0.5789 & 0.15 & 2\\
$1070$ & 73.54593 & -3.02394 & 0.5893 & 0.1 & 3\\
$910$ & 73.55169 & -3.02980 & 0.5898 & 0.16 & 3\\
$sing15_2$ & 73.53104 & -3.00940 & 0.6196 & 0.31 & 3\\
$1190$ & 73.54370 & -3.01997 & 0.6471 & 0.13 & 3\\
$1158$ & 73.54088 & -3.02108 & 0.6542 & 0.3 & 2\\
$1078$ & 73.54804 & -3.02422 & 0.6650 & 0.12 & 3\\
$sing48_2$ & 73.54825 & -3.00572 & 0.6784 & 0.2 & 2\\
$1871$ & 73.54791 & -3.00464 & 0.6785 & 0.08 & 2\\
$1869$ & 73.53563 & -3.00445 & 0.7942 & 0.15 & 3\\
$1502$ & 73.54016 & -3.01108 & 0.9164 & 0.24 & 2\\
$1567$ & 73.54055 & -3.01105 & 0.9164 & 0.4 & 3\\
$1031$ & 73.54826 & -3.02564 & 0.9205 & 0.73 & 3\\
$1076$ & 73.54913 & -3.02424 & 0.9211 & 1.55 & 2\\
$1162$ & 73.54427 & -3.02071 & 0.9211 & 0.26 & 2\\
$1521$ & 73.53300 & -3.01178 & 0.9215 & 0.29 & 2\\
$1118$ & 73.54839 & -3.02263 & 0.9215 & 0.23 & 2\\
$1421$ & 73.55546 & -3.01548 & 1.1610 & 0.21 & 2\\
$1082$ & 73.55141 & -3.02366 & 1.1801 & 0.31 & 3\\
$1184$ & 73.55631 & -3.02025 & 1.2116 & 0.13 & 3\\
$1029$ & 73.55455 & -3.02617 & 1.3103 & 0.2 & 2\\
$1074$ & 73.55860 & -3.02379 & 1.3970 & 0.68 & 2\\
$1390$ & 73.55696 & -3.01620 & 2.6306 & 0.59 & 3\\
$1084$ & 73.54630 & -3.02401 & 2.9100 & 3.0 & 2\\
$1436$ & 73.55396 & -3.01479 & 2.9101 & 1.95 & 3\\
$1415$ & 73.55395 & -3.01521 & 2.9135 & 1.72 & 3\\
$1396$ & 73.55387 & -3.01593 & 2.9140 & 3.09 & 3\\
$1408$ & 73.55391 & -3.01570 & 2.9169 & 6.99 & 3\\
$1139$ & 73.55270 & -3.02121 & 2.9204 & 1.48 & 3\\
$1120$ & 73.55069 & -3.02258 & 2.9232 & 6.3 & 3\\
$sing1_2$ & 73.54602 & -3.01140 & 2.9237 & 0.84 & 3\\
$1525$ & 73.55592 & -3.01190 & 2.9247 & 8.1 & 3\\
$sing11_2$ & 73.53656 & -3.00472 & 3.6005 & 0.6 & 2\\
$sing4$ & 73.54980 & -3.02711 & 3.6540 & 0.29 & 2\\
\textbf{$cont29_2$} & 73.53618 & -3.01331 & 3.6950 & 0.80 & 3 \\
$sing12_2$ & 73.53632 & -3.01233 & 3.7644 & 0.68 & 3\\
\textbf{$R.3$} & 73.54229 & -3.01263 & 3.7670 & 0.84 & 2 \\ 
$sing6$ & 73.54722 & -3.01284 & 4.4513 & 0.53 & 3\\
\textbf{$S.2$} & 73.54627 & -3.01263 & 4.4540 & 1.21 & 3 \\ 
$sing18$ & 73.55104 & -3.02616 & 4.8483 & 0.23 & 2\\
\hline
\end{supertabular}
\end{center}

\newpage
\begin{table}
\begin{tabular}{cc}
\, & \,\\
\end{tabular}
\caption{List of measured redshifts in MACS\,J2129. The column are the same as in Table\,\ref{rxj2129_z}.}
\label{macs2129_z}
\end{table}
\scriptsize
\begin{center}
\tablefirsthead{\hline
                                \multicolumn{1}{c}{ID} &
                                \multicolumn{1}{c}{R.\,A.} &
                                \multicolumn{1}{c}{Decl.} &
                                \multicolumn{1}{c}{$z$} &
                                \multicolumn{1}{c}{$z_{\rm err}$ [$\times 10^{-4}$]} &
                                \multicolumn{1}{c}{QF}
                                \\ \hline  }
\tablehead{\hline \multicolumn{6}{l}{\small\sl continued from previous page}\\
           \hline \multicolumn{1}{c}{ID} &
                                \multicolumn{1}{c}{R.\,A.} &
                                \multicolumn{1}{c}{Decl.} &
                                \multicolumn{1}{c}{$z$} &
                                \multicolumn{1}{c}{$z_{\rm err}$ [$\times 10^{-4}$]} &
                                \multicolumn{1}{c}{QF}
                                \\ \hline  }
\tabletail{\hline\multicolumn{6}{r}{\small\sl continued on next page}\\\hline}
\tablelasttail{\hline}
\par
\begin{supertabular}{cccccc}
$1387$ & 322.35269 & -7.69278 & -0.0001 & 0.64 & 3\\
$1706$ & 322.37128 & -7.68706 & -0.0002 & 1.05 & 3\\
$1763$ & 322.34305 & -7.68218 & -0.0005 & 0.08 & 3\\
$1335$ & 322.35529 & -7.69425 & -0.0007 & 0.52 & 3\\
$1691$ & 322.35846 & -7.68730 & -0.0011 & 0.18 & 3\\
$1263$ & 322.36911 & -7.69526 & 0.0000 & 0.18 & 3\\
$1454$ & 322.35287 & -7.69254 & 0.0000 & 1.90 & 3\\
$1605$ & 322.35022 & -7.68940 & 0.0000 & 0.99 & 2\\
$1570$ & 322.35553 & -7.69030 & 0.0001 & 0.44 & 3\\
$1148$ & 322.36786 & -7.69608 & 0.0002 & 0.29 & 3\\
$1071$ & 322.36948 & -7.69470 & 0.0002 & 0.09 & 3\\
$1645$ & 322.37402 & -7.68854 & 0.0008 & 0.91 & 3\\
$1927$ & 322.35565 & -7.68223 & 0.0013 & 0.73 & 3\\
$1438$ & 322.36545 & -7.68853 & 0.0016 & 0.11 & 3\\
$1918$ & 322.35388 & -7.68261 & 0.2149 & 0.10 & 3\\
$1924$ & 322.34756 & -7.68232 & 0.2661 & 0.11 & 3\\
$1193$ & 322.35391 & -7.69657 & 0.2971 & 0.11 & 3\\
$1409$ & 322.36011 & -7.69323 & 0.2992 & 0.4 & 2\\
$1773$ & 322.36298 & -7.68628 & 0.3101 & 0.12 & 3\\
$1195$ & 322.37247 & -7.69483 & 0.3114 & 0.08 & 3\\
$1463$ & 322.36722 & -7.69237 & 0.3510 & 0.16 & 3\\
$1183$ & 322.36951 & -7.69665 & 0.4150 & 3.63 & 2\\
$1750$ & 322.34833 & -7.68313 & 0.4170 & 0.15 & 3\\
$1544$ & 322.34427 & -7.69091 & 0.4313 & 0.65 & 3\\
$1789$ & 322.35919 & -7.68599 & 0.4330 & 0.69 & 2\\
$1580$ & 322.36270 & -7.69031 & 0.4362 & 0.18 & 3\\
$1457$ & 322.35794 & -7.69255 & 0.4369 & 0.39 & 3\\
$1349$ & 322.36258 & -7.69408 & 0.4371 & 0.18 & 3\\
$1552$ & 322.36386 & -7.68784 & 0.4371 & 0.13 & 3\\
$1838$ & 322.36682 & -7.68462 & 0.4373 & 0.11 & 3\\
$1146$ & 322.37332 & -7.69744 & 0.4464 & 3.55 & 2\\
$1626$ & 322.37378 & -7.68912 & 0.4579 & 0.13 & 3\\
$1600$ & 322.37375 & -7.68974 & 0.4579 & 0.13 & 3\\
$1049$ & 322.36740 &  -7.69949 & 0.4669 & 0.14 & 3\\
$1274$ & 322.36349 & -7.69507 & 0.4728 & 1.08 & 2\\
$1258$ & 322.35397 & -7.69569 & 0.5297 & 0.12 & 3\\
$1669$ & 322.34607 & -7.68804 & 0.5326 & 0.1 & 3\\
$1351$ & 322.36176 & -7.69399 & 0.5510 & 0.11 & 3\\
$1698$ & 322.34726 & -7.68733 & 0.5523 & 0.11 & 3\\
$1846$ & 322.35754 & -7.68381 & 0.5720 & 0.28 & 3\\
$1537$ & 322.37415 & -7.69017 & 0.5736 & 2.41 & 2\\
$1415$ & 322.37039 & -7.69209 & 0.5737 & 1.26 & 3\\
$1868$ & 322.35773 & -7.68334 & 0.5737 & 0.15 & 3\\
$1768$ & 322.35776 & -7.68621 & 0.5740 & 2.76 & 2\\
$1938$ & 322.35876 & -7.68250 & 0.5746 & 0.52 & 3\\
$1197$ & 322.36127 & -7.69643 & 0.5759 & 1.37 & 3\\
$1589$ & 322.36716 & -7.68846 & 0.5762 & 0.15 & 3\\
$1662$ & 322.35352 & -7.68804 & 0.5782 & 1.35 & 3\\
$1642$ & 322.35919 & -7.68868 & 0.5784 & 3.21 & 2\\
$1478$ & 322.36310 & -7.69129 & 0.5789 & 0.92 & 3\\
$1612$ & 322.35629 & -7.68902 & 0.5803 & 1.8 & 3\\
$1603$ & 322.35623 & -7.68937 & 0.5806 & 2.99 & 2\\
$1549$ & 322.34891 & -7.69023 & 0.5811 & 1.95 & 3\\
$1215$ & 322.36050 & -7.69639 & 0.5814 & 0.56 & 2\\
$1548$ & 322.35751 & -7.69071 & 0.5830 & 1.11 & 2\\
$1496$ & 322.35373 & -7.69140 & 0.5830 & 4.16 & 2\\
$1693$ & 322.35718 & -7.68687 & 0.5831 & 0.93 & 3\\
$1128$ & 322.36618 & -7.69722 & 0.5832 & 1.34 & 3\\
$1807$ & 322.37036 & -7.68483 & 0.5833 & 1.9 & 3\\
$1806$ & 322.35995 & -7.68527 & 0.5834 & 0.77 & 3\\
$1458$ & 322.34671 & -7.69157 & 0.5834 & 1.17 & 3\\
$1485$ & 322.36526 & -7.69087 & 0.5835 & 1.41 & 3\\
$1599$ & 322.35712 & -7.68909 & 0.5835 & 1.75 & 3\\
$1822$ & 322.36389 & -7.68384 & 0.5836 & 1.08 & 3\\
$1267$ & 322.36090 & -7.69336 & 0.5842 & 0.86 & 3\\
$1738$ & 322.37073 & -7.68601 & 0.5843 & 1.79 & 3\\
$1777$ & 322.36072 & -7.68576 & 0.5846 & 1.71 & 3\\
$1364$ & 322.36838 & -7.69373 & 0.5849 & 6.2 & 2\\
$sing17$ & 322.34783 & -7.68908 & 0.5853 & 0.51 & 3\\
$1720$ & 322.34698 & -7.68537 & 0.5855 & 0.53 & 3\\
$1735$ & 322.35837 & -7.68670 & 0.5857 & 2.85 & 3\\
$1347$ & 322.36859 & -7.69387 & 0.5858 & 0.96 & 3\\
$1559$ & 322.35193 & -7.68910 & 0.5858 & 0.57 & 3\\
$1567$ & 322.36057 & -7.68923 & 0.5860 & 1.28 & 3\\
$1787$ & 322.35110 & -7.68505 & 0.5860 & 0.76 & 3\\
$1817$ & 322.34610 & -7.68450 & 0.5860 & 0.86 & 3\\
$1765$ & 322.35818 & -7.68646 & 0.5860 & 1.41 & 2\\
$1276$ & 322.35010 & -7.69443 & 0.5864 & 0.94 & 3\\
$1425$ & 322.35574 & -7.69189 & 0.5867 & 0.96 & 3\\
$1522$ & 322.36194 & -7.69083 & 0.5873 & 0.58 & 3\\
$1452$ & 322.36026 & -7.69195 & 0.5874 & 1.35 & 3\\
$1352$ & 322.34616 & -7.69360 & 0.5875 & 1.49 & 3\\
$1430$ & 322.35880 & -7.69105 & 0.5881 & 0.8 & 3\\
$1508$ & 322.36923 & -7.69123 & 0.5881 & 2.35 & 2\\
$1433$ & 322.34970 & -7.69246 & 0.5881 & 2.57 & 3\\
$1781$ & 322.37024 & -7.68588 & 0.5887 & 1.10 & 2\\
$1524$ & 322.34262 & -7.68952 & 0.5888 & 0.77 & 3\\
$1624$ & 322.35852 & -7.68911 & 0.5889 & 2.24 & 3\\
$1301$ & 322.34546 & -7.69476 & 0.5889 & 4.12 & 3\\
$1484$ & 322.35962 & -7.69079 & 0.5891 & 0.58 & 3\\
$1418$ & 322.36261 & -7.69238 & 0.5893 & 1.27 & 3\\
$1515$ & 322.35941 & -7.69050 & 0.5896 & 2.82 & 3\\
$1397$ & 322.35675 & -7.69250 & 0.5897 & 1.6 & 3\\
$1747$ & 322.34860 & -7.68656 & 0.5898 & 1.11 & 2\\
$1348$ & 322.35495 & -7.6918 & 0.5900 & 0.83 & 3\\
$1199$ & 322.35843 & -7.69649 & 0.5901 & 0.22 & 3\\
$1718$ & 322.35464 & -7.68643 & 0.5902 & 0.56 & 3\\
$1828$ & 322.35291 & -7.68506 & 0.5904 & 1.51 & 3\\
$1110$ & 322.37140 & -7.69753 & 0.5905 & 0.95 & 3\\
$1309$ & 322.36017 & -7.69409 & 0.5906 & 1.77 & 3\\
$1762$ & 322.35156 & -7.68638 & 0.5910 & 0.58 & 2\\
$1136$ & 322.35794 & -7.69713 & 0.5913 & 0.96 & 3\\
$1573$ & 322.35974 & -7.69008 & 0.5917 & 5.0 & 2\\
$1654$ & 322.36929 & -7.68824 & 0.5919 & 1.79 & 3\\
$1610$ & 322.35474 & -7.68899 & 0.5925 & 0.99 & 3\\
$1103$ & 322.35095 & -7.69715 & 0.5930 & 0.78 & 3\\
$1165$ & 322.35486 & -7.69697 & 0.5931 & 0.67 & 3\\
$1346$ & 322.34598 & -7.69394 & 0.5951 & 1.96 & 2\\
$1834$ & 322.36456 & -7.68481 & 0.5957 & 1.29 & 3\\
$1631$ & 322.37442 & -7.68875 & 0.5959 & 1.28 & 3\\
$1790$ & 322.35806 & -7.68574 & 0.5965 & 0.49 & 2\\
$1360$ & 322.35126 & -7.69368 & 0.5980 & 3.01 & 3\\
$1543$ & 322.34665 & -7.69053 & 0.5980 & 0.17 & 3\\
$1709$ & 322.34641 & -7.68711 & 0.5983 & 0.14 & 3\\
$1417$ & 322.34802 & -7.69209 & 0.5983 & 0.1 & 3\\
$1601$ & 322.35968 & -7.68939 & 0.5984 & 1.61 & 3\\
$1361$ & 322.36017 & -7.69370 & 0.5994 & 1.48 & 2\\
$1473$ & 322.35751 & -7.69135 & 0.5997 & 0.95 & 3\\
$1290$ & 322.35123 & -7.69514 & 0.6002 & 0.40 & 2\\
$1115$ & 322.37045 & -7.69776 & 0.6004 & 1.20 & 2\\
$1476$ & 322.36533 & -7.69156 & 0.6010 & 0.87 & 2\\
$1337$ & 322.34451 & -7.69362 & 0.6010 & 1.63 & 3\\
$1359$ & 322.36926 & -7.69367 & 0.6012 & 3.3 & 2\\
$1681$ & 322.35608 & -7.68804 & 0.6020 & 1.62 & 2\\
$1759$ & 322.35101 & -7.68596 & 0.6024 & 0.83 & 3\\
$1770$ & 322.34607 & -7.68621 & 0.6032 & 2.3 & 2\\
$1656$ & 322.35663 & -7.68760 & 0.6047 & 0.91 & 3\\
$1541$ & 322.35672 & -7.69006 & 0.6085 & 0.51 & 3\\
$1697$ & 322.36032 & -7.68779 & 0.7251 & 0.25 & 3\\
$1592$ & 322.34378 & -7.68991 & 0.7252 & 2.08 & 2\\
$sing53$ & 322.34876 & -7.69481 & 0.7344 & 0.18 & 2\\
$1837$ & 322.34564 & -7.68508 & 0.7858 & 0.17 & 3\\
$1657$ & 322.36642 & -7.68825 & 0.8716 & 0.4 & 3\\
$1401$ & 322.36804 & -7.69306 & 0.8721 & 0.21 & 3\\
$1131$ & 322.34683 & -7.69803 & 0.9426 & 1.00 & 3\\
$1122$ & 322.36899 & -7.69783 & 0.9431 & 1.45 & 2\\
$1799$ & 322.36133 & -7.68589 & 0.9432 & 1.46 & 2\\
$1159$ & 322.36166 & -7.69632 & 0.9476 & 0.26 & 3\\
$1363$ & 322.36679 & -7.69373 & 0.9540 & 0.34 & 3\\
$1488$ & 322.35486 & -7.69069 & 1.0480 & 0.21 & 3\\
$1377$ & 322.35538 & -7.69329 & 1.0483 & 0.16 & 3\\
$1306$ & 322.34613 & -7.69495 & 1.0507 & 0.47 & 3\\
$1795$ & 322.36032 & -7.68577 & 1.0690 & 3.09 & 2\\
$1880$ & 322.34387 & -7.68363 & 1.0864 & 0.84 & 3\\
$1943$ & 322.34561 & -7.68235 & 1.0871 & 1.98 & 2\\
$1102$ & 322.36435 & -7.69865 & 1.1954 & 1.16 & 3\\
$1835$ & 322.36703 & -7.68440 & 1.1960 & 0.21 & 3\\
$1879$ & 322.35132 & -7.68351 & 1.2022 & 0.64 & 2\\
$1296$ & 322.37250 & -7.69428 & 1.2138 & 0.27 & 3\\
$1078$ & 322.36398 & -7.69898 & 1.2233 & 0.31 & 3\\
$1809$ & 322.34601 & -7.68510 & 1.2375 & 0.19 & 3\\
$1771$ & 322.37024 & -7.68637 & 1.2486 & 0.45 & 2\\
$1307$ & 322.34521 & -7.69515 & 1.2748 & 0.6 & 2\\
$1882$ & 322.36148 & -7.68340 & 1.3556 & 0.37 & 3\\
$1342$ & 322.35718 & -7.69431 & 1.3569 & 0.2 & 3\\
$sing61$ & 322.35620 & -7.69170 & 1.3570 & 0.94 & 2\\
$1820$ & 322.35672 & -7.68557 & 1.3571 & 1.61 & 3\\
\textbf{$13.2$} & 322.35391 & -7.68758 & 1.3582 & 0.52 & 2 \\
$cont43$ & 322.35330 & -7.69113 & 1.3585 & 0.54 & 2\\
\textbf{$13.3$} & 322.35443 & -7.69441 & 1.3600 & 2.0 & 2  \\
$1948$ & 322.35086 & -7.68188 & 1.3591 & 0.24 & 3\\
$1302$ & 322.35861 & -7.69489 & 1.3617 & 0.52 & 2\\
$1474$ & 322.35712 & -7.69109 & 1.3619 & 0.63 & 3\\
$1521$ & 322.35925 & -7.69095 & 1.3620 & 2.83 & 3\\
$1782$ & 322.35797 & -7.68589 & 1.3628 & 1.02 & 3\\
$1863$ & 322.36484 & -7.68367 & 1.3683 & 0.49 & 2\\
\textbf{$14.3$} & 322.36259 & -7.69360 & 1.4518 & 1.21 & 2 \\
\textbf{$14.1$} & 322.36131 & -7.68590 & 1.4520 & 5.10 & 1 \\
$1426$ & 322.36823 & -7.69250 & 1.5447 & 0.77 & 3\\
$1780$ & 322.34647 & -7.68590 & 1.8674 & 1.47 & 3\\
$1221$ & 322.34933 & -7.69591 & 2.1497 & 1.29 & 3\\
$1563$ & 322.35022 & -7.68887 & 2.2427 & 4.35 & 3\\
$1424$ & 322.34818 & -7.69300 & 2.3895 & 3.11 & 2\\
$1137$ & 322.36508 & -7.69731 & 2.5339 & 3.14 & 2\\
$1269$ & 322.35245 & -7.69558 & 2.9089 & 1.29 & 2\\
$1126$ & 322.34351 & -7.69795 & 2.9640 & 0.64 & 2\\
$1140$ & 322.34363 & -7.69778 & 2.9647 & 0.22 & 2\\
$1272$ & 322.34506 & -7.69561 & 2.9686 & 1.12 & 2\\
$sing5$ & 322.36334 & -7.69707 & 3.1081 & 0.43 & 3\\
$sing6$ & 322.36491 & -7.69010 & 3.1082 & 0.78 & 3\\
\textbf{$11.3$} & 322.36167 & -7.68362 & 3.1082 & 1.02 & 2 \\
$1729$ & 322.37363 & -7.68699 & 3.2427 & 1.97 & 2\\
$sing7$ & 322.37405 & -7.68731 & 3.4518 & 0.28 & 2\\
$1674$ & 322.35278 & -7.68841 & 3.8272 & 0.13 & 2\\
$1836$ & 322.35455 & -7.68518 & 3.8977 & 0.68 & 2\\
$sing10$ & 322.367420 & -7.69659 & 4.1518 & 0.18 & 2\\
$1739$ & 322.37216 & -7.68699 & 4.3834 & 0.09 & 2\\
$cont30$ & 322.35419 & -7.68876 & 4.4056 & 0.08 & 3\\
\textbf{$10.3$} & 322.35499 & -7.68896 & 4.4076 & 0.25 & 3 \\
$1842$ & 322.35861 & -7.68491 & 4.4085 & 0.1 & 2\\
$1703$ & 322.36163 & -7.68806 & 4.4087 & 0.61 & 3\\
$1625$ & 322.35699 & -7.68924 & 4.4092 & 0.62 & 2\\
$sing23$ & 322.35760 & -7.68471 & 4.4093 & 0.15 & 2\\
$1742$ & 322.34583 & -7.68704 & 4.4800 & 0.31 & 2\\
$sing15$ & 322.37130 & -7.69542 & 4.9189 & 0.21 & 2\\
\hline
\end{supertabular}
\end{center}

\section{Multiple image systems with MUSE spectroscopy}
\label{A_A}
\begin{figure*}
\begin{center}
        \includegraphics[width=0.24\textwidth]{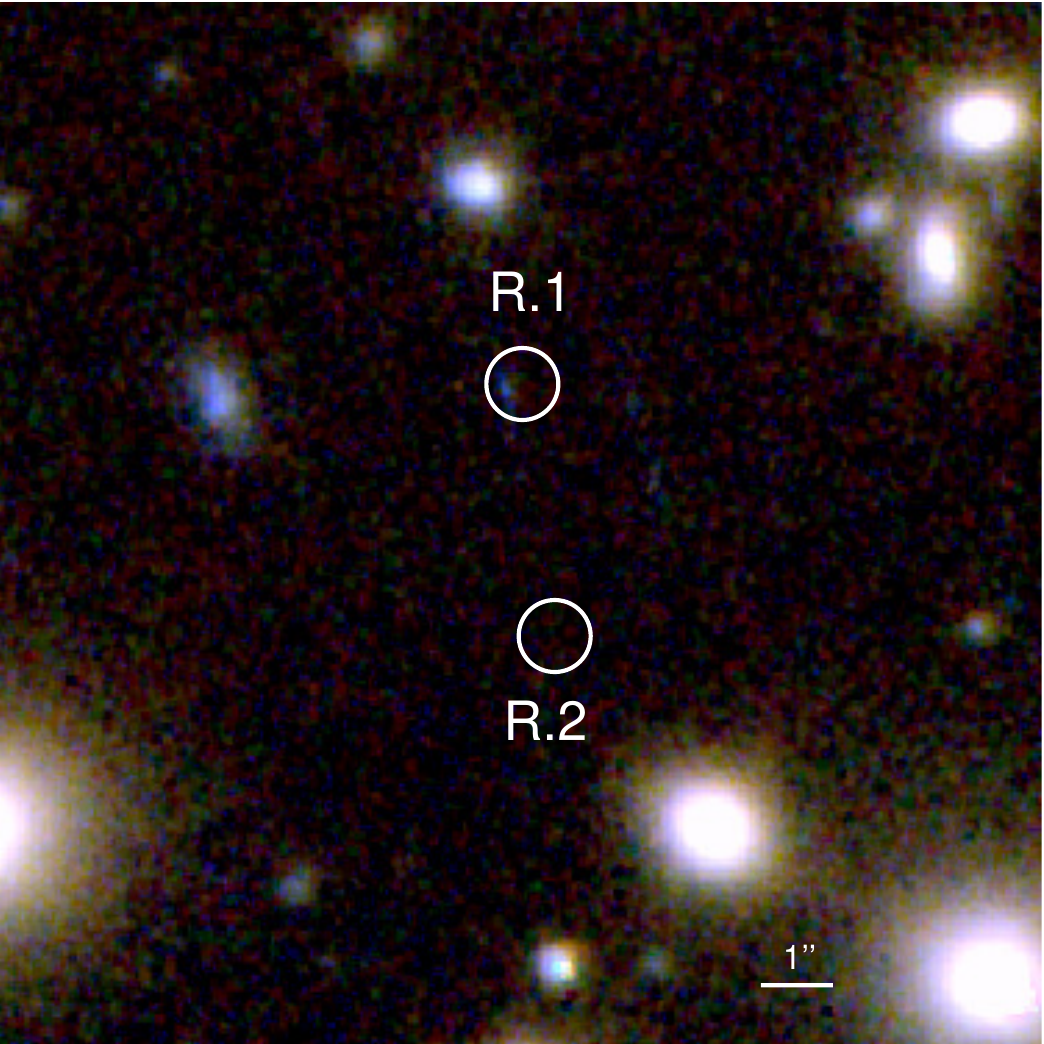}\ 
         \includegraphics[width=0.24\textwidth]{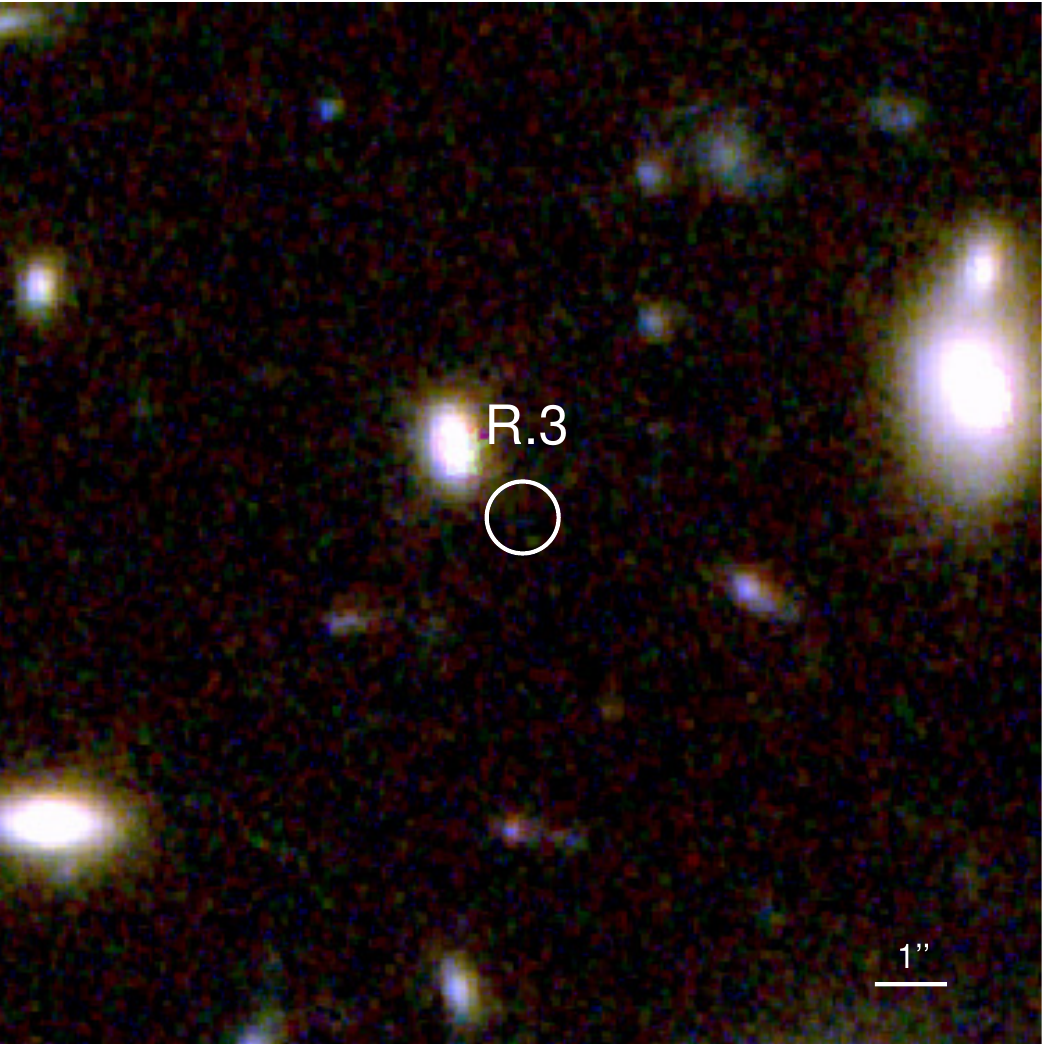}\ 
         \includegraphics[width=0.48\textwidth]{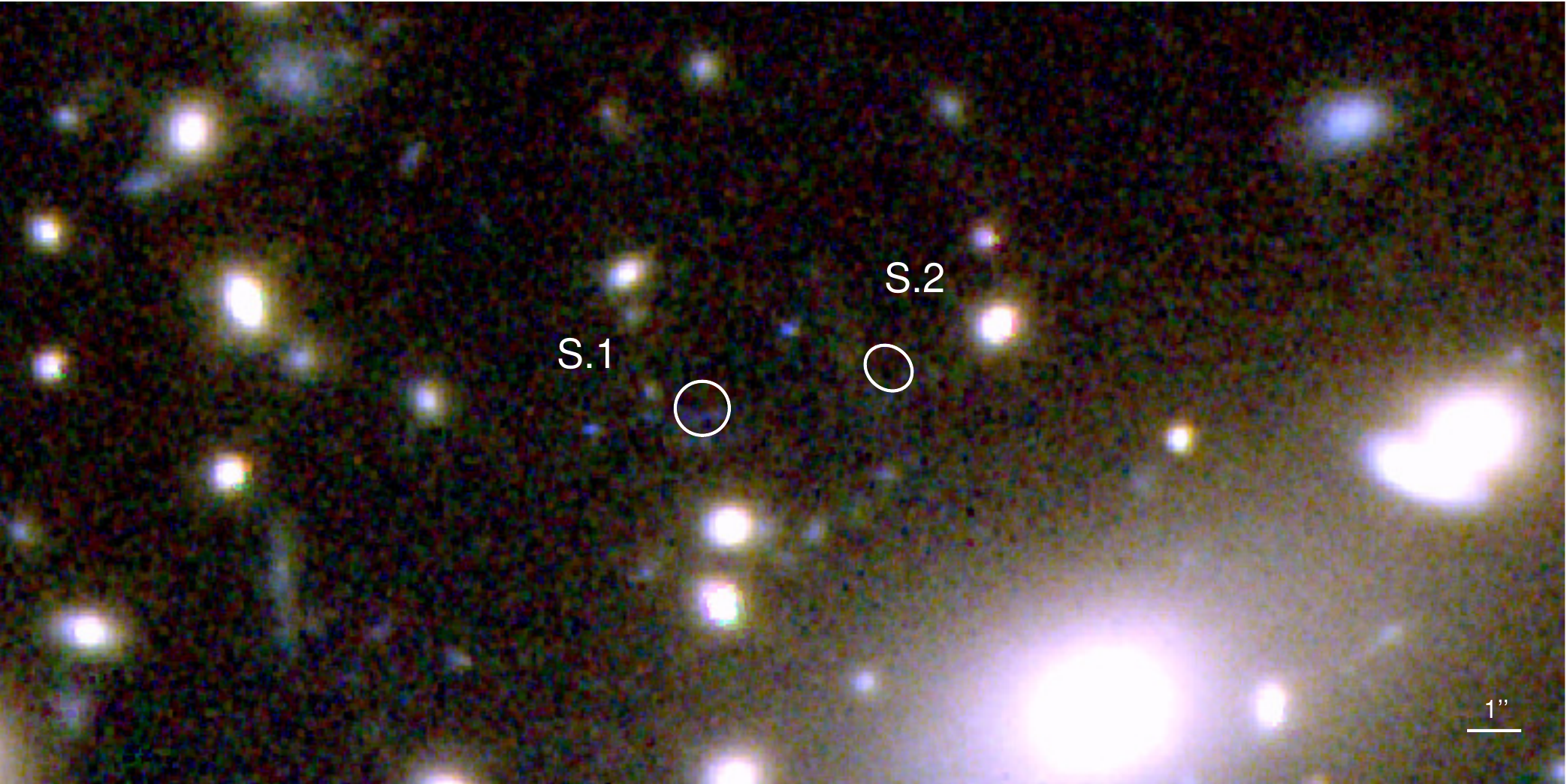}\\
        \includegraphics[width=0.24\textwidth]{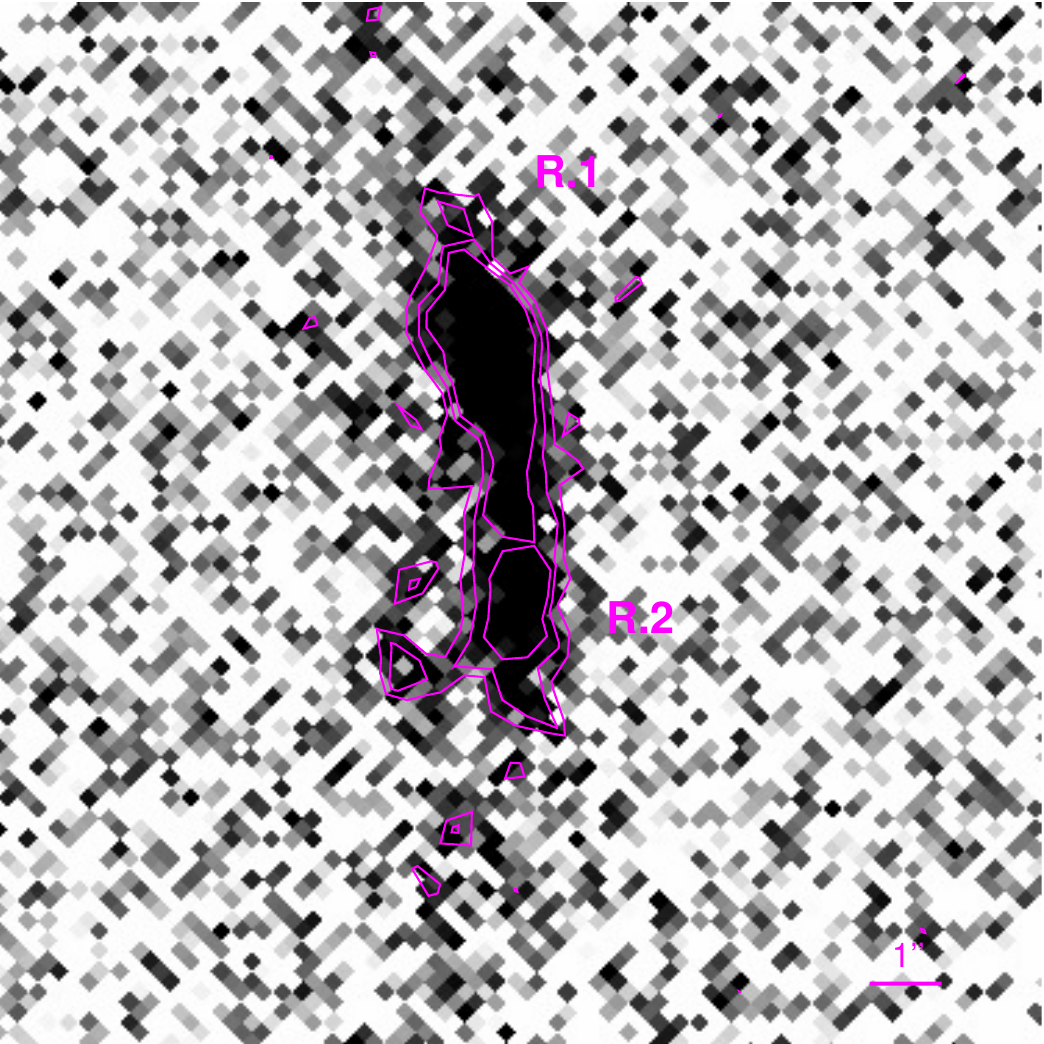}\ 
         \includegraphics[width=0.24\textwidth]{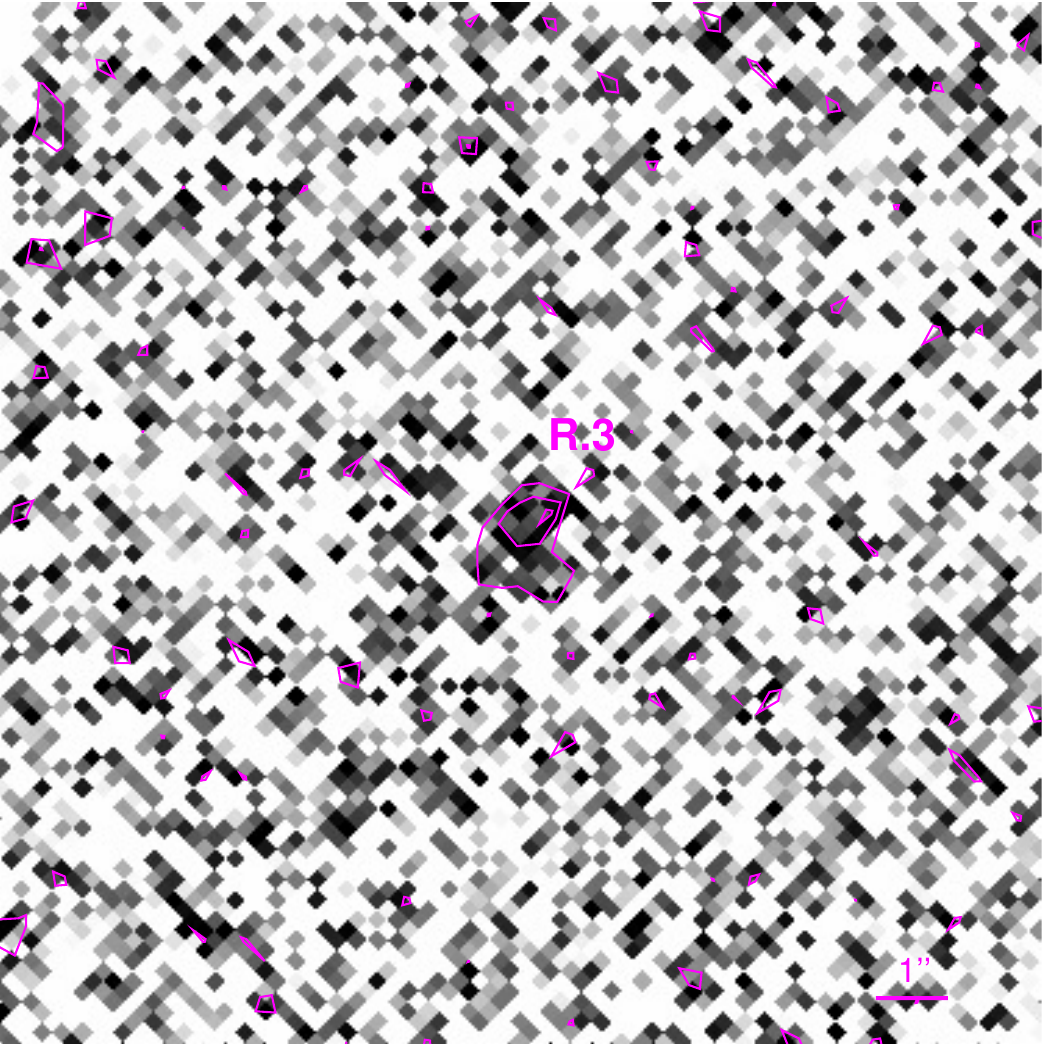}\ 
         \includegraphics[width=0.48\textwidth]{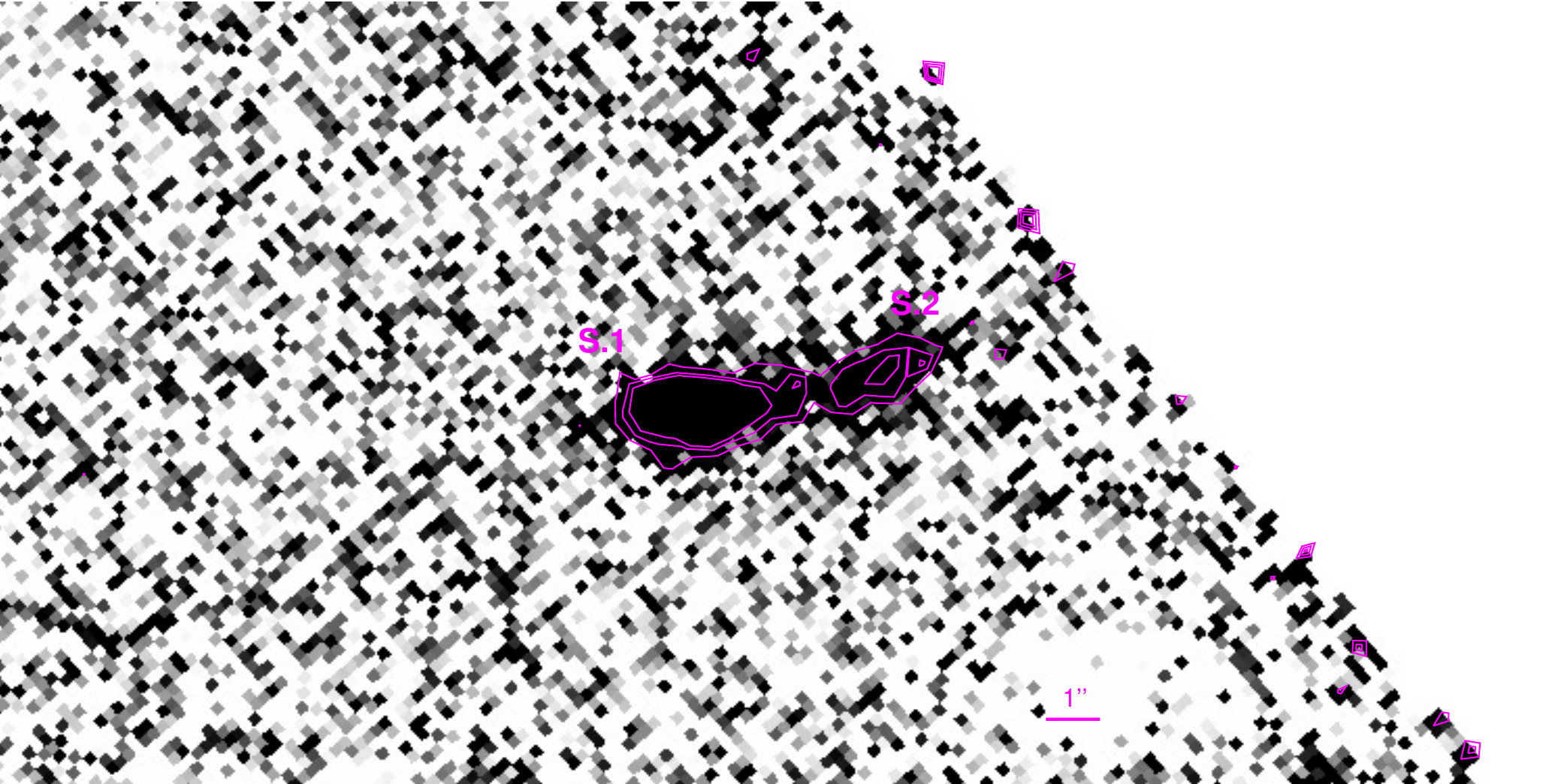}\\
        \includegraphics[width=0.5\textwidth]{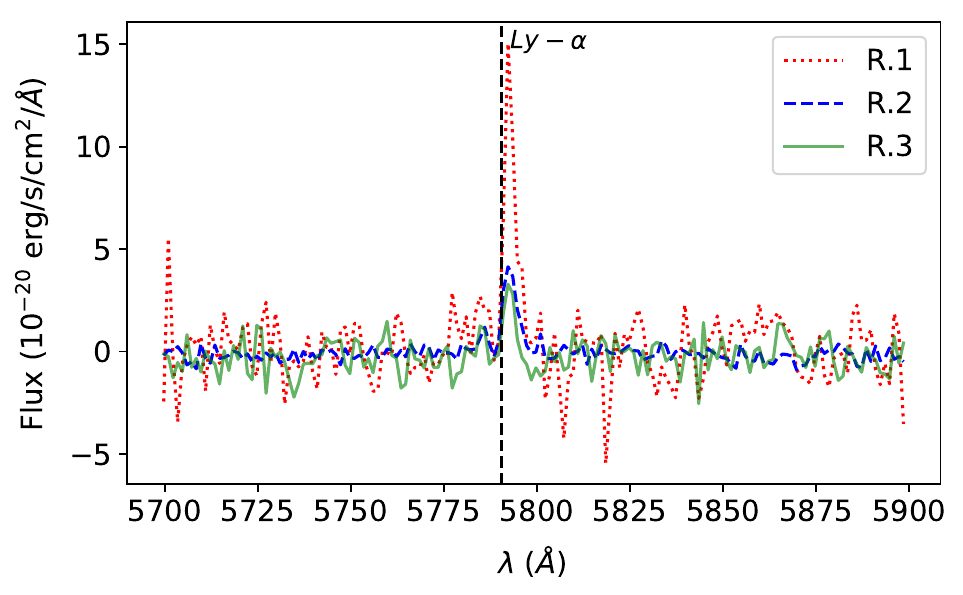}\ 
         \includegraphics[width=0.49\textwidth]{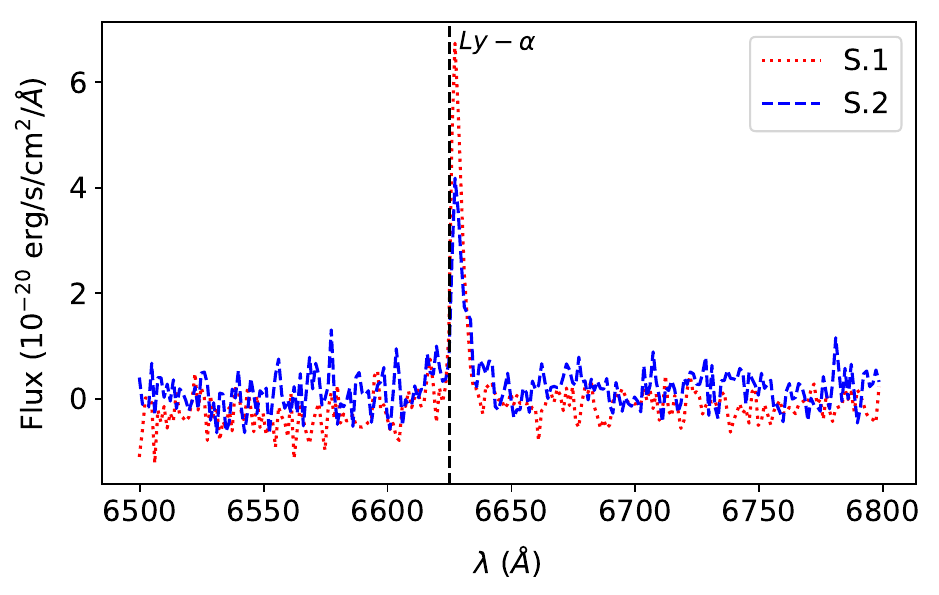}
\caption{\textbf{MS\,0451 --} Newly identified multiple image systems with MUSE. \emph{Top:} Systems R (left, $z = 3.7645$) and S (right, $z = 4.4514$) on a composite colour \emph{HST} image. 
\emph{Middle:} \textsc{muselet} narrow-band datacube at the wavelength corresponding to the maximum emission of the source, i.e. $\lambda = 5793$\,{\AA} for System\ R (left) and  $\lambda=6628$\,{\AA} for System\ S (right). Magenta contours are computed at intensities of 2.0, 3.0, and 4.0$\times10^{-20}$\,erg/s/cm$^{2}$ for both systems, except for image R.3, where the contours are computed for an intensity of 1.0$\times10^{-20}$\,erg/s/arcsec$^{2}$.
\emph{Bottom:} MUSE spectra of the multiple images centered on the most prominent line.}
\label{detail_spec_RS}
\end{center}
\end{figure*}

\begin{figure*}
\begin{center}
        \includegraphics[width=0.24\textwidth]{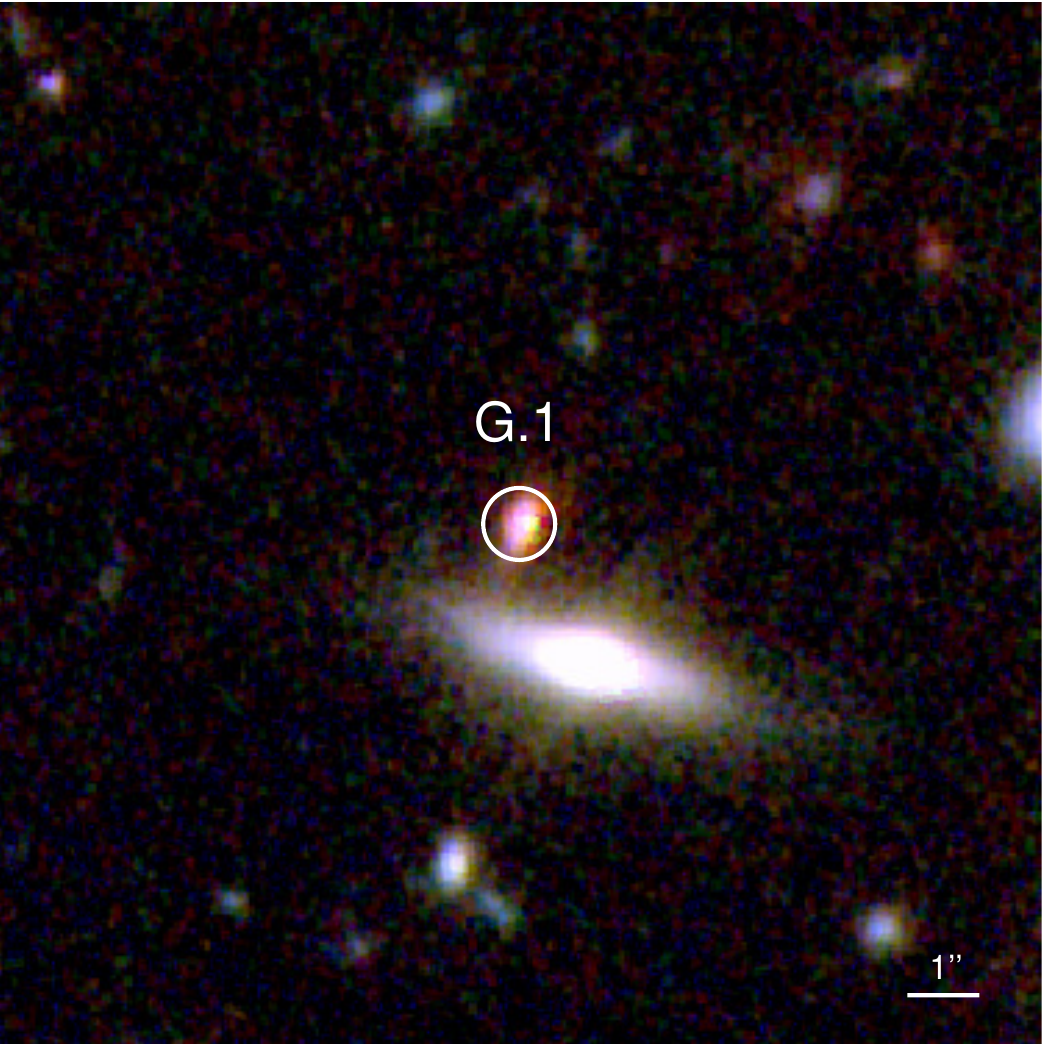}\ 
        \includegraphics[width=0.24\textwidth]{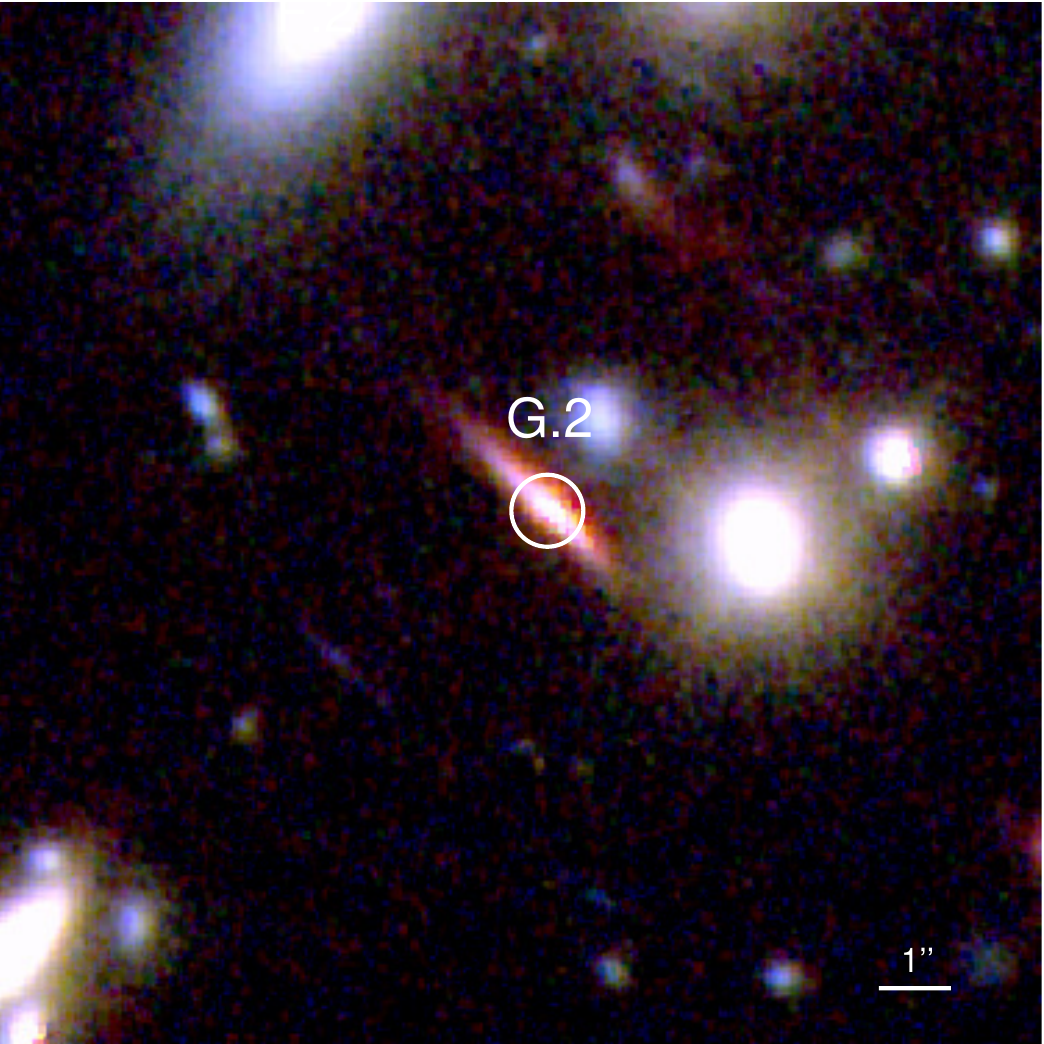}\ 
         \includegraphics[width=0.24\textwidth]{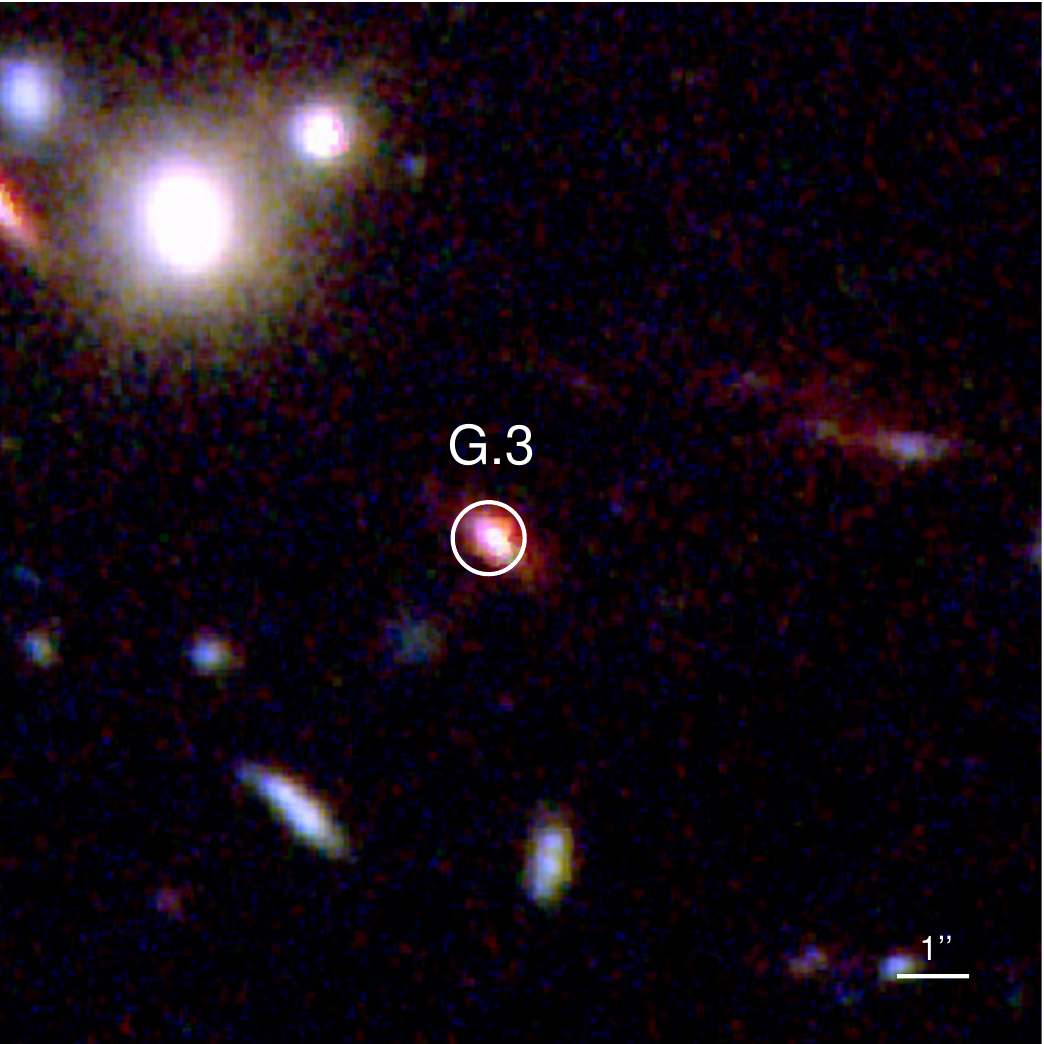}\\
                 \includegraphics[width=0.24\textwidth]{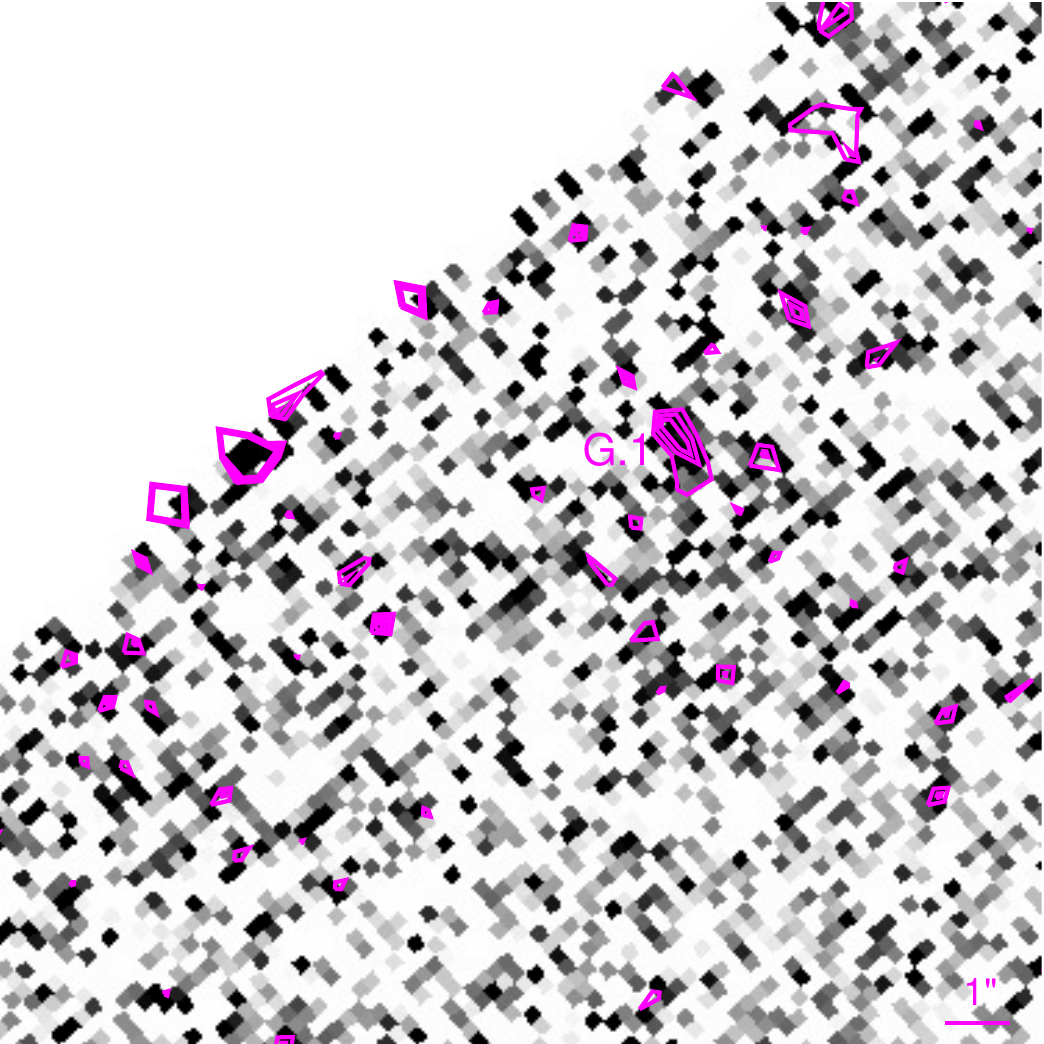}\ 
        \includegraphics[width=0.24\textwidth]{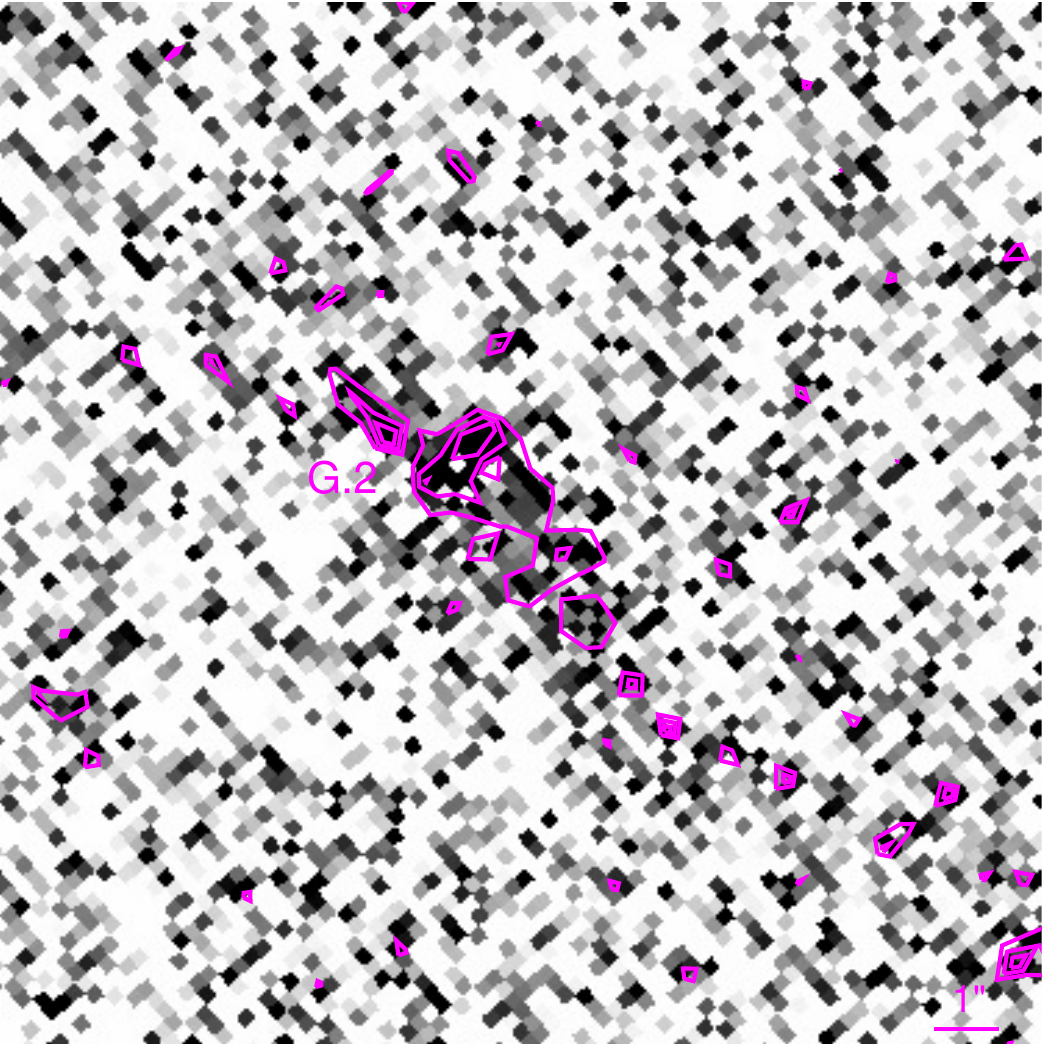}\ 
         \includegraphics[width=0.24\textwidth]{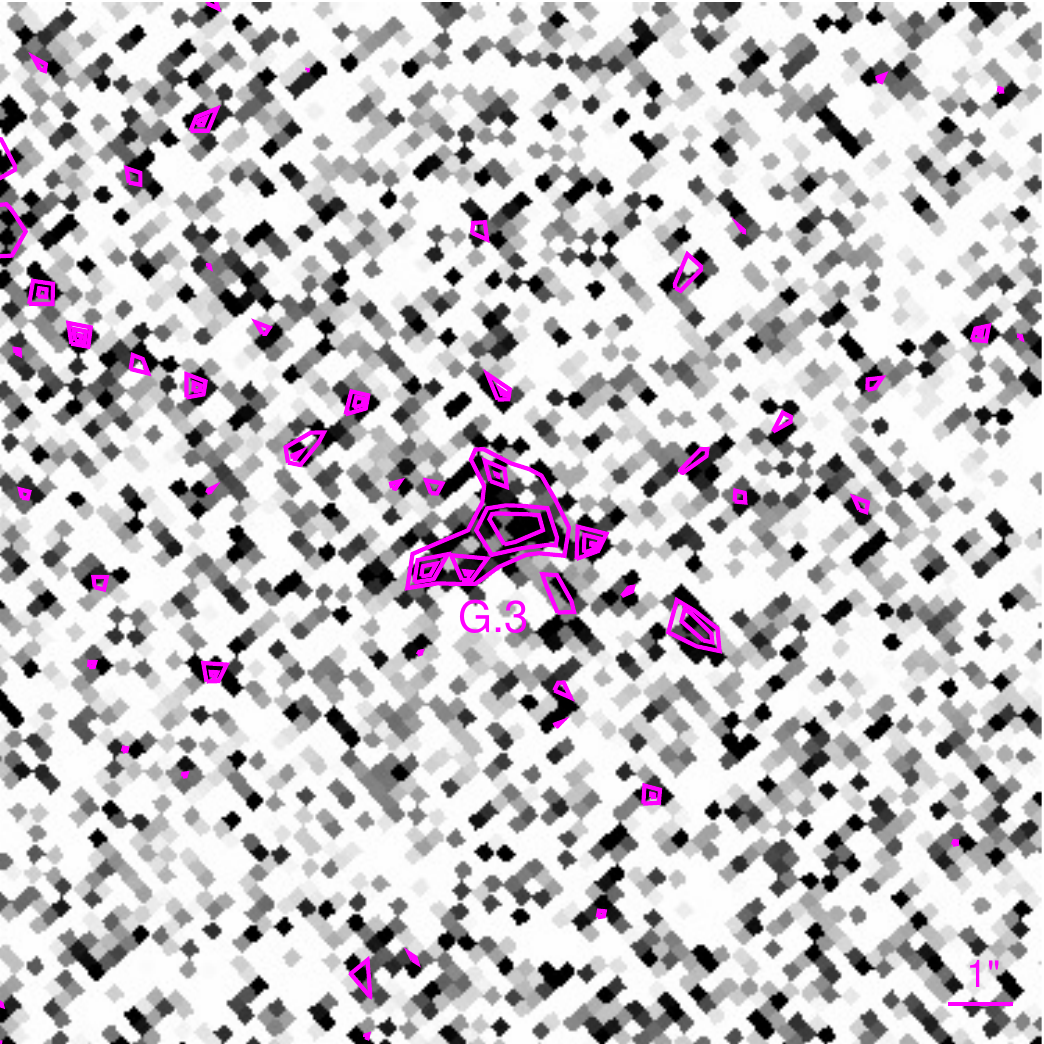}\\
		\includegraphics[width=0.7\textwidth]{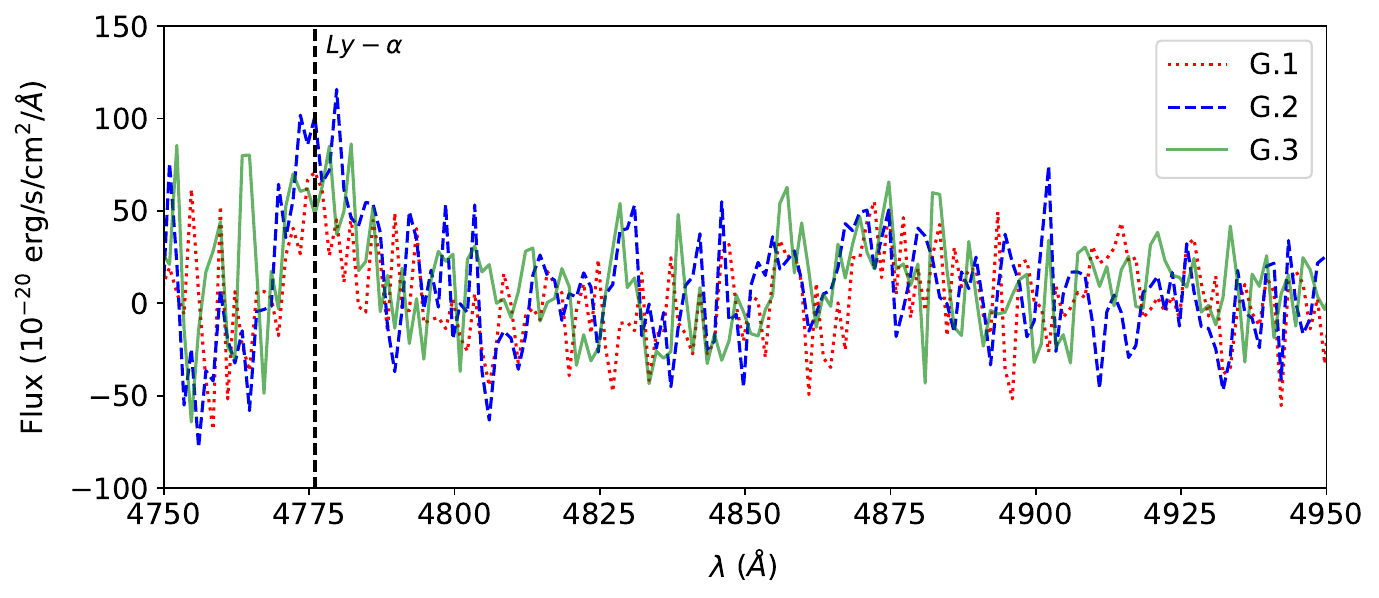}
\caption{\textbf{MS\,0451 --} System\ G ($z = 2.93$) identified by \citet{takata2003}, and spectroscopically confirmed with MUSE.
\emph{Top:} Composite colour \emph{HST} images. 
\emph{Middle:} \textsc{muselet} narrow-band datacube stamp at the wavelength of the maximum emission of the source, i.e. $\lambda=4780$\,{\AA}. The contours are displayed in magenta for 2, 3 and 4\,$\times10^{-20}$\, erg/s/arcsec$^{2}$.
\emph{Bottom:} MUSE extracted spectra of the multiple images.}
\label{detail_spec_G}
\end{center}
\end{figure*}

\normalsize
We here provide the MUSE spectra of the newly identified multiple image systems in MS\,0451, System\ R and S, and display each system on a composite colour \emph{HST} stamp in Fig.\,\ref{detail_spec_RS}. We provide similar information for System\ G in Fig.\,\ref{detail_spec_G}. We refer the reader to \cite{caminha2019} for RX\,J2129 and MACS\,J2129.

\bsp	
\label{lastpage}
\end{document}